\RequirePackage{ifpdf}
\documentclass[paper,notoc]{JHEP3}
\pdfoutput=1
\usepackage{amssymb}   % for math
\usepackage{amsmath}
\usepackage{graphicx}
\usepackage{booktabs} % for book-like tables

\allowdisplaybreaks
\newcommand{\ord}{\begin{cal}O\end{cal}}

\def\beq{\begin{equation}}
\def\eeq{\end{equation}}
\def\bea{\begin{eqnarray}}  
\def\eea{\end{eqnarray}} 
\def\bsp#1\esp{\begin{split}#1\end{split}}
\def\dd{\textrm{d}}

\def\beq{\begin{equation}}
\def\eeq{\end{equation}}
\def\bsp#1\esp{\begin{split}#1\end{split}}

\newcommand{\nnnlo}[0]{N$^3$LO }

\title{
High precision determination of the gluon fusion Higgs boson cross-section at the LHC
}

\author{Charalampos Anastasiou$^a$, Claude Duhr$^{b,c}$\footnote{On leave from the `Fonds National de la Recherche Scientifique' (FNRS), Belgium.}, Falko Dulat$^a$,
  Elisabetta Furlan$^{a}$, Thomas Gehrmann$^e$, 
Franz Herzog$^f$, 
Achilleas Lazopoulos$^a$, 
Bernhard Mistlberger$^b$\\
{}$^a$Institute for Theoretical Physics, ETH Z\"urich,
  8093 Z\"urich, Switzerland 
\\
{}$^b$Theoretical Physics Department, CERN, Geneva, Switzerland\\
{}$^c$Center for Cosmology, Particle Physics and Phenomenology (CP3),\\
\phantom{{}$^c$}Universit\'e catholique de Louvain,\\  
\phantom{{}$^c$}Chemin du Cyclotron 2, 1348 Louvain-La-Neuve, Belgium
\\
{}$^e$Physik-Institut, Universit\"at Z\"urich, 
Winterthurerstrasse 190, 8057 Z\"urich, Switzerland 
{}$^f$Nikhef, Science Park 105, NL-1098 XG Amsterdam, The Netherlands
}

\preprint{CP3-16-01, ZU-TH 27/15, NIKHEF 2016-004, CERN-TH-2016-006}

\abstract{ 
We present the most precise value for the Higgs boson cross-section in 
the gluon-fusion production mode at the LHC. Our result is based on a perturbative 
expansion through N$^3$LO in QCD, in an effective theory where the 
top-quark is assumed to be infinitely heavy, while all other Standard Model 
quarks are massless. We combine this result with QCD corrections to the 
cross-section where all finite quark-mass effects are included exactly through NLO. 
In addition, electroweak corrections  and the first corrections in the inverse mass of the top-quark 
are incorporated at three loops. We also investigate the effects of threshold 
resummation, both in the traditional QCD framework and following a SCET approach, 
which resums a class of $\pi^2$ contributions to all orders.  We assess the uncertainty 
of the cross-section from missing higher-order corrections due to both
perturbative QCD effects
beyond  N$^3$LO and unknown mixed QCD-electroweak effects. 
In addition, we determine the sensitivity of the cross-section to the choice of parton 
distribution function (PDF) sets and to the parametric uncertainty in the strong coupling constant
and quark masses. For a Higgs mass of $m_H = 125~{\rm GeV}$ and an LHC center-of-mass energy of 
$13~{\rm TeV}$, our best prediction for the gluon fusion cross-section is
\[
\boxed{
\sigma = 48.58\,{\rm pb} {}^{+2.22\, {\rm pb}\, (+4.56\%)}_{-3.27\, {\rm pb}\, (-6.72\%)} \mbox{ (theory)} 
\pm 1.56 \,{\rm pb}\, (3.20\%)  \mbox{ (PDF+$\alpha_s$)} 
}
\]
}


\begin{document}

\section{Introduction}
\label{sec:introduction}

With the discovery of the Higgs boson~\cite{Aad:2012tfa,Chatrchyan:2012xdj}, the Large Hadron Collider (LHC) has achieved a major landmark 
in science. 
Indeed, we now have conclusive evidence that space is filled with the Higgs field and 
that the mass of elementary particles is not an ad-hoc concept, but an elaborate 
outcome of the mechanism of spontaneous symmetry breaking.    
Moreover, with the Higgs boson the Standard Model is a mathematically self-consistent 
theory, and it can be used to formulate physically credible predictions at extremely high energies, many orders 
of magnitude higher than what we can probe with man-made experiments. 
This is a great triumph of theoretical physics. 

Besides the success of the Standard Model as a theory of electroweak interactions, it is a phenomenologically incomplete theory, and it needs to be extended in order to obtain
a satisfying description of all known physics, including cosmology. It is unclear at 
what energy the Standard Model will stop being a good theory and it will require 
the introduction of new laws of physics. If open questions, such as
for example the origin of dark matter, are related to the question
 of the origin of mass of elementary particles,  then it is 
likely that Higgs phenomena will differ quantitatively from Standard Model expectations. 
% new text
%Despite the success of the Standard Model as a theory describing the properties 
%and interactions of all visible matter, many outstanding issues remain unaddressed --
%the unnaturally light mass of the Higgs, a description of dark matter, the origin 
%of neutrino masses just to mention a few. 
%A larger, unified theoretical framework could exist, that accounts for a broader set 
%of observed phenomena and from which the Standard Model arises as a low-energy 
%theory. The scale at which the Standard Model needs to be superseded by 
%this new model is still unknown. The fine-tuning problem of the Higgs
%mass seems however to point towards a scale of a few TeV, potentially within the 
%reach of the LHC. The Higgs boson itself could be a tool for exploring the 
%new beyond-the Standard Model sector.
%Indeed, if open questions such as the origin of dark matter 
%are related to the question
%of the origin of mass of elementary particles, it is 
%likely that the Higgs boson also plays a role in the new sector. 
%Interactions with new particles, or a more complicated nature of the Higgs boson 
%than the one given by the minimal interpretation of electroweak symmetry breaking 
%in the Standard Model, would imply that Higgs
%phenomena  differ quantitatively from Standard Model expectations. 
%
We should
therefore view the Higgs boson discovery as the foundation of a long-term precision physics 
program measuring the properties of the Higgs boson.  
This program may yield direct or indirect evidence of physics beyond the Standard Model, 
and it requires the measurement of the mass, spin/parity, width, branching ratios and production 
rates of the Higgs boson.  All of the above are predicted or constrained in the Standard Model 
and its  viable extensions. 
The success of the program will rely crucially on the combination of highly precise experimental data with 
equally accurate theoretical predictions.

The purpose of this article is to supply a key ingredient to upcoming high-precision studies of the 
 Higgs boson by providing  the most accurate determination of the Higgs production 
cross-section in gluon fusion.
Higgs production in gluon fusion is mediated mainly through a top-quark loop~\cite{Georgi:1977gs}. 
Despite the absence of a tree-level contribution (which makes gluon fusion a pure quantum process), 
it is the dominant production mode of the Higgs boson due to the large gluon luminosity and the size of the top-quark Yukawa coupling. 
Reliable predictions of this process require the inclusion of higher-order corrections, both from the QCD and the electroweak sectors of the Standard Model. 
The phenomenological importance of these corrections can be seen from the large
size of the next-to-leading order (NLO) QCD
corrections~\cite{Dawson:1990zj,Graudenz:1992pv,Djouadi:1991tka,Spira:1995rr,Harlander:2005rq,Aglietti:2006tp,Bonciani:2007ex,Anastasiou:2006hc,Anastasiou:2009kn}, 
which almost double the original leading order (LO) prediction~\cite{Georgi:1977gs}. 
The large size of the NLO corrections indicate potentially significant 
contributions from even higher perturbative orders, thus resulting in a substantial 
theoretical uncertainty on the gluon-fusion cross-section. 
This uncertainty is difficult to quantify conventionally
by varying nuisance parameters in the theoretical prediction such as
renormalization and factorization scales. 

The past twenty years have seen substantial theoretical advances in the perturbative description of Higgs 
production in gluon fusion, using a multitude of techniques and aiming 
for various directions of improvement. 
The theoretical description of gluon fusion is rendered particularly
challenging by the fact that the Born process is 
already a one-loop process involving two mass scales (the masses of
the top quark and the Higgs boson), 
such that higher-order corrections will involve multi-scale multi-loop amplitudes. This challenge can be overcome by 
integrating out the top quark at the level of the Standard Model Lagrangian. This procedure results in an effective field theory (EFT)~\cite{Wilczek:1977zn,Shifman:1978zn,Inami:1982xt,Spiridonov:1988md} 
containing a tree-level coupling of the Higgs boson to gluons.
 This EFT can be matched onto the full
 Standard Model in a systematic manner, resulting in corrections from higher orders in perturbation theory to the Wilson coefficients~\cite{Chetyrkin:1997un,Schroder:2005hy,Chetyrkin:2005ia} and 
from subleading terms in the large mass expansion~\cite{Harlander:2009mq,Pak:2009dg}. 
In the EFT framework the coefficient function for inclusive Higgs production in gluon fusion depends only on the ratio of Higgs boson mass $m_H$ to the partonic 
center-of-mass energy $\sqrt{{s}}$, usually expressed through the variable $z=m_H^2/s$. By 
comparing the full NLO QCD 
expression for the gluon fusion cross-section with the EFT result, 
one observes that a very good approximation of the full NLO prediction
can be obtained by re-weighting the EFT prediction with the ratio
of LO predictions in the full and effective theories. 
This re-weighting is commonly applied to all predictions obtained within the EFT framework. 

NNLO corrections in the EFT~\cite{Anastasiou:2002yz,Harlander:2002wh,Ravindran:2003um} 
turn out to be substantial, albeit smaller than at NLO.
This slow convergence pattern entails the risk that the estimation of
the uncertainty at NNLO, obtained by varying the
renormalization and factorization scales,
may be misleading.  
To settle the size of the QCD corrections, additional information about the behavior of the perturbative expansion beyond
NNLO is necessary.

A part of  these corrections arises from the region of phase-space
where the Higgs boson is produced at or near to its kinematical
threshold, $z\to 1$. In this region, contributions from soft and collinear gluon emission can be resummed to all orders in
the coupling constant, either using Mellin space methods
~\cite{%css,
Sterman:1986aj,Collins:1988ig,%catani
Catani:1996yz,Moch:2005ba,Catani:2014uta} or within soft-collinear effective field theory 
(SCET)~\cite{Bauer:2002nz,Beneke:2002ph,Becher:2006mr,Idilbi:2006dg,Becher:2014oda}. 
Always working in the EFT framework, either method enables the resummation of logarithmically-enhanced threshold corrections  to the gluon fusion process to high logarithmic order~\cite{%resum
Catani:2003zt,Bonvini:2014joa,%scethiggs
Ahrens:2008nc}. 
Both methods agree to a given formal logarithmic accuracy, but their
results can in principle differ~\cite{Sterman:2013nya,Bonvini:2014qga}
by non-logarithmic terms. 

The combination of predictions at fixed order with threshold
resummation (together with electroweak
corrections~\cite{Aglietti:2004nj,Actis:2008ug,Actis:2008ts,Anastasiou:2008tj},
bottom and charm quark contributions through NLO and subleading mass corrections
at NNLO~\cite{Harlander:2009mq, Pak:2009dg, Harlander:2009bw, Harlander:2009my}) provided the default predictions for the interpretation of 
Higgs production data in the LHC 
Run 1~\cite{Dittmaier:2011ti,Dittmaier:2012vm,Heinemeyer:2013tqa}. 
Besides uncertainties from the parametrization of  parton 
distributions and the values of the strong coupling and quark masses, 
these predictions were limited in accuracy by missing
terms at N$^3$LO in the fixed-order expansion.  
By expanding resummed predictions in powers of the coupling constant, logarithmic terms in perturbative orders 
beyond NNLO can be extracted. These were used (often combined with the knowledge of the high-energy $z\to 0$ 
behaviour of the coefficient function~\cite{Hautmann:2002tu}) to obtain estimates of the fixed-order gluon fusion cross-section at 
N$^3$LO and beyond~\cite{Ball:2013bra,Bonvini:2014jma,deFlorian:2014vta}. Although individual results for these estimates typically quoted a very small residual 
uncertainty, the scatter among different estimates was quite substantial, thereby putting serious 
doubts on the reliability of any such estimation procedure. These ambiguities were resolved by the recent calculation of the full N$^3$LO QCD 
corrections to the gluon fusion process~\cite{Anastasiou:2015ema}, which are the key ingredient to the work
presented here. 

The gluon-fusion cross-section in N$^3$LO QCD in the EFT approach receives (besides a network of lower order 
renormalization and mass-factorization terms~\cite{%loword
Anastasiou:2012kq, Hoschele:2012xc, Buehler:2013fha,%mvv
Moch:2004pa,Vogt:2004mw})
contributions from four types of 
processes, ranging at fixed sum of loops and external legs from three-loop virtual corrections to the $ggH$-vertex to triple 
real radiation corrections from processes like $gg\to H ggg$, and denoted 
respectively as VVV, (RV)$^2$, RVV, RRV, RRR. While the three-loop virtual corrections were already known 
for quite some time~\cite{Baikov:2009bg,Gehrmann:2010ue}, new technical advances were needed in order to evaluate all the contribution from the different real-radiation subprocesses, 
either in closed from or as a high-order expansion around the threshold limit that is sufficient to precisely account for the 
full $z$-dependence. Based on the two-loop matrix elements for Higgs-plus-jet production~\cite{Gehrmann:2011aa}, closed 
expressions for the RVV contributions could be obtained by direct 
integration~\cite{Duhr:2014nda,Dulat:2014mda}. In the same way, it was possible to derive closed expressions for the (RV)$^2$ contribution~\cite{Anastasiou:2013mca,Kilgore:2013gba}. The major challenge in the 
RRV and RRR processes are the very intricate phase space integrals for double real radiation at one loop and triple real 
radiation at tree level. These phase-space integrals  can be related to specific cuts of loop integrals using reverse-unitarity~\cite{Anastasiou:2002yz,Anastasiou:2003ds,Anastasiou:2002qz,Anastasiou:2003yy}, 
allowing the application of modern integral reduction techniques~\cite{Tkachov:1981wb,Chetyrkin:1981qh,Laporta:2001dd} that express all relevant phase-space integrals by a limited set of master integrals, which are functions of $z$. With the same integral reduction techniques,
differential equations~\cite{Remiddi:1997ny,Gehrmann:1999as,Henn:2013pwa} 
can be derived for the master integrals. By solving these differential equations (either in closed 
form~\cite{Anzai:2015wma}, or as an expansion in $z$) 
for appropriate 
boundary conditions, the direct integration of the master integrals can be circumvented. 
The use of these techniques enabled the computation of the 
RRV and RRR~\cite{Anastasiou:2013srw} contributions at threshold. Combining them with the two-loop correction to the 
 soft-gluon current~\cite{Duhr:2013msa,Li:2013lsa} enabled first breakthroughs with the N$^3$LO threshold cross-section~\cite{Anastasiou:2014vaa,Li:2014bfa,Li:2014afw} and the 
 first beyond-threshold term~\cite{Anastasiou:2014lda}. More recently, the
  systematic expansion of the RRV~\cite{Anastasiou:2015yha} and RRR contributions to 
 very high orders in $z$ enabled the calculation of the full  N$^3$LO
 gluon fusion cross-section~\cite{Anastasiou:2015ema}. 

In this publication, we combine the N$^3$LO cross-section in the EFT
with the previously available state-of-the-art predictions for other types of 
 corrections (electroweak, mass effects, resummation) to obtain a highly precise theoretical description of 
 the inclusive Higgs production cross-section in gluon fusion.
We also assess remaining theoretical 
uncertainties on 
 the cross-section. 
Our results will form a cornerstone to precision studies of the Higgs boson in the upcoming high-energy and high-luminosity data taking periods of the CERN LHC. 
 
 Our paper is structured as follows: in Section~\ref{sec:setup} we present our setup and summarize the different contributions that we include into our prediction. In Section~\ref{sec:EFT} we study the phenomenological impact of QCD corrections through N$^3$LO in the large $m_t$-limit. We investigate the missing higher-order effects and threshold resummation in the EFT in Section~\ref{sec:MHO}. Effects due to quark masses and electroweak corrections are studied in Sections~\ref{section:top_quark_mass_effects} and~\ref{sec:EW}, and we assess the uncertainty on the cross-section due to PDFs and the strong coupling constant in Section~\ref{sec:pdfs}. In Section~\ref{sec:recommendation} we combine all effects and present our recommendation for the most precise theoretical prediction of the inclusive Higgs cross-section. In Section~\ref{sec:conclusion} we draw our conclusions. We include appendices where we present the coefficients appearing in the threshold expansion of the N$^3$LO coefficient, as well as tables summarizing our results for a variety of different Higgs masses and collider energies.

%%% Local Variables: 
%%% mode: latex
%%% TeX-master: "paper"
%%% End: 

%\section{Input parameters}
%\input{input_parameters.tex}
\section{Setup}
\label{sec:setup}
The inclusive hadronic cross-section $\sigma$ for Higgs production in gluon fusion can be calculated as the convolution integral
\begin{equation}
\label{eq:sigma}
%\sigma = \sum_{ij} \int dx_1\, dx_2\, f_i(x_1)\,f_j(x_2)\, \hat{\sigma}_{ij}(z)\,,
\sigma = \tau \sum_{ij} \left( f_i \otimes f_j  \otimes 
\frac{\hat{\sigma}_{ij}(z)}{z}
\right)\left(\tau \right) \,,
\end{equation} 
where  $\hat{\sigma}_{ij}$ are the partonic cross-sections for
producing a Higgs boson from a scattering of partons $i$ and $j$, and $f_i$ and
$f_j$ are the corresponding parton densities. We have defined the ratios
\beq
\tau = \frac{m_H^2}{S} {\rm~~and~~} z = \frac{m_H^2}{s}\,,
\eeq
where $m_H$, $s$ and $S$ denote the Higgs mass and the  
squared partonic and hadronic center-of-mass energies.
The convolution of two functions is  defined as
\begin{equation}\label{eq:convolution}
(h \otimes g)(\tau) = \int_0^1 \dd x \,\dd y\, h(x)\,g(y)\, \delta(\tau - x y)\,.
\end{equation}
In the Standard Model (SM) the Higgs boson is predominantly produced through the annihilation of virtual top and bottom quarks, as 
well as $W$ and $Z$ bosons, produced in gluon fusion. All of these channels are greatly enhanced by QCD corrections, and also electroweak corrections are important. Hence, having good control over higher-order corrections in perturbation theory, both in the QCD and electroweak sectors, is of paramount importance to make precision predictions for Higgs production in the framework of the SM. We note that non-perturbative contributions to the inclusive Higgs boson production cross-section 
are suppressed by powers of $(\Lambda/m_H)$, with $\Lambda$ being the QCD scale. For the Drell-Yan process, 
the linear non-perturbative correction could be shown to vanish~\cite{Beneke:1995pq,Akhoury:1997pb}, such that the 
leading power correction term is quadratic and potentially relevant at low invariant masses. 
For the inclusive Higgs production cross-section, no thorough investigation of the power corrections has 
been performed up to now, but even the linear term would be 
a per-mille-level correction. 

The goal of this paper is to provide the most precise predictions for the inclusive hadronic Higgs production cross-section in the SM. We use state-of-the-art precision computations for electroweak and QCD corrections to inclusive Higgs production and combine them into the most precise theoretical prediction for the Higgs cross-sections available to date. The master formula that summarizes all the ingredients entering our prediction for the (partonic) cross-sections is
\beq\label{eq:master}
\hat{\sigma}_{ij} \simeq R_{LO}\,\left(\hat{\sigma}_{ij,EFT} + \delta_{t}\hat{\sigma}_{ij,EFT}^{NNLO} + \delta\hat{\sigma}_{ij,EW}\right) + \delta\hat{\sigma}_{ij,ex;t,b,c}^{LO}+\delta\hat{\sigma}_{ij,ex;t,b,c}^{NLO}\,.
\eeq

Equation~\eqref{eq:master} includes QCD corrections to the production cross-section in an effective theory where the top quark is infinitely heavy and has been integrated out. In this limit, the Higgs boson couples directly to the gluons via an effective operator of dimension five,
\beq\label{eq:L_eff}
\mathcal{L}_{\text{eff}}=\mathcal{L}_{\textrm{SM},5}-\frac{1}{4}C \, H\, G_{\mu\nu}^a G_a^{\mu\nu},
\eeq
where $H$ is the Higgs boson field, $G_{\mu\nu}^a$ is the gluon field strength tensor and $\mathcal{L}_{SM,5}$ denotes the SM Lagrangian with $N_f=5$ massless quark flavours. The Wilson coefficient $C$ is obtained by matching the effective theory to the full SM in the limit where the top quark is infinitely heavy. In Appendix~\ref{app:WC_MSbar-OS} 
we give its analytic expression through \nnnlo in the $\overline{\textrm{MS}}$ and OS schemes~\cite{Chetyrkin:1997un,Schroder:2005hy}, 
in the five-flavour effective theory with the strong coupling constant decoupled. 

QCD corrections to the production cross-section $\hat{\sigma}_{ij,EFT}$ in the heavy-top limit have been computed at NLO~\cite{Dawson:1990zj,Graudenz:1992pv,Djouadi:1991tka} and NNLO~\cite{Anastasiou:2002yz,Harlander:2002wh,Ravindran:2003um}. Recently also the N$^3$LO corrections have become available~\cite{Anastasiou:2015ema}. One of the main goals of this work is to combine the N$^3$LO corrections in the large-$m_t$ limit with other effects that can provide corrections at a similar level of accuracy, in particular  quark-mass effects and electroweak corrections. We also investigate the impact of the resummation of threshold logarithms, both within the frameworks of exponentiation of large logarithms in Mellin space and using soft-collinear effective theory (SCET).

While the production cross-section is known to high accuracy in the framework of the
 effective theory, reaching a similar level of accuracy when including quark-mass effects (also from bottom and charm quarks) is currently beyond our technical capabilities. Nonetheless, various quark-mass effects have been computed, which we consistently include into our prediction~\eqref{eq:master}. First, it was already observed at LO and NLO that the validity of the effective theory can be greatly enhanced by rescaling the effective theory by the exact LO result. We therefore rescale the cross-section $\hat{\sigma}_{ij,EFT}$ in the effective theory by the ratio
\beq
R_{LO} \equiv \frac{\sigma_{ex;t}^{LO}}{\sigma_{EFT}^{LO}}\,,
\label{eq:KLO}
\eeq
where $\sigma_{ex;t}^{LO}$ denotes the exact (hadronic) LO cross-section in the SM with a massive top quark and $N_f=5$ massless quarks. Moreover, at LO and NLO we know the exact result for the production cross-section in the SM, including all mass effects from top, bottom and charm quarks. We include these corrections into our prediction via the terms $\delta\hat{\sigma}_{ij,ex;t,b,c}^{(N)LO}$ in eq.~\eqref{eq:master}, consistently matched to the contributions from the effective theory to avoid double counting. As a consequence, eq.~\eqref{eq:master} agrees with the exact SM cross-section (with massless $u$, $d$ and $s$ quarks) through NLO in QCD. Beyond NLO, we only know the value of the cross-section in the heavy-top effective theory. We can, however, include subleading corrections at NNLO in the effective theory as an expansion in the inverse top mass~\cite{Harlander:2009mq, Pak:2009dg, Harlander:2009bw, Harlander:2009my}. These effects are taken into account through the term $\delta_{t}\hat{\sigma}_{ij,EFT}^{NNLO}$ in eq.~\eqref{eq:master}, rescaled by $R_{LO}$.

Finally, we also include electroweak corrections to the gluon-fusion cross-section (normalised to the exact LO cross-section) through the term $\delta\hat{\sigma}_{ij,EW}$ in eq.~\eqref{eq:master}. Unlike QCD corrections, electroweak corrections have only been computed through NLO in the electromagnetic coupling constant $\alpha$~\cite{Aglietti:2004nj,Actis:2008ug,Actis:2008ts}. Moreover, mixed QCD-electroweak corrections, i.e., corrections proportional to $\alpha\,\alpha_s^3$, are known in an effective 
theory~\cite{Anastasiou:2008tj} valid in the limit where not only the top quark but also the electroweak bosons are much heavier than the Higgs boson. In this limit the interaction of the Higgs boson with the $W$ and $Z$ bosons is described via a point-like vertex coupling the gluons to the Higgs boson. Higher-order corrections in this limit can thus be included into the Wilson coefficient in front of the dimension-five operator in eq.~\eqref{eq:L_eff}.

In the remainder of this paper we give a detailed account of all the ingredients that enter our best prediction for the inclusive gluon-fusion cross-section. Furthermore, we carefully analyze the residual uncertainty associated with all of these contributions. In this way we obtain the most precise theoretical prediction for the Higgs production cross-section available to date.

We conclude this section by summarizing, for later convenience, the default values of the input parameters and the concrete choices for PDFs and quark-mass schemes used in our numerical studies. In particular, we investigate three different setups, which are summarized in Tab.~\ref{tab:setup1}--\ref{tab:setup3}. Note that we use NNLO PDFs even when we refer to lower order terms of the cross-section, unless stated otherwise. The values for the quark masses used are in accordance with the recommendations of the Higgs Cross Section Working Group~\cite{Denner:2047636}, wherein the top-quark mass was selected to facilitate comparisons with existing experimental analyses at LHC, Run 1\footnote{Note that the current world average $m_t^{\textrm{OS}}=173.2$ GeV is within the recommended uncertainty of $1$ GeV from the proposed $m_t^{\textrm{OS}}=172.5$ GeV that we use here.}.

%------------------------------------------------------- pretty tables
\begin{table}[!thb]
\tiny\setlength{\tabcolsep}{1pt}
%
%    table 1
%
\begin{minipage}{.33\linewidth}
\centering

\caption{Setup 1}
\label{tab:setup1}

\medskip

\begin{tabular}{cc}
    \toprule
    
    $\sqrt{S}$	&  13\,TeV                      \\
$m_h$  		&  125\,GeV                      \\
PDF 			&  {\tt PDF4LHC15\_nnlo\_100}                      \\
$a_s(m_Z)$ 	&  0.118                    \\
$m_t(m_t)$	&  162.7 \,GeV ($\overline{\textrm{MS}}$)\\
$m_b(m_b)$	&  4.18 \,GeV ($\overline{\textrm{MS}}$)\\
$m_c(3GeV)$	&  0.986\,GeV  ($\overline{\textrm{MS}}$)\\
$\mu=\mu_R=\mu_F$	&  62.5\,GeV  ($=m_H/2$)\\
    \bottomrule
\end{tabular}
\end{minipage}\hfill
%
%    table 2
%
\begin{minipage}{.33\linewidth}
\centering

\caption{Setup 2}
\label{tab:setup2}

\medskip

\begin{tabular}{cc}
    \toprule
    
    $\sqrt{S}$	&  13\,TeV                      \\
$m_h$  		&  125\,GeV                      \\
PDF 			&  {\tt PDF4LHC15\_nnlo\_100}                      \\
$a_s(m_Z)$ 	&  0.118                    \\
$m_t$	&  172.5\,GeV  (OS)\\
$m_b$	&  4.92 \,GeV (OS)\\
$m_c$	&  1.67 \,GeV (OS)\\
$\mu=\mu_R=\mu_F$	&  62.5 \,GeV ($=m_H/2$)\\
    \bottomrule
\end{tabular}
\end{minipage}\hfill
%
%    table 3
%
\begin{minipage}{.33\linewidth}
\centering

\caption{Setup 3}
\label{tab:setup3}

\medskip

\begin{tabular}{cc}
    \toprule
    
    $\sqrt{S}$	&  13\,TeV                      \\
$m_h$  		&  125\,GeV                      \\
PDF 			&  {\tt abm12lhc\_5\_nnlo}                      \\
$a_s(m_Z)$ 	&  0.113                    \\
$m_t(m_t)$	&  162.7 \,GeV  ($\overline{\textrm{MS}}$)\\
$m_b(m_b)$	&  4.18\,GeV  ($\overline{\textrm{MS}}$)\\
$m_c(3GeV)$	&  0.986 \,GeV ($\overline{\textrm{MS}}$)\\
$\mu=\mu_R=\mu_F$	&  62.5\, GeV ($=m_H/2$)\\
    \bottomrule
\end{tabular}
\end{minipage}\hfill
\end{table}

\section{The cross-section through N$^3$LO in the infinite top-quark limit}
\label{sec:EFT}

\subsection{The partonic cross-section at N$^3$LO in the heavy-top limit}
In this section we discuss the contribution $\hat{\sigma}_{ij,EFT}$ in eq.~\eqref{eq:master} from the effective theory where the top quark is infinitely heavy. This contribution can be expanded into a perturbative series in the strong coupling constant,
\beq
\label{eq:sigmaeff}
\frac{\hat{\sigma}_{ij,EFT}}{z} =\frac{\pi\,|C|^2}{8V}  \sum_{n=0}^\infty \eta^{(n)}_{ij}(z) \,a_s^n\,,
\eeq
where $V\equiv N_c^2-1$ is the number of adjoint $SU(N_c)$ colours, $a_s \equiv \alpha_s(\mu^2)/\pi$ denotes the strong coupling constant evaluated at a scale $\mu$ and $C$ is the Wilson coefficient introduced in eq.~\eqref{eq:L_eff}, which admits itself a perturbative expansion in the strong coupling~\cite{Chetyrkin:1997un,Schroder:2005hy,Chetyrkin:2005ia},
\beq\label{eq:WL_series}
C=a_s\,\sum_{n=0}^\infty C_n\, a_s^n\,.
\eeq
Here both the coefficients $C_n$ and the strong coupling are functions of a common scale $\mu$.
At LO in $a_s$ only the gluon-gluon initial state contributes, and we have
\beq
\eta_{ij}^{(0)}(z)~=~\delta_{ig}\,\delta_{jg}\,\delta(1-z)\,.
\eeq
QCD corrections beyond LO are also known. In particular, the perturbative coefficients $\eta^{(n)}_{ij}$ are known at NLO~\cite{Dawson:1990zj,Graudenz:1992pv,Djouadi:1991tka} and NNLO~\cite{Anastasiou:2002yz,Harlander:2002wh,Ravindran:2003um} in QCD. Recently, also the N$^3$LO corrections $\eta^{(3)}_{ij}$ have been computed~\cite{Anastasiou:2015ema}. As they are the main new addition in our computation, we briefly review the N$^3$LO corrections to the inclusive gluon fusion cross-section in the heavy-top limit in this section.

We follow the notation of ref.~\cite{Anastasiou:2014lda} and
we split the partonic cross-sections into a singular and a regular part,
\begin{equation}
\eta_{ij}^{(3)}(z)  = \delta_{ig}\,\delta_{jg}\,\eta_{gg}^{(3),\textrm{sing}}(z) + \eta_{ij}^{(3),\textrm{reg}}(z)\,.
\end{equation}
The singular contribution is precisely the cross-section at
threshold, also known as the soft-virtual cross-section. It contains the contributions from purely virtual three-loop corrections as well as from the emission of soft gluons~\cite{Li:2014bfa,Anastasiou:2014vaa,Li:2014afw,Moch:2005ky,Laenen:2005uz}. 
The regular term takes the form of a polynomial in $\log(1-z)$,
\beq\label{eq:eta_reg_log}
\eta_{ij}^{(3),\textrm{reg}}(z) = \sum_{m=0}^5\log^m(1-z)\,\eta_{ij}^{(3,m),\textrm{reg}}(z) \,,
\eeq
where the $\eta_{ij}^{(3,m),\textrm{reg}}(z)$ are holomorphic in a 
neighbourhood of $z=1$.
The functions $\eta_{ij}^{(3,m),{\rm reg}}$ for $m=5,4,3$ have been given in closed 
analytic form in ref.~\cite{Anastasiou:2014lda}.  
For $m=2,1,0$ no closed analytic expression is available in the literature so far (except for the $qq'$ channel~\cite{Anzai:2015wma}). In ref.~\cite{Anastasiou:2015ema} these coefficients were computed as an expansion around threshold to order 30 in $\bar{z}\equiv 1-z$. 
In Appendix~\ref{app:expansioncoefficients_numeric} we present the numerical values 
for the first 37 coefficients of the expansion, setting the renormalization and
factorization scales equal to the Higgs mass and substituting  $N_f=5$ for the number of 
light quark flavors and $N_c=3$ for the number of quark colours. Moreover, 
it was shown in ref.~\cite{Anastasiou:2015ema} that a truncation of the series
 at order ${\cal O}({\bar z}^5)$ yields a good
approximation to the hadronic cross-section. The first few terms of the 
expansion may be insightful for theoretical studies of perturbative QCD and 
discovering universality patterns in subleading terms of the soft expansion. We therefore provide 
the analytic results for the coefficients in the threshold expansion up to $\ord(\bar{z}^5)$ in  Appendix~\ref{app:expansioncoefficients_analytic}. 

In the rest of this section we study the numerical impact of the N$^3$LO corrections to the inclusive gluon fusion cross-section in the heavy-top limit. We start by studying the validity of approximating the cross-section by its threshold expansion and we quantify the uncertainty introduced by truncating the expansion after only a finite number of terms. We then move on and investigate the perturbative stability of $\hat{\sigma}_{ij,EFT}$ by studying the scale variation of the gluon-fusion cross-section at N$^3$LO in the heavy-top limit.

\subsection{Convergence of the threshold expansion at N$^3$LO}
As parts of the \nnnlo coefficient functions
$\eta_{ij}^{(3,m),\textrm{reg}}(z)$  have not yet been derived in closed analytic form and are
only known as  truncated series expansions in $\bar z $, it is important to assess how well these truncated 
power series approximate the exact result. In other words, we need to establish how well the threshold expansion
converges. Indeed, the partonic cross-sections $\hat{\sigma}_{ij,EFT}$
need to be convoluted with the partonic luminosities,
eq.~\eqref{eq:sigma}, and the convolution integrals receive 
in principle contributions down to values of $z\simeq \tau\simeq 10^{-4}$. Hence, assessing the residual uncertainty due to the truncation of the series is of utmost importance.

\begin{figure}[!t]
\begin{center}
\includegraphics[width=0.9\textwidth]{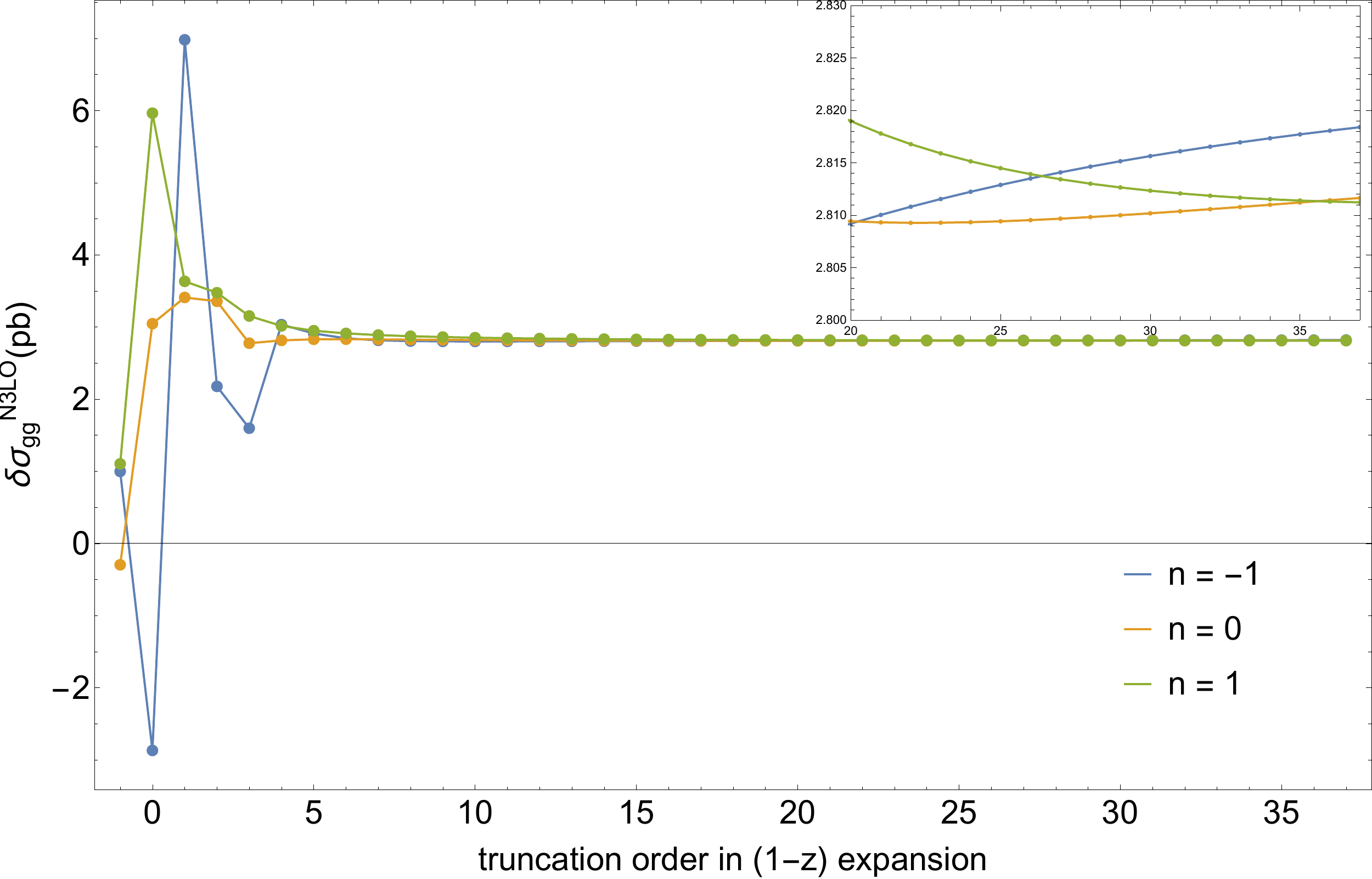}
\end{center}
\caption{
\label{fig:sigma_deformed}
The numerical effect in Setup 1 (see Tab.~\ref{tab:setup1}) of the \nnnlo
  corrections in the gluon-gluon channel as a function of the
  truncation order of the threshold expansion and for various values
  of the parameter $n$ in eq.~\eqref{eq:sigma_deformed}.
}
\end{figure}

\begin{figure}[!t]
\begin{center}
\includegraphics[width=0.8\textwidth]{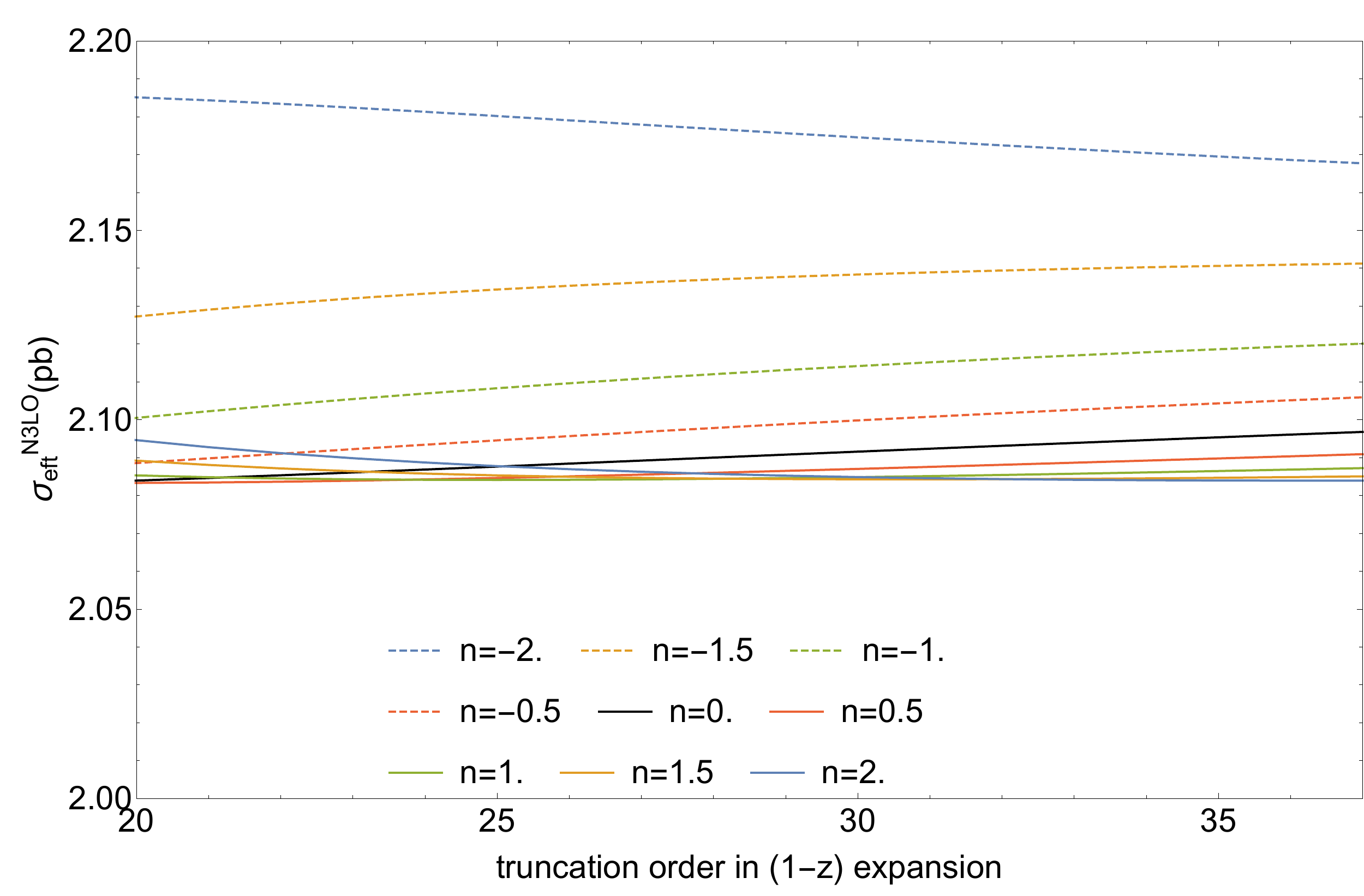}
\end{center}
\caption{ 
\label{fig:truncation_all_channels_variousN} 
The numerical effect in Setup 1 (see Tab.~\ref{tab:setup1}) of the \nnnlo
  corrections as a function of the
  truncation order of the threshold expansion and for various values
  of the parameter $n$ in eq.~\eqref{eq:sigma_deformed}. All channels are included.}
\end{figure}
In ref.~\cite{Anastasiou:2014lda,Herzog:2014wja} a method was introduced to study the convergence of the threshold expansion. 
We start by casting the hadronic cross-section in the large-$m_t$ limit in
the form
\begin{equation}
\label{eq:sigma_deformed}
%\sigma = \sum_{ij} \int dx_1\, dx_2\, f_i(x_1)\,f_j(x_2)\, \hat{\sigma}_{ij}(z)\,,
\sigma_{EFT} = \tau^{1+n} \sum_{ij} \left( f_i^{(n)} \otimes f_j^{(n)}  \otimes 
\frac{\hat{\sigma}_{ij,EFT}}{z^{1+n}}
\right)\left(\tau \right) \,,
\end{equation}
where 
\begin{equation}
f_i^{(n)}(z) \equiv \frac{f_i(z)}{z^n}\,.
\end{equation}
For $n=0$, we recover precisely the usual QCD factorization formula. For $n\neq0$, however, eq.~\eqref{eq:sigma_deformed} is a deformed, but equally valid and equivalent, formulation of the usual QCD factorization formula~\eqref{eq:sigma}. Indeed, it is easy to check that the hadronic cross-section $\sigma_{EFT}$ is independent of the arbitrary parameter $n$.
Expanding ${\hat{\sigma}_{ij,EFT}}/{z^{1+n}}$ into a series around $z=1$, however, introduces a dependence on $n$ order by order in the expansion, which only cancels once infinitely many terms in the series are summed up. Hence, if a truncated series is used to evaluate ${\hat{\sigma}_{ij,EFT}}/{z^{1+n}}$, the result will in general depend on $n$, and we can use the spread of the $n$-dependence as a quantifier for the convergence of the series.
In Fig.~\ref{fig:sigma_deformed} we show the \nnnlo 
contribution to the hadronic cross-section from the $gg-$channel. 
We observe that the hadronic cross-section is very stable with respect to the choice of the arbitrary parameter
$n$ after the first $\sim 5$ terms in the threshold
expansion. In ref.~\cite{Anastasiou:2015ema}  we observed a mild
growth of the cross-section at 
high orders of the threshold
expansion (see inlay of Fig.~\ref{fig:sigma_deformed} in ref.~\cite{Anastasiou:2015ema}). This is attributed to the presence of $\log z$ terms~\cite{Hautmann:2002tu} (and for $n>-1$ also global factors of $1/z^n$) 
which, after threshold expansion and convolution with the parton distributions, 
yield a small part of the cross-section. In Fig.~\ref{fig:truncation_all_channels_variousN} we show the convergence of the total cross-section, 
including all partonic channels, for a variety of different values of $n$, 
from the 20th term onwards in the expansion. While we observe good apparent convergence for $n>-1$, 
there remains a relatively large spread between the different curves for $n\le -1$. 
The qualitative difference between these two cases can be understood as follows: For $n>-1$, 
we absorb additional
factors of $1/z$ into the partonic cross-sections and expand them around $z=1$. This may result in a slower convergence of the partonic threshold expansion for small values of $z$. At the same time, however, the luminosities are multiplied with powers of $z$ which suppress the contribution from the region $z\sim0$ in the convolution~\eqref{eq:sigma_deformed}. The net effect is then a fast apparent convergence for $n>-1$. This has to be contrasted with the case $n\le -1$, where the luminosities are multiplied 
by factors of $1/z$, which enhance the contribution from the region $z\sim0$ in the convolution~\eqref{eq:sigma_deformed}. This leads to a slower apparent convergence, 
at least in the case where only a few terms are taken into account in the threshold expansion. While the spread between the different curves gives a measure for the quality of the convergence of the threshold expansion, we know of no compelling argument why any of this curves should be preferable over others at this order of the expansion. 
We observe, however, that the different curves agree among each other within a range of 0.1 pb, 
thereby corroborating our claim that the threshold expansion provides reliable results for the N$^3$LO cross-section.

\begin{figure}[!t]
\begin{center}
\includegraphics[width=0.9\textwidth]{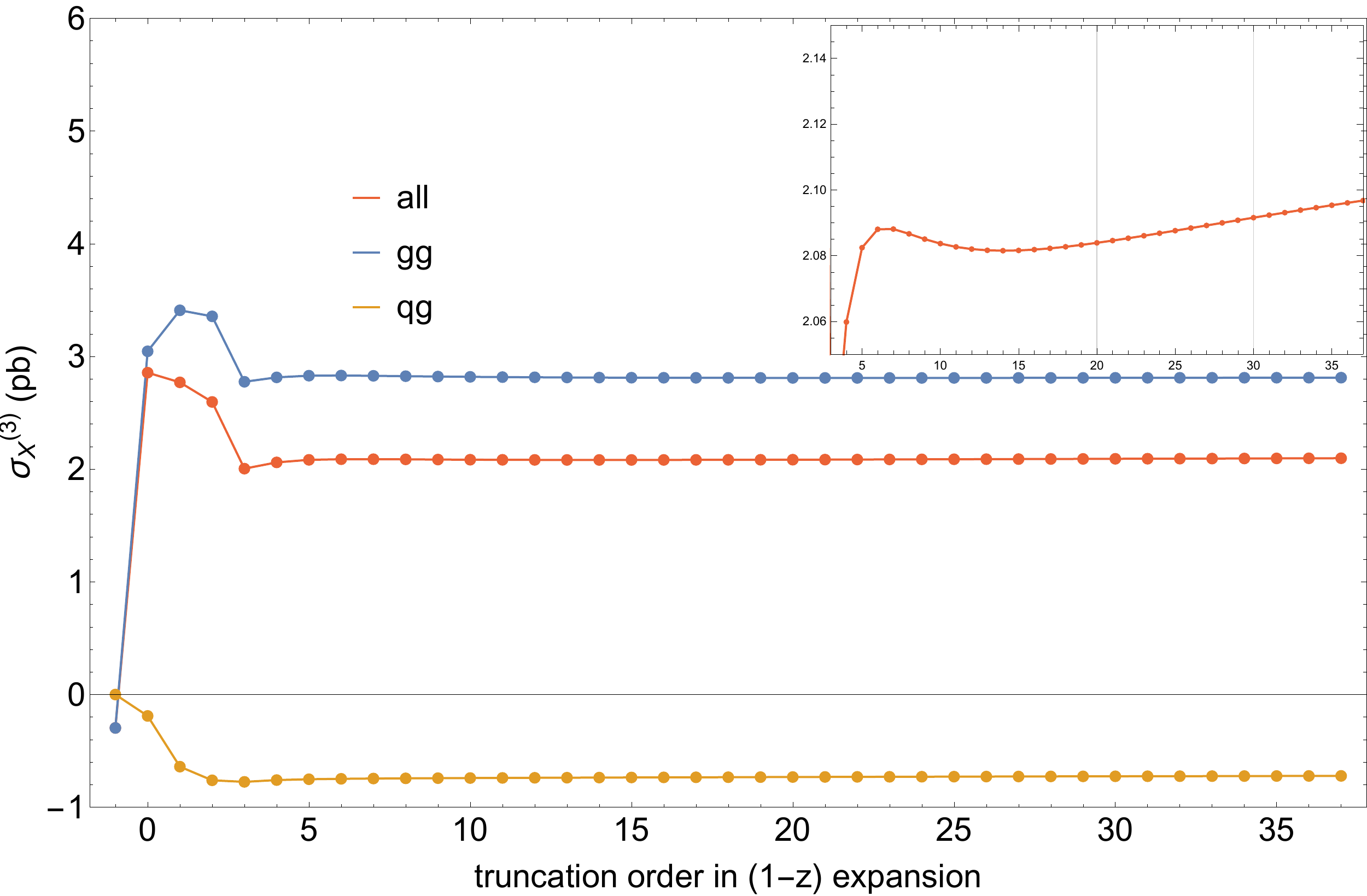}
\end{center}
\caption{
\label{fig:truncation_all_channels_absolute}
The numerical effect in Setup 1 of the \nnnlo\,
  correction in the main partonic channels and the total cross-section 
as a function of the truncation order in the threshold expansion, for $n=0$ in eq.~\eqref{eq:sigma_deformed}.
}
\end{figure}
\begin{figure}[!t]
\begin{center}
\includegraphics[width=0.8\textwidth]{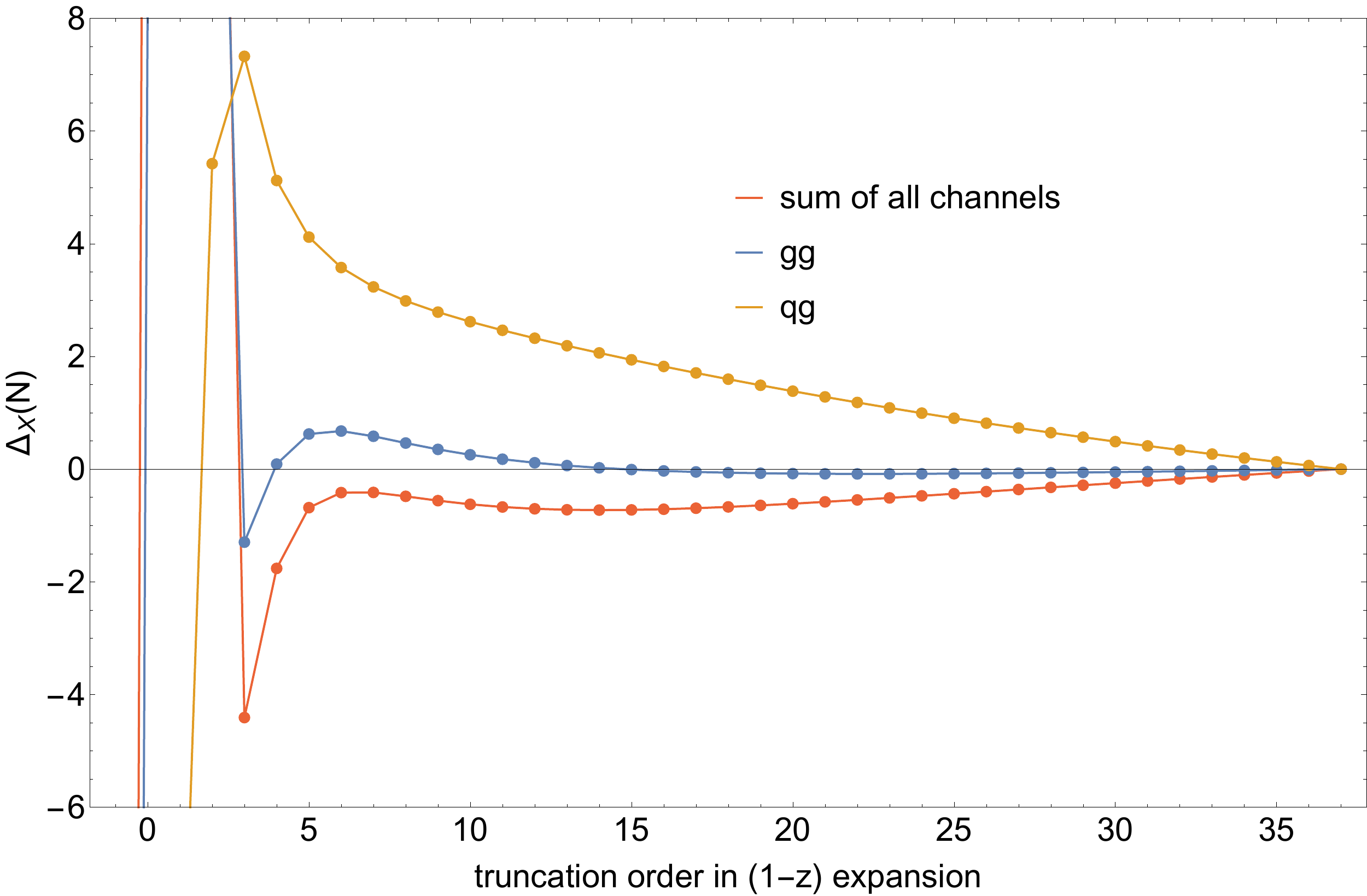}
\end{center}
\caption{ 
\label{fig:truncation_all_channels_relative} 
The ratios of eq.~\eqref{eq:ratio_convergence} for the
  convergence for the threshold expansion at \nnnlo for individual partonic channels, as well as for the full hadronic cross-section. The $qq$ and $qq'$ channels are negligible and are not shown in the plot.}
\end{figure}

In Fig.~\ref{fig:truncation_all_channels_absolute} we plot the \nnnlo 
corrections for the $gg$ and $qg$ 
channels\footnote{We sum of course over all possible quark and anti-quark flavours.}, as well as the total inclusive cross-section, as a function of the
truncation order (for $n=0$).  The quark-initiated channels contribute only a 
small fraction to the inclusive cross-section. 
The convergence of the threshold expansion for these channels is 
less rapid than for the dominant gluon-gluon channel. 
This is better demonstrated in Fig.~\ref{fig:truncation_all_channels_relative},
where we plot the ratio 
\begin{equation}
\label{eq:ratio_convergence}
\Delta_{X}(N) \equiv \frac{\sigma_{X,EFT}^{(3)}(N) -  \sigma_{X,EFT}^{(3)}(N_{\rm last}) }{
\sigma_{X,EFT}^{(3)}(N_{\rm last}) 
} \, 100\%\,.
\end{equation} 
Here,
$\sigma_{X,EFT}^{(3)}(N)$ 
denotes the contribution of the partonic channel $X$ to the \nnnlo
correction to the hadronic cross-section when computed through $\ord(\bar{z}^N)$ 
in the threshold expansion. $N_{\rm last}$ (equal to 37) is the highest
truncation  order used in our current computation.
Although the convergence of the quark-gluon and the quark channels is
rather slow, the total cross-section and the
convergence rate of the threshold expansion are dominated by the 
gluon-gluon channel. This enables us to obtain a reliable estimate of
the cross-section for Higgs production via gluon fusion, even though we have only included a finite number of terms in the threshold expansion. We remark,
however, that for quark-initiated processes such as Drell-Yan production  
a computation in closed form will most likely be necessary. 

Besides studying the $n$-dependence of the truncated power series, we have another way to assess the convergence of the expansion. In ref.~\cite{Anastasiou:2014lda} it was shown that the knowledge of the single-emission contributions at N$^3$LO~\cite{Anastasiou:2013mca,Kilgore:2013gba,Dulat:2014mda,Duhr:2014nda} and the three-loop splitting functions~\cite{Moch:2004pa,Vogt:2004mw} is sufficient to determine the coefficients $\eta_{ij}^{(3,m)}$ in the \nnnlo cross-section~\eqref{eq:eta_reg_log} exactly for $m=5,4,3$. Recently, also the double-emission contribution at one-loop has been computed in closed form~\cite{Dulat:2015xyz}. Using a similar analysis as for $m=5,4,3$ in ref.~\cite{Anastasiou:2014lda}, it has now been possible to determine also the coefficients with $m=2,1$ exactly for all partonic subchannels. As a consequence, we know all the logarithmically-enhanced terms in eq.~\eqref{eq:eta_reg_log} in closed form, and we only need to resort to a truncated threshold expansion for the constant term, $m=0$.
We can thus study the convergence of the threshold expansion for the coefficients of $\eta_{ij}^{(3,m)}$, $m\ge1$.
In particular, the use of the exact expressions instead of a truncated expansion 
for the logarithmically-enhanced contributions changes the \nnnlo correction to the
cross-section by
\begin{equation}
\left.  \sigma^{(3)}_{EFT} \right|_{{\rm expansion}} - \left.  \sigma^{(3)}_{EFT}
\right|_{{\rm full \, logs}} = 0.004\,\textrm{pb}\,.
\end{equation}
Hence, the difference between exact expressions or truncated power series for the coefficients with $m\ge1$ in eq.~\eqref{eq:eta_reg_log} is at the sub-per mille level, and thus completely negligible.

To summarize, we have investigated the convergence of the threshold expansion at N$^3$LO using two different methods. Both methods confirm our expectation that the threshold expansion provides a very good approximation to the exact result. 
The result of our analysis can be quantified by assigning a (conservative) uncertainty estimate to the truncation of the threshold expansion.
We assign an uncertainty due to the truncation of the threshold expansion which is as large 
as\footnote{In the estimate of the various components of the theoretical uncertainty 
that we carry out in these sections, we always give numerical results for Setup I. When considering different parameters 
(Higgs mass or collider energy, for example), we re-assess these uncertainties. For example, $\delta({\rm trunc})$ increases
from 0.11\% at 2 TeV to 0.38\% at 14 TeV.}. 
\begin{equation}\label{eq:trunc_error}
\delta({\rm trunc}) = 10 \times \frac{\sigma^{(3)}_{EFT}(37)  -  \sigma^{(3)}_{EFT}(27)    }{\sigma^{\textrm{N$^3$LO}}_{EFT} } = 0.37\%\,.
\end{equation}
The factor $10$ is a conservative estimator of the progression of the
series beyond the first $37$
%, respectively $37$, 
terms. Note that the complete N$^3$LO cross-section appears in the denominator of eq.~\eqref{eq:trunc_error}, i.e., the uncertainty applies to the complete  N$^3$LO result, not just the coefficient of $a_s^5$.

\subsection{Scale variation at \nnnlo and the omission of \nnnlo effects
in parton densities}
\label{sec:scale}

\begin{figure}[!t]
\begin{center}
\includegraphics[width=0.8\textwidth]{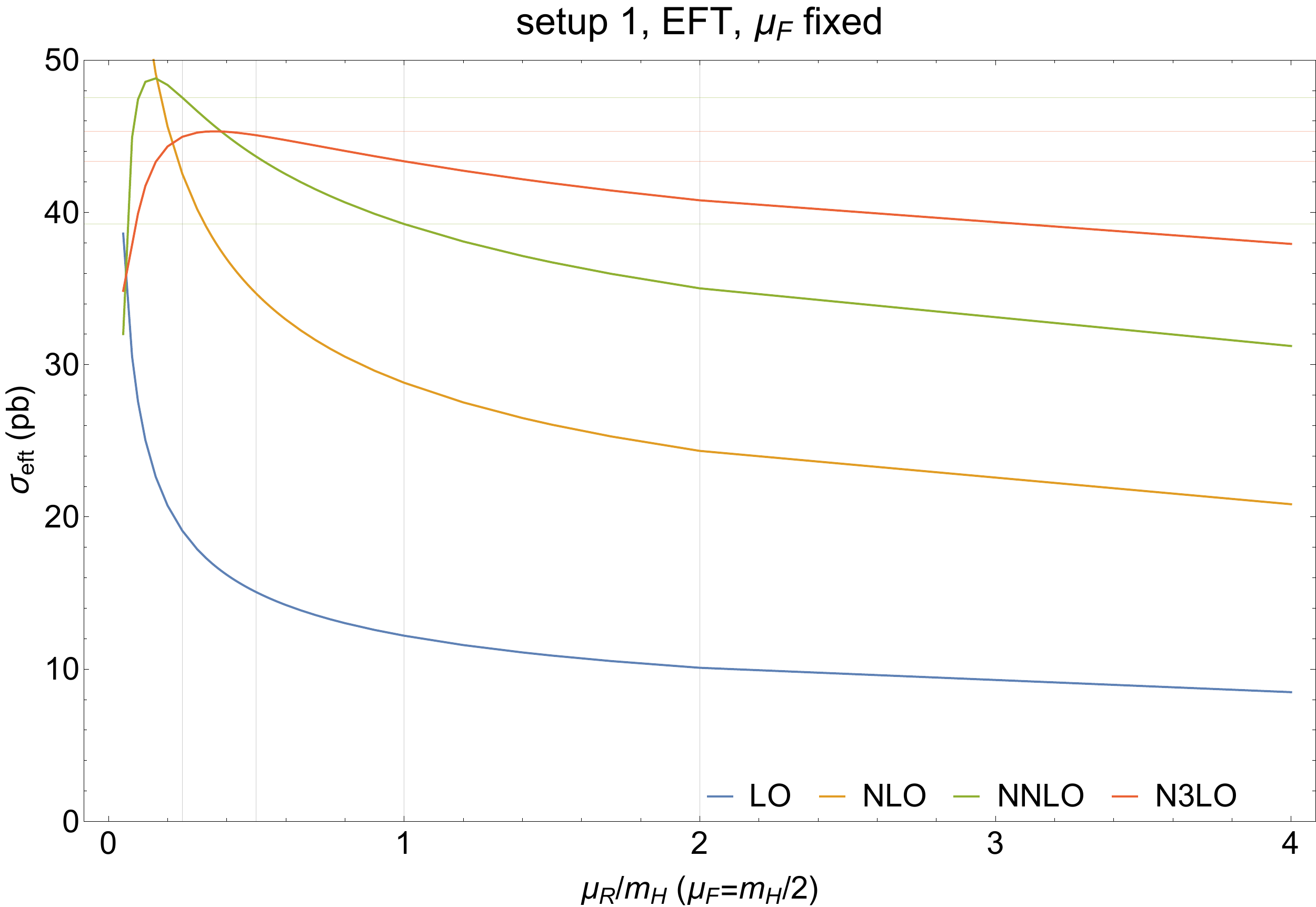}
\end{center}
\caption{
\label{fig:muf_fixed_eft} 
The dependence of the cross-section on the renormalization
  scale for a fixed value of the factorization scale.}
\end{figure}

Having established that the threshold expansion provides a reliable estimate of the N$^3$LO cross-section,
we proceed to study the dependence of the cross-section on the
renormalization and factorization scales $\mu_R$ and $\mu_F$. 

In Fig.~\ref{fig:muf_fixed_eft} we fix the factorization scale to
$\mu_F = {m_H}/2$ and vary the renormalization scale.  
We observe that the perturbative series in the strong coupling 
converges faster for small values of the renormalization scale. 
It is well known that the scale variation is very large at LO and NLO, 
and it is still significant at NNLO. To emphasize this point, we indicate in
Fig.~\ref{fig:muf_fixed_eft} by horizontal lines the range of predictions 
for the cross-section at each perturbative order when $\mu_R$ varies in the
interval $[{m_H}/{4}, m_H]$. This interval seems to capture the 
characteristic physical scales of the process, as indicated 
by the convergence pattern of the series. 
We quantify the renormalization scale variation by looking at the spread around the average value 
of the cross-section in this interval, i.e., we define
\begin{equation}\label{eq:Delta_scale}
\Delta_{EFT,k}^{\textrm{scale}} = \pm  \frac{\sigma^{\rm max}_{EFT,k} - \sigma^{\rm min}_{EFT,k}}{\sigma^{\rm max}_{EFT,k} +
  \sigma^{\rm min}_{EFT,k}}\, 100\%\,,
\end{equation}
with
\beq
\sigma^{\rm max}_{EFT,k} = \max_{\mu_R\in[m_H/4,m_H]}\sigma^{\textrm{N$^k$LO}}_{EFT}(\mu_R)\,,
\eeq 
and similarly for $\sigma^{\rm min}_{EFT,k}$. The results are shown in Tab.~\ref{tab:Delta_scale_mu_F_fixed}.
\begin{table}[!t]
\begin{center}
\begin{tabular}{ ll| c}
\hline
\multicolumn{3}{c}{$\displaystyle\Delta_{EFT,k}^{\textrm{scale}}$ ($\mu_F = m_H/2$)}\\
\hline
LO &($k=0$)&  $\pm 22.0\%$ \\
NLO &($k=1$)& $\pm 19.2\%$ \\ 
NNLO &($k=2$)& $\pm 9.5\%$ \\ 
\nnnlo  &($k=3$)& $\pm 2.2\%$  \\
\hline
\end{tabular}
\caption{\label{tab:Delta_scale_mu_F_fixed}Renormalization scale variation of the cross-section as defined in eq.~\eqref{eq:Delta_scale}. The factorization scale is fixed to $\mu_F = m_H/2$.}
\end{center}
\end{table}

Before we move on to study the dependence of the cross-section on the factorization scale, we note that 
we evolve the strong coupling $\alpha_s(\mu_R)$ at N$^3$LO, and we use 
NNLO parton densities at all perturbative orders. The scale
variation
differs quantitatively from the above table and the convergence of the
perturbative 
series is faster than what is displayed in
Fig.~\ref{fig:muf_fixed_eft} if one uses LO or NLO PDFs and $\alpha_s$ evolution at 
the corresponding orders.

\begin{figure}[!t]
\begin{center}
\includegraphics[width=0.8\textwidth]{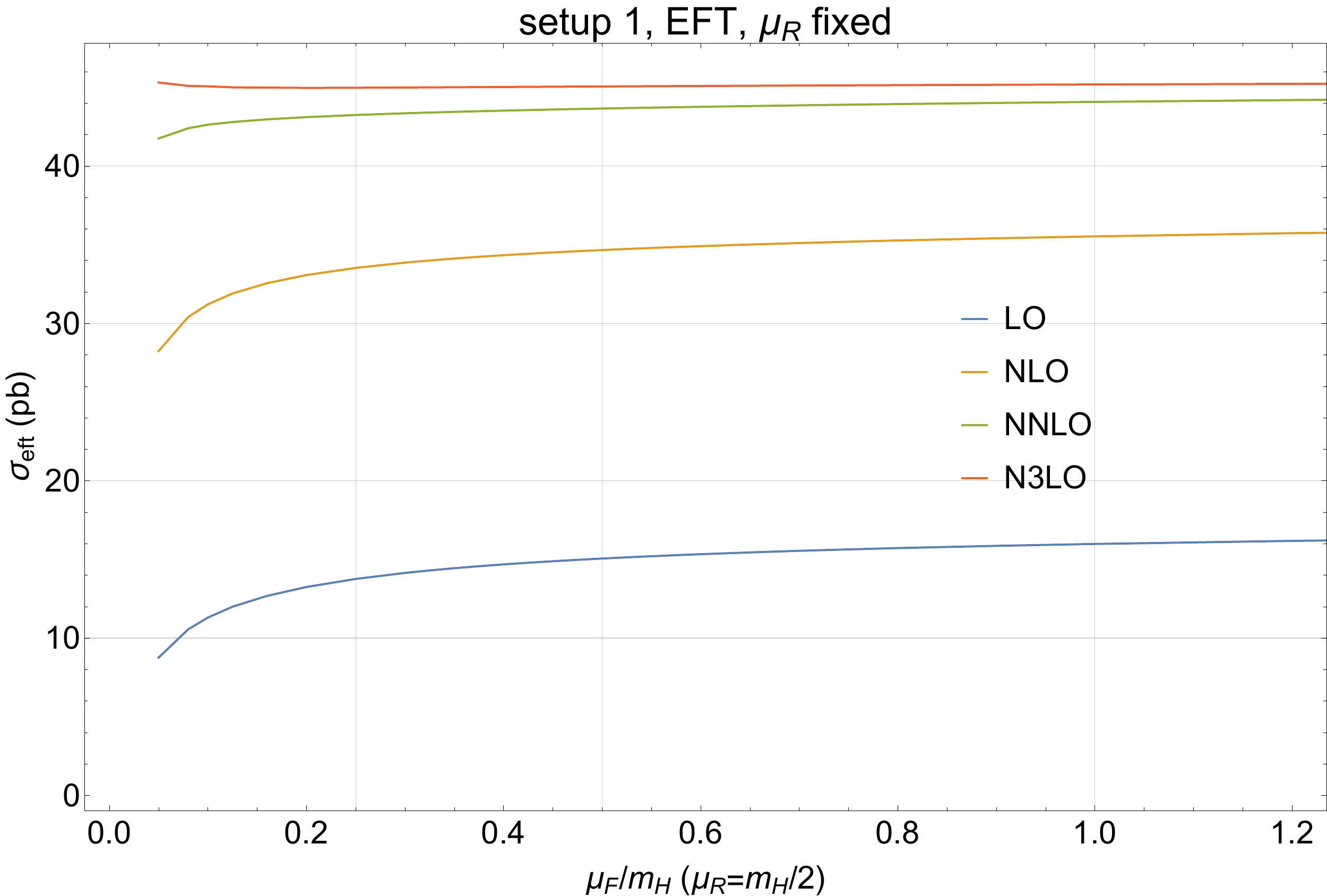}
\end{center}
\caption{
\label{fig:mur_fixed_eft} 
The dependence of the cross-section on the factorization
  scale for a fixed value of the renormalization scale.}
\end{figure}

Let us now turn to the study of the factorization scale dependence of the N$^3$LO cross-section.
In Fig.~\ref{fig:mur_fixed_eft} we fix the renormalization scale to
$\mu_R =  {m_H}/2$ and we vary the factorization scale.  
We observe that at all perturbative orders the variation with the
factorization scale is much smaller than with the corresponding variation
of the renormalization scale. At N$^3$LO, the factorization scale
dependence is practically constant over a wide range of values of $\mu_F$. 

\begin{figure}[!th]
\begin{center}
\includegraphics[width=0.8\textwidth]{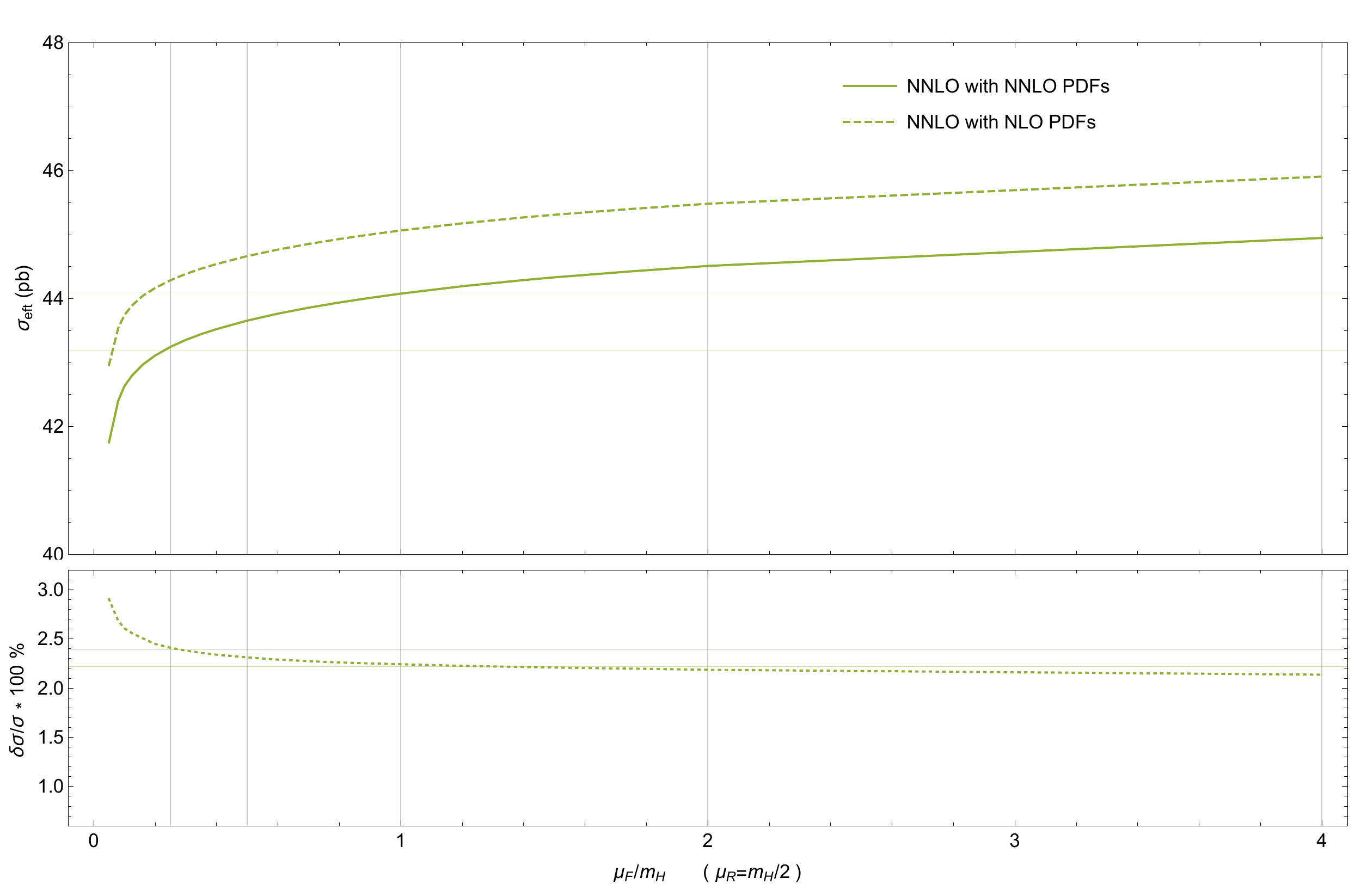}
\end{center}
\caption{
\label{fig:mur_fixed_eft_NNLO} 
The effect of using NLO or NNLO PDFs for the NNLO cross-section in the 
effective theory as a
function of the factorization scale and for a fixed value of the
renormalization scale. A shift is observed which varies little with
the factorization scale.}
\end{figure}

A comment is in order concerning the self-consistency of the
factorization scale variation at N$^3$LO. Traditionally, in  
a LO computation of a hadronic cross-section the parton-densities are
not taken to be constant, but they are evolved with the
one-loop Altarelli-Parisi splitting functions $P^{(0)}$. Similarly, at
NLO and NNLO the $P^{(1)}$ and $P^{(2)}$ corrections to the splitting
functions are included. Following this approach, one would be
compelled to  include the yet unknown $P^{(3)}$ corrections to the splitting functions in the
evolution of the parton densities for our \nnnlo Higgs cross-section
computation, which is of course not possible at this point. Nevertheless, our \nnnlo
computation with corrections only through $P^{(2)}$ in the DGLAP
evolution is consistent in fixed-order perturbation theory, since this is the 
highest-order splitting function term appearing in the mass factorization contributions. 
 Including
the $P^{(3)}$ corrections would be merely a phenomenological improvement 
(which is necessary for LO calculations in order to obtain qualitatively
the physical energy dependence of hadronic cross-sections) but it is
not formally required. An inconsistency may only arise due to the extraction
of the parton densities from data for which there are no \nnnlo
predictions. In fact, this problem has already arisen at NNLO where in
global fits of parton distributions jet observables are fitted with
NLO coefficient functions. When additional processes are computed at
N$^3$LO, it is expected that the gluon and other parton densities
will be extracted with different values. To our understanding, the
uncertainties assigned to the parton densities do not presently account for
missing higher-order corrections, but merely incorporate the experimental 
uncertainties of the data from which they were extracted. 

To assess this uncertainty we resort to the experience from the
previous orders and present in Fig.~\ref{fig:mur_fixed_eft_NNLO} the NNLO
gluon-fusion cross-section using either NNLO or NLO parton densities
as a function of the factorization scale (for a fixed renormalization
scale). We notice that the shape of the two predictions is very
similar, indicating that differences in the evolution
kernels of the DGLAP equation beyond NLO have a small impact. 
However, in the mass range $\left[{m_H}/{4},m_H\right]$  
the NNLO cross-section decreases by about $2.2-2.4\%$  when NNLO
PDFs are used instead of NLO PDFs. We can attribute this shift mostly to differences
in the extraction of the parameterization of the parton densities at
NLO and NNLO. Similarly, we can expect a shift to
occur when the N$^3$LO cross-section gets evaluated in the future
with N$^3$LO parton densities rather than the currently available
NNLO sets.  The magnitude of  the potential shift will be determined from 
the magnitude of the unknown N$^3$LO corrections in standard candle cross-sections
used in the extraction of parton densities. Given that N$^3$LO
corrections are expected to be milder in general than their
counterparts at NNLO, we anticipate that they will induce a smaller
shift than what we observe in
Fig.~\ref{fig:mur_fixed_eft_NNLO}. Based on these considerations, we assign a
conservative
uncertainty estimate due to missing higher orders in the extraction of the
parton densities obtained as\footnote{An alternative way to estimate this uncertainty, based on the Cacciari-Houdeau (CH) method,  was presented in ref.~\cite{Forte:2013mda}. The uncertainty obtained form the CH method is sizeably smaller than the uncertainty in eq.~\eqref{eq:delta_PDF-TH}, and we believe that the CH method may underestimate the size of the missing higher-order effects.}
\begin{equation}\label{eq:delta_PDF-TH}
\delta({\rm PDF-TH}) = \frac{1}{2}\,\left|\frac{\sigma_{EFT}^{(2),NNLO}-\sigma_{EFT}^{(2),NLO}}{\sigma_{EFT}^{(2),NNLO}}\right|=\frac{1}{2}\, 2.31\% = 1.16\%\,,
\end{equation}
where $\sigma_{EFT}^{(2),(N)NLO}$ denotes the NNLO cross-section evaluated with (N)NLO PDFs 
at the central scale $\mu_F=\mu_R=m_H/2$.
In the above,  we assumed conservatively that the size of the N$^3$LO corrections is
about half of the corresponding NNLO corrections.  
 This estimate is supported by the magnitude of the third-order corrections 
to the coefficient functions for deep inelastic scattering~\cite{Vermaseren:2005qc} and a 
related gluonic scattering process~\cite{Soar:2009yh}, which are the 
only two coefficient functions 
that were computed previously to this level of accuracy. 

\begin{figure}[ht]
\begin{center}
\includegraphics[width=0.8\textwidth]{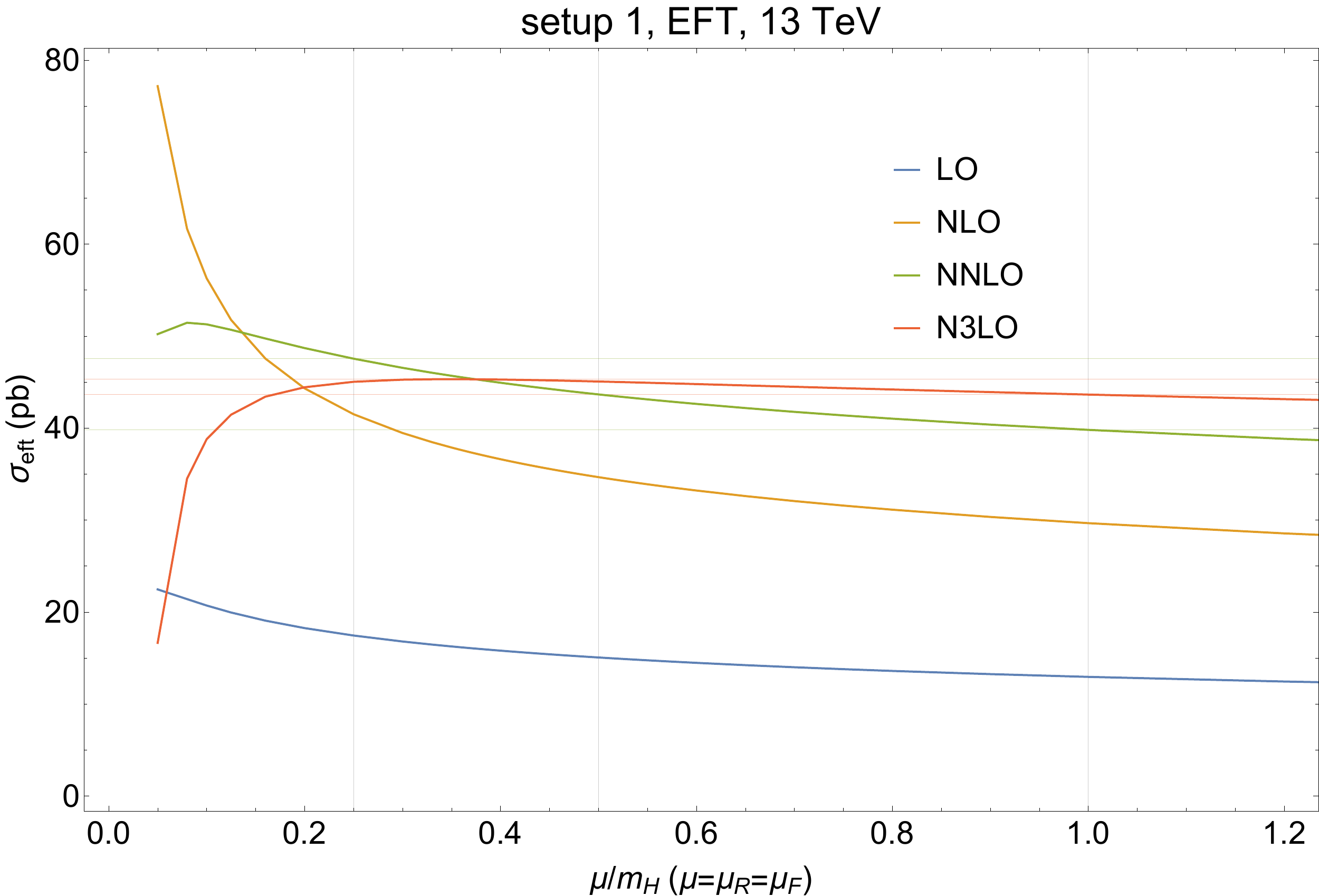}
\end{center}
\caption{
\label{fig:mu_common_eft} 
The dependence of the cross-section on a common renormalization and 
 factorization scale $\mu = \mu_F = \mu_R$.}
\end{figure}

So far we have only studied the scale variation from varying $\mu_F$ and $\mu_R$ 
separately. 
The separation into a renormalization and factorization scale is to a
certain extent conventional and somewhat artificial. Indeed, only
one regulator and one common scale is required for the treatment 
of both infrared and ultraviolet singularities. 
For a physical process such as inclusive Higgs
production, where one cannot identify very disparate physical scales,
large separations between the renormalization from the factorization scale entail 
the risk of introducing unnecessarily large logarithms. 
In Fig.~\ref{fig:mu_common_eft} we present the 
dependence of the cross-section on a common renormalization and
factorization scale $\mu=\mu_R=\mu_F$. 
Through N$^3$LO, the behaviour is very close to the scale-variation pattern 
observed when varying only the renormalization scale with the
factorization scale held fixed. More precisely, using the same quantifier as introduced in
eq.~\eqref{eq:Delta_scale} for the variation of the renormalization scale only, the variation of the cross-section in the range 
$[m_H/4,  m_H]$ for the common scale $\mu$ is shown in Tab.~\ref{tab:Delta_scale}. We observe that the scale variation with $\mu_R=\mu_F$ is
slightly reduced compared to varying only the
renormalization scale at NLO and NNLO, and this difference becomes indeed
imperceptible at N$^3$LO. 
\begin{table}[!t]
\begin{center}
\begin{tabular}{ ll| c}
\hline
\multicolumn{3}{c}{$\displaystyle\Delta_{EFT,k}^{\textrm{scale}}$}\\
\hline
LO &($k=0$)&  $\pm 14.8\%$ \\
NLO &($k=1$)& $\pm 16.6\%$ \\ 
NNLO &($k=2$)& $\pm 8.8\%$ \\ 
\nnnlo  &($k=3$)& $\pm 1.9\%$  \\
\hline
\end{tabular}
\caption{\label{tab:Delta_scale}Scale variation of the cross-section as defined in eq.~\eqref{eq:Delta_scale} for a common renormalization and factorization scale $\mu=\mu_F=\mu_R$.}
\end{center}
\end{table}

The scale variation is the main tool for estimating the theoretical
uncertainty of a cross-section in perturbative QCD, and it has been
successfully applied to a multitude of processes. However, in Higgs
production via gluon fusion it underestimates the uncertainty both at 
LO and NLO.  It is therefore a critical question to assess whether the
scale variation uncertainty is a reliable estimate of the true
uncertainty due to missing higher orders in perturbative QCD. 
We believe that  this is most likely the case, because, at least for natural choices of the scales
in the interval $[m_H/4,m_H]$, the \nnnlo 
cross-section takes values within the
corresponding range of cross-section values at NNLO. Therefore, the 
progression of the perturbative series from NNLO to \nnnlo
corroborates the uncertainty obtained by the scale variation. Indeed, 
for the central scale $\mu = {m_H}/2$ the \nnnlo cross-section is 
only $\sim 3.1\%$ higher than at NNLO, i.e., the shift from NNLO to N$^3$LO is of the same size as
the scale variation uncertainty at N$^3$LO. We will therefore take the
scale variation uncertainty as our uncertainty estimate for missing
higher-order QCD corrections at N$^4$LO and beyond. In Section~\ref{sec:MHO}, we
will also discuss the effect of missing higher orders
through resummation methods. This will give additional support to our claim that the scale variation
at N$^3$LO provides a reliable estimate of missing higher orders beyond N$^3$LO.

\begin{figure}[!t]
\begin{center}
\includegraphics[width=0.8\textwidth]{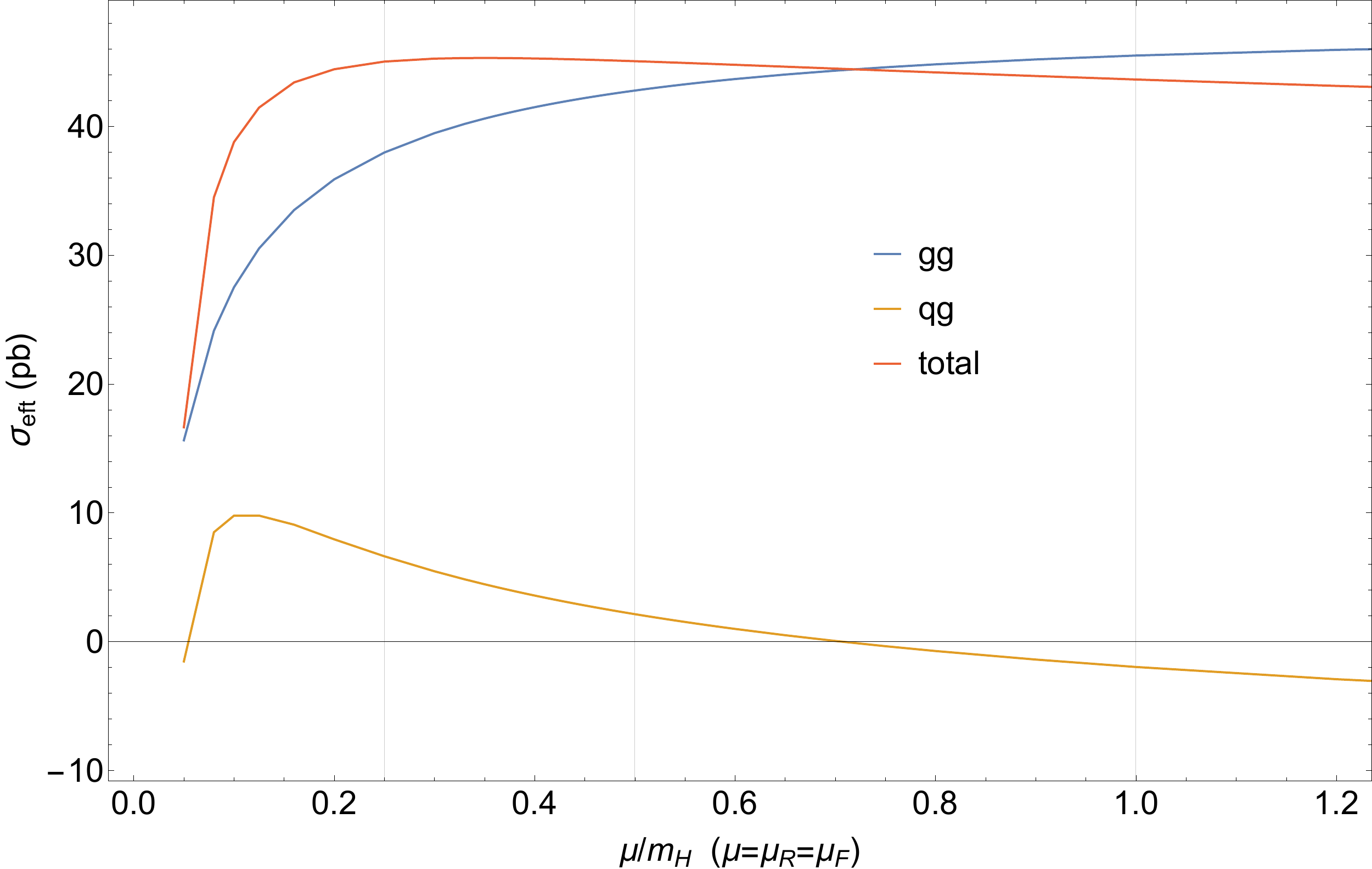}
\end{center}
\caption{
\label{fig:mu_common_eft_channels} 
The dependence of the cross-section on a common renormalization and 
 factorization scale $\mu = \mu_F = \mu_R$ per partonic channel.}
\end{figure}

So far we have only discussed the scale variation for the total hadronic cross-section.
It is also interesting and instructive to analyze the scale dependence of the cross-section
for individual partonic channels. In Fig.~\ref{fig:mu_common_eft_channels} we
present the scale dependence at \nnnlo of the gluon-gluon channel,
the quark-gluon channel and the total cross-section.  The quark-quark
and quark-antiquark channels are very small and are not shown
explicitly in the plot. We see that, while the gluon-gluon channel dominates over the
quark-gluon channel, the latter is important in stabilizing
the scale dependence of the total cross-section. Indeed, with the
exception of extremely small values of $\mu$, the quark-gluon channel has the
opposite slope as the gluon-gluon channel, and therefore a somewhat
larger scale variation of the gluon-gluon channel is getting cancelled
in the total cross-section. This behaviour can be qualitatively understood from the fact that a change in the 
factorization scale modifies the resolution on quark-gluon splitting processes, therefore turning quarks into gluons and vice versa.
We remark that this feature is not
captured by approximate predictions of the cross-section based on the
soft-approximations, which only include the gluon-gluon channel.

To summarize, we have identified in this section two sources of
uncertainty for the N$^3$LO cross-section in the limit of infinite top mass.  
We observe that the dependence on the factorization scale 
is flat over wide ranges of values of $\mu_F$, and the scale variation is dominated
by the $\mu_R$ variation. Moreover, we see that the inclusion of the quark-gluon channel plays an important role
in stabilizing the scale dependence at N$^3$LO. Our
scale variation estimate of the uncertainty is $1.9\%$ (according to our prescription in eq.~\eqref{eq:Delta_scale}). We believe
that at this order in perturbation theory this uncertainty gives a
reliable estimate of missing higher-order corrections from N$^4$LO and beyond. In the next section we give further support to this claim by analyzing the effect of various resummations beyond N$^3$LO.

%%% Local Variables: 
%%% mode: latex
%%% TeX-master: "paper"
%%% End: 

\section{Corrections at N$^4$LO and beyond in the infinite top-quark limit}
\label{sec:MHO}
In the previous section we have argued that the scale variation at \nnnlo gives a reliable estimate for the missing higher-order corrections to the hadronic gluon-fusion cross-section. In this section we corroborate this claim by investigating various other sources of terms beyond N$^3$LO. We check that, if we restrict the analysis to the natural choice of scales from the interval $[m_H/4,m_H]$, the phenomenological effect of these terms is always captured by the scale variation at N$^3$LO. 
We start by investigating higher-order terms generated by using an alternative prescription to include the Wilson coefficient $C$ into a perturbative computation, and we turn to the study of higher-order effects due to resummation in subsequent sections. 
We note at this point that the effect of missing higher-order terms beyond N$^3$LO was already investigated in ref.~\cite{deFlorian:2014vta} by analysing the numerical impact of the leading N$^4$LO threshold logarithms. The conclusions of ref.~\cite{deFlorian:2014vta} are consistent with the findings in this section.

\subsection{Factorization of the Wilson coefficient}
%In the previous section we have argued that the scale variation at \nnnlo gives a reliable estimate for the missing higher-order corrections to the hadronic gluon-fusion cross-section. In this section we corroborate this claim by investigating various other sources of terms beyond N$^3$LO, and we check that, if we restrict the analysis to the natural choice of scales from the interval $[m_H/4,m_H]$, the effect of these terms is always included in the uncertainty band coming from the scale variation at N$^3$LO. We start by investigating higher-order terms generated by using an alternative prescription to include the Wilson coefficient $C$ into a perturbative computation, and we turn to the study of higher-order effects due to the resummation of threshold logarithms in subsequent sections.

The (partonic) cross-section in the effective theory is obtained by multiplying (the square of) the Wilson coefficient by the perturbative expansion of the coefficient functions $\eta_{ij}$, see eq.~\eqref{eq:sigmaeff}. As the Wilson coefficient itself admits a perturbative expansion, eq.~\eqref{eq:WL_series}, eq.~\eqref{eq:sigmaeff} takes the following form up to N$^3$LO in perturbation theory,
\beq\label{eq:Wilson_expand}
\frac{\hat{\sigma}_{ij,EFT}}{z} =\sigma_0\,|1+\ldots+a_s^3\,C_3+\ord(a_s^4)|^2\, \sum_{i,j} (1+\ldots +a_s^3\, \eta^{(3)}_{ij}(z)+\ord(a_s^4))\,,
\eeq
where $\sigma_0$ denotes the Born cross-section.
Conventionally in fixed-order perturbation theory through N$^3$LO, one only includes corrections up to $\ord(a_s^5)$ from the product in eq.~\eqref{eq:Wilson_expand} and drops all terms of higher order (the Born cross-section is proportional to $a_s^2$). This is also the approach adopted in Section~\ref{sec:EFT}, where consistently only terms up to $\ord(a_s^5)$ had been included. In this section we analyze how the cross-section changes if all the terms shown in eq.~\eqref{eq:Wilson_expand} are included. In this way, we obviously include terms into our prediction that are beyond the reach of our fixed-order N$^3$LO computation. We stress that the inclusion of these terms does not spoil the formal N$^3$LO accuracy. They can lead, however, to sizeable effects that can be used as a quantifier for missing higher-order terms.

\begin{figure}[!t]
\begin{center}
\includegraphics[width=1.0\textwidth]{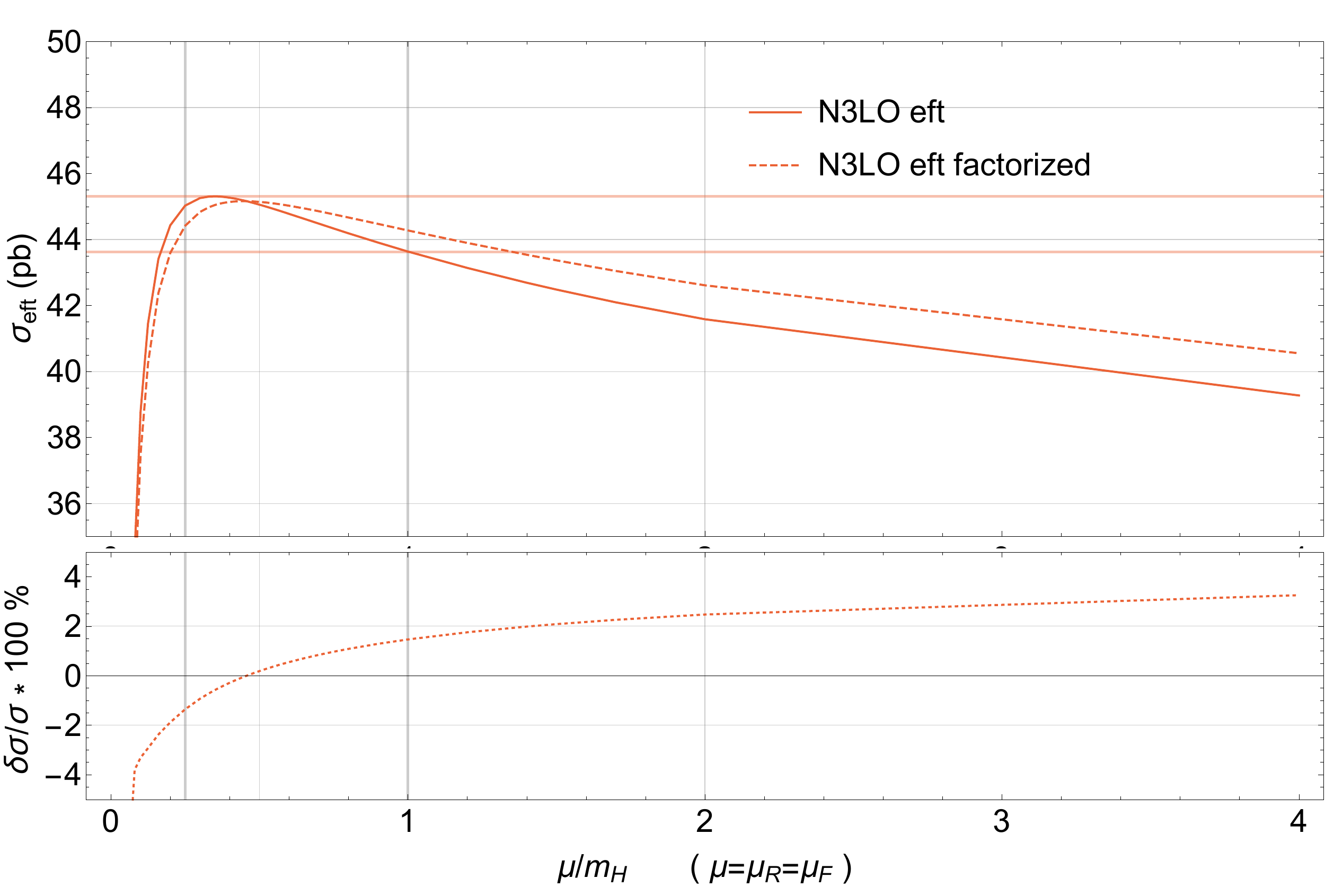}
\end{center}
\caption[Scale dependence of the fixed-order cross-section and the factorized cross-section.]{
\label{fig:wilson}
Scale variation with $\mu = \mu_R=\mu_F$ at N$^3$LO within Setup 1 (solid line), compared to the factorized form of the cross-section where the Wilson coefficient and the coefficient functions are separately truncated to ${\cal O}(a_s^5)$ (dashed line).
}
\end{figure}

In Fig.~\ref{fig:wilson} we show the value of the cross-section as a function of a common renormalization and factorization scale $\mu$, obtained by either truncating the full cross-section (solid) or by multiplying the truncated Wilson coefficient and truncated coefficient function (dashed). We stress that the difference between the two curves stems entirely from terms at N$^4$LO and beyond. However, we observe that if the scale is chosen to lie in our preferred range $\mu\in[m_H/4,m_H]$, then the two curves agree within the scale uncertainty at fixed order N$^3$LO, and the difference between the two cross-section values is always well below 2\%. In particular, the two curves intersect for $\mu\simeq m_H/2$. Hence, if we choose the scale $\mu$ in the range $[m_H/4,m_H]$, both approaches give phenomenologically equivalent answers, and higher-order terms generated by the factorization of the Wilson coefficient only have a very mild phenomenological impact, which is captured by the fixed-order scale variation. We stress that this also supports the claim in Section~\ref{sec:EFT} that the scale variation at N$^3$LO gives a reliable estimate of missing higher-order terms in perturbation theory.

\subsection{Threshold resummation in Mellin space}
Fixed-order computations beyond N$^3$LO are currently beyond our
technical capabilities. Nevertheless, we can get some information on
corrections at N$^4$LO and beyond from resummation formul\ae, which
allow one to resum certain logarithmically-enhanced terms to all
orders in perturbation theory.

In this section we look in particular at higher-order corrections
generated by the resummation of threshold logarithms in Mellin space. 
Before studying the phenomenology of the resummed inclusive Higgs cross-section 
at N$^3$LO+N$^3$LL, we give a short review of the formalism that
allows one to resum large threshold logarithms in Mellin space.

The Mellin transform of the hadronic cross-section with respect to $\tau = m_H^2/S$ is 
\beq
\sigma(N) = \int_0^1\dd \tau\, \tau^{N-1}\,\frac{\sigma(\tau)}{\tau}\,.
\eeq
In the following, we always work in the effective theory with an infinitely-heavy top quark. 
Since the Mellin transform maps a convolution of the type~\eqref{eq:convolution} to an ordinary product, the QCD factorization formula~\eqref{eq:sigma} takes a particularly simple form in Mellin space,
\beq
\sigma(N) = \sum_{ij} f_i(N)\,f_j(N)\,\hat{\sigma}_{ij}(N)\,,
\eeq
with the Mellin moments
\beq\bsp
f_i(N) &= \int_0^1 \dd z\, z^{N-1}\, f_i(z)\,,\\
\hat{\sigma}_{ij}(N) &= \int_0^1 \dd z\, z^{N-1} \frac{\hat{\sigma}_{ij}(z)}{z}\,,
\esp\eeq
where we suppressed the dependence of the PDFs and the partonic cross-sections on the scales.
The Mellin transform is invertible, and its inverse is given by
\beq\label{eq:inverse_Mellin}
\sigma(\tau) = \sum_{ij} \int_{c-i\infty}^{c+i\infty}\frac{\dd N}{2\pi i}\,\tau^{1-N}\,f_i(N)\,f_j(N)\,\hat{\sigma}_{ij}(N)\,,
\eeq
where the contour of integration is chosen such that it lies to the right
of all possible singularities of the Mellin moments in the complex $N$ plane.

From the definition of the Mellin transform it is apparent that
the limit $z\to1$ of the partonic cross-sections corresponds to the limit
$N\to\infty$ of the Mellin moments of $\hat{\sigma}_{ij}(N)$.
In the limit $N\to\infty$ the partonic cross-section in Mellin space can be written as~\cite{Catani:2003zt}
\beq
\bsp
\label{eq:reslim}
\hat{\sigma}_{ij}(N) &= \delta_{ig}\,\delta_{jg}\,\hat{\sigma}_{\textrm{res}}(N) + \mathcal{O}\left(\frac{1}{N}\right)\\
&= \delta_{ig}\,\delta_{jg}\,a_s^2\,\sigma_0\,\left[1+\sum_{n=1}^{\infty}a_s^n\sum_{m=0}^{2n}\hat{\sigma}_{n,m}\log^mN\right]
+\mathcal{O}\left(\frac{1}{N}\right)\,,
\esp
\eeq
where $\sigma_0$ denotes the LO cross-section in the large-$m_t$ limit and $\hat{\sigma}_{\textrm{res}}(N)$ is related to the Mellin transform of the soft-virtual cross-section.
The constant and logarithmically-divergent contributions in the limit $N\to\infty$
can be written in terms of an all-order resummation formula~\cite{Sterman:1986aj,Catani:2003zt,Catani:1989ne,Catani:1990rp},
\beq
\label{eq:resform}
\hat{\sigma}_{\textrm{res}}(N) = a_s^2\, \sigma_0\,C_{gg}(a_s)
\exp\left[\mathcal{G}_H(a_s,\log N)\right]\,,
\eeq
where the function $C_{gg}$ contains all contributions that are constant for $N\to\infty$.
The function $\mathcal{G}_H$ exponentiates
the large logarithmic contributions to all orders and can be written as
\beq
\bsp
\mathcal{G}_H(a_s,\log N) &= \log N\,g_H^{(1)}(\lambda)+\sum_{n=2}^{\infty}a_s^{n-2}\,g_H^{(n)}(\lambda)\,,\qquad \lambda\equiv \beta_0\,a_s\,\log N\,,
\esp
\eeq
where $\beta_0$ denotes the LO coefficient in the QCD $\beta$ function.
The functions $g_H^{(n)}$ are known exactly up to
NNLL accuracy~\cite{Moch:2005ba}, i.e., up to $g_H^{(3)}$, which requires knowledge
of the cusp anomalous dimension up to three loops~\cite{Catani:2003zt,Moch:2005ky}.
In order to perform resummation at N$^3$LL
accuracy~\cite{Catani:2014uta,Bonvini:2014joa}, the function $g_H^{(4)}$ is needed. This function depends on the four-loop
cusp anomalous dimension, which is not yet known in QCD. 
We employ the Pad\'e approximation of ref.~\cite{Moch:2005ba} for the four-loop cusp anomalous
dimensions to obtain a numerical estimate for $g_H^{(4)}$. The numerical impact of this
approximation has been studied, e.g., in ref.~\cite{Moch:2005ba} and we checked that by varying the Pad\'e approximation up and down by a factor of 10, our results do not change\footnote{We note, though, that the Pad\'e approximation was obtained under the assumption of Casimir scaling of the cusp anomalous dimension, an assumption which is likely to break down at four loops.}.

Let us now turn to the phenomenological implications of resummation at N$^3$LL. We obtain N$^3$LO $+$ N$^3$LL predictions for the cross-section by matching the resummation formula~\eqref{eq:resform} to the fixed-order N$^3$LO cross-section, i.e., by subtracting from eq.~\eqref{eq:resform} its expansion through $\ord(a_s^5)$. In this way we make sure that the resummation only starts at $\ord(a_s^6)$, which is
beyond the reach of our fixed-order calculation.
We present our numerical method to perform the inverse Mellin transform~\eqref{eq:inverse_Mellin} in Appendix~\ref{app:inverse_Mellin}.
In Fig.~\ref{fig:resummationsv} we show the scale dependence
of the resummed cross-section in comparison to the fixed-order cross-section.
We see that the resummation stabilizes
the scale dependence of the cross-section in comparison to the fixed-order result.
At N$^3$LO+N$^3$LL, the value of the cross-section is essentially independent of the scale choice,
and roughly equal to the value of the fixed-order cross-section at $\mu\equiv\mu_F=\mu_R=m_H/2$.  
In particular, at $\mu={m_H}/{2}$ the effect of the resummation on the
N$^3$LO cross-section is completely negligible, and in the range $\mu\in[m_H/4,m_H]$, the effect of the resummation is captured by the scale uncertainty at fixed order (albeit at the upper end of the uncertainty band). Hence, the fixed-order result at N$^3$LO for $\mu\in[m_H/4,m_H]$ contains the value of the cross-section at N$^3$LO+N$^3$LL. This corroborates our claim made at the end of Section~\ref{sec:EFT} that at N$^3$LO the scale uncertainty provides a reliable estimate of higher-order corrections at N$^4$LO and beyond. 

\begin{figure}[!t]
\begin{center}
\includegraphics[width=1.0\textwidth]{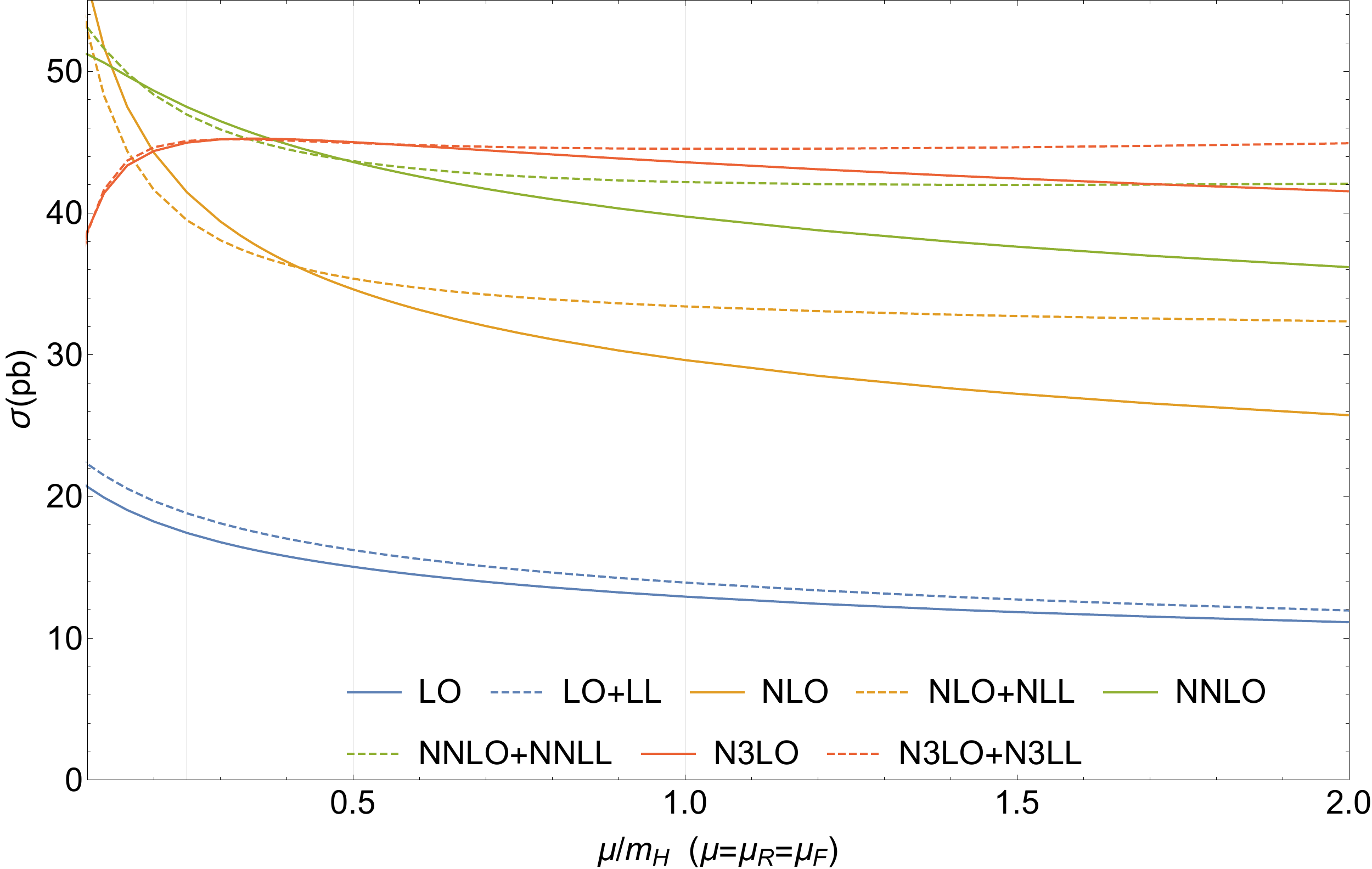}
\end{center}
\caption[Scale dependence of the resummed cross-section.]{
\label{fig:resummationsv}
Scale variation with $\mu = \mu_R=\mu_F$ at all perturbative
orders through N$^3$LO within Setup 1, resummed at the corresponding
logarithmic accuracy. The fixed-order cross-sections are shown for comparison.
}
\end{figure}

\begin{figure}[!t]
\begin{center}
\includegraphics[width=1.0\textwidth]{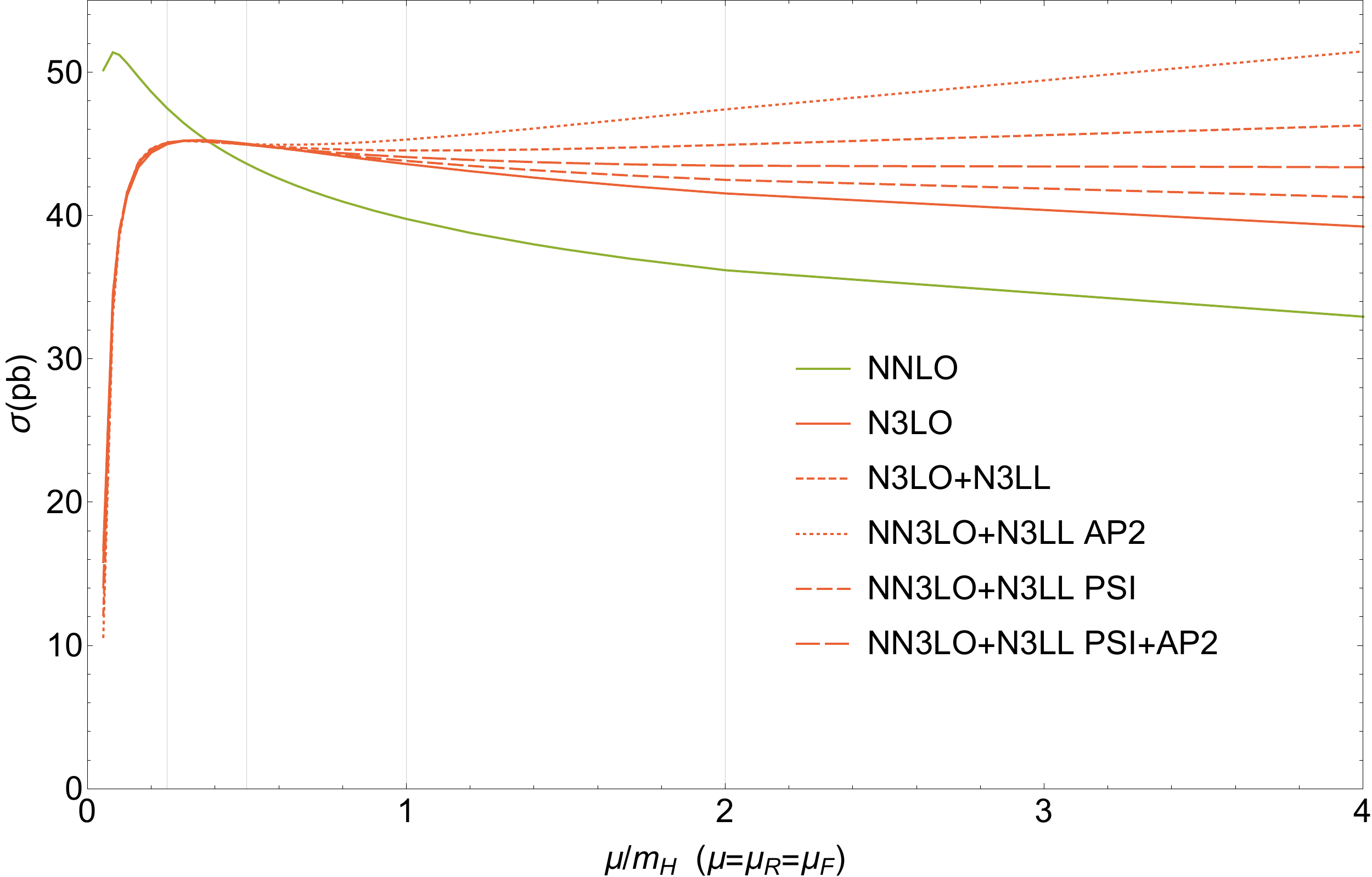}
\end{center}
\caption[Scale dependence of the resummed cross-section.]{
\label{fig:resummationschemes}
Scale variation with $\mu = \mu_R=\mu_F$ of the N$^3$LO+N$^3$LL cross-section within Setup 1 for different resummation schemes. The fixed-order cross-sections are shown for comparison.
}
\end{figure}

We conclude this section by studying the impact of changing the prescription of which terms are exponentiated in Mellin space. 
The resummation formula eq.~\eqref{eq:resform} exponentiates the large-$N$ limit
of the fixed-order cross-section eq.~\eqref{eq:reslim}, and as such it is only defined
up to subleading terms in this limit. It is therefore possible to construct
different resummation schemes that formally agree in the limit $N\to\infty$, but that differ
by terms that are suppressed by $1/N$. In particular, we may change the exponent in eq.~\eqref{eq:resform} to
$\mathcal{G}_H(a_s,L(N))$, where $L(N)$ is any function on Mellin space such that $L(N) = \log N+\ord(1/N)$. In the remainder of this section we study the impact on the Higgs cross-section of different choices for $L(N)$ that have been considered in the literature (see, e.g., ref.~\cite{Bonvini:2014joa}):
\begin{enumerate}
\item (PSI): $L(N) = \psi(N)$, where $\psi(N) = \frac{\dd}{\dd N}\log\Gamma(N)$ denotes the digamma function. 
This choice is motivated by the fact that the threshold
logarithms appear as $\psi(N)$ in the Mellin transform of the soft-virtual partonic cross-section and that the Mellin transform of the partonic cross-section is supposed to exhibit poles in Mellin space rather than branch cuts.
\item (AP2): A different resummation scheme can be obtained by exponentiating the Mellin transform
of the Altarelli-Parisi splitting kernel. In particular, the function $L(N)\equiv \mathcal{AP}_2[\log N] = 2\log N-3\log(N+1)+2\log(N+2)$ allows one to exponentiate the first two subleading terms as $z\to 1$ coming from the Altarelli-Parisi splitting function $P_{gg}^{(0)}(z)$.
\item (PSI+AP2): Combining the two previous variants, we obtain a new variant, corresponding to $L(N)\equiv \mathcal{AP}_2[\psi(N)]=2\psi(N)-3\psi(N+1)+2\psi(N+2)$.
\end{enumerate}
All these schemes are formally equivalent resummation schemes, because they agree in the 
large-$N$ limit. However, the formally subleading corrections can have a significant 
numerical impact. In Fig.~\ref{fig:resummationschemes} we show the cross-section predictions for the  four different resummation schemes discussed in this section.
We observe that within our preferred range of scales, $\mu\in[m_H/4,m_H]$, all four schemes considered in this paper give results that agree within the fixed-order scale variation at N$^3$LO, giving further support to our claim that the scale variation at N$^3$LO provides a reliable estimate of the remaining missing perturbative orders. We note, however, that outside this range of scales the different prescriptions may differ widely, and we know
of no compelling argument why any one of these schemes should be more correct or reliable than the 
others. Based on these two observations, we are led to conclude that threshold resummation does not modify our result
beyond its nominal theory error interval
over the fixed-order N$^3$LO prediction when the scales are chosen in the range $[m_H/4,m_H]$, and we will therefore not include the effects of threshold resummation in Mellin space into our final cross-section prediction.

\subsection{Threshold and $\pi^2$-resummation in Soft-Collinear Effective Theory}
In this section we discuss an alternative way to represent the soft-virtual cross-section in Higgs production, based on ideas from Soft-Collinear Effective Theory (SCET)~\cite{Bauer:2002nz,Beneke:2002ph,Bauer:2000yr,Becher:2007ty,Bauer:2001yt}. 
%While threshold resummation in SCET is formally equivalent in the strict threshold limit $z\to 1$ to the results obtained in the %previous section, the two approaches are conceptually distinct and can lead to sizeable phenomenological differences.
Just like in the case of threshold resummation in Mellin space, 
we start by introducing the necessary terminology and review the main ideas, in particular
the resummation formula of ref.~\cite{Ahrens:2008nc,Ahrens:2008qu,Ahrens:2010rs}. 
At the end, we combine the N$^3$LO coefficient functions with the SCET resummation and study its phenomenological impact. In our analysis we closely follow ref.~\cite{Ahrens:2008qu} (for a pedagogical review see ref.~\cite{Becher:2014oda}). For a comparison to Mellin space resummation see for example refs.~\cite{Sterman:2013nya,Bonvini:2014qga}.

Using SCET factorization theorems, the partonic cross-sections at threshold can be factorized into a product of a hard function $H$, a soft function $\tilde S$ and the effective theory Wilson coefficient $C$, multiplied by the Born cross-section $\sigma_{0}$. 
%In the SCET formalism the hadronic cross-section at the production threshold of the Higgs boson can be factorized via a Laplace transformation into parton distribution functions, 
%\beq
%\label{eq:factheorem}
%\hat\sigma_{ij,\, EFT}^{} (z,\mu^2)=\hat\sigma^{0}(\mu^2) \left| H(m_h^2,\mu^2)\right|^2 S(\hat s (1-z)^2,\mu^2) C_t^2(m_t^2,\mu^2)\,+\mathcal{O}((1-z)^0).
%\eeq
SCET provides a field theoretical description of these individual functions. They arise when effective field theory is systematically applied and 
degrees of freedom corresponding to various energy scales are integrated out. 
%In SCET each coefficient of this product corresponds to different kinematic degrees of freedom that are separated at distinct energy scales. 
The individual coefficients are defined at the respective energy scales, but the total cross-section is  independent of these scales.
The idea of the SCET formalism is to exploit the factorization of degrees of freedom to derive and solve an evolution equation for each coefficient in the cross-section separately. Consequently, one solves the renormalization group equation for the hadronic cross-section, with the explicit aim to cure the dependence of the cross-section on the various scales. 
%A byproduct of this procedure is that the resulting formulae allow to predict the analytic coefficient of 4$^{th}$ leading threshold logarithm (N$^3$LL) as in the case of Mellin space resummation.
We refer to the scheme outlined above as \emph{SCET resummation}. 

A formula that achieves the aforementioned goals has been derived in ref.~\cite{Ahrens:2008nc,Ahrens:2008qu}. It reads,
\bea
\label{eq:resformula}
\hat\sigma_{ij,\, EFT}^{SCET,\text{thr}} (z,\mu^2)&=&z^{\frac{3}{2}}\,\sigma_0\, |C(m_t^2,\mu_t^2)|^2 \left| H(m_H^2,\mu_h^2) \right|^2U(m_H^2,\mu^2,\mu_t^2,\mu_h^2,\mu_s^2)  \nonumber\\
&\times& \frac{z^{-\xi} }{(1-z)^{1-2\xi}}\tilde{S}\left(\log \left(\frac{m_H^2(1-z)^2}{z\mu_s^2}\right)+\partial_\xi , \mu_s^2\right)\frac{e^{-2\xi \gamma_E}}{\Gamma(2\xi)}\Bigg|_{\xi=C_A A_{\gamma_{cusp}}(\mu^2,\mu_s^2)}\,,
\eea
with
\bea
\label{eq:UFunc}
U(m_H^2,\mu^2,\mu_t^2,\mu_h^2,\mu_s^2) &=&\frac{\alpha_S(\mu^2)^2}{\alpha_S(\mu_t^2)^2}\Bigg| \left(\frac{-m_H^2}{\mu_h^2}\right)^{-C_A/2  A_{\gamma_{cusp}}(\mu_s^2,\mu_h^2)}\Bigg|^2
%\frac{\alpha_S(\mu_s^2)^2}{\alpha_S(m_H^2)^2}
\left(\frac{\beta(\mu_s^2)\alpha_S(\mu_t^2)^2}{\beta(\mu_t^2)\alpha_S(\mu_s^2)^2}\right)^2\nonumber\\
&\times&\Big|\text{exp}\left[C_A S_{\gamma_{cusp}}(\mu_s^2,\mu_h^2)-A_{\gamma_V}(\mu_s^2,\mu_h^2)+2 A_{\gamma_g}(\mu^2,\mu_s^2)\right]\Big|\,.
\eea
Here, $C_A=N_c$ is the quadratic Casimir of the adjoint representation of $SU(N_c)$. The hard function was computed through fourth order in refs.~\cite{Ahrens:2008nc,Gehrmann:2010ue} (in particular, see ref.~\cite{Gehrmann:2010ue} eq. (7.6), (7.7) and (7.9)). The soft function was recently computed through N$^3$LO in ref.~\cite{Li:2014afw,Bonvini:2014tea}. 
We recomputed the soft function up to N$^3$LO based on the soft-virtual Higgs cross-section at N$^3$LO of ref.~\cite{Anastasiou:2014vaa}, and we confirm the result of ref.~\cite{Li:2014afw,Bonvini:2014tea}. 
The definition of $A_\gamma$ and $S_\gamma$ are given for example in ref.~\cite{Ahrens:2008nc}. $\gamma_{\text{cusp}}$ is the cusp anomalous dimension~\cite{Moch:2005ba,Korchemsky:1987wg,Korchemsky:1985xj,Moch:2005tm}. The anomalous dimension $\gamma_V$ can be extracted from the QCD form factor~\cite{Gehrmann:2010ue} and $\gamma_g$ corresponds to the coefficient of $\delta(1-z)$ of the $g\rightarrow g$ splitting function~\cite{Moch:2005tm} . 
%$\sigma^0$ is the Born cross-section. 
%We slightly modified the definition of our Wilson coefficient eq.~\eqref{eq:L_eff} by normalizing its leading order coefficient to one and absorbing the normalization into our definition of $\hat \sigma^{0}$.
In the previous expression, the soft scale $\mu_s$, hard scale $\mu_h$ and top-quark scale $\mu_t$ are the energy scales of the soft function, the hard function and the Wilson coefficient. 
The function $U$ mediates the evolution of the individual coefficient functions to the common perturbative scale $\mu$. 
%The functions $A_\gamma$ and $S_\gamma$ can be found ref.~\cite{Ahrens:2008nc} and the hard function was computed up %to fourth order in ref.~\cite{Gehrmann:2010ue}. 
%The soft function was recently computed through N$^3$LO in ref.~\cite{Bonvini:2014tea,Li:2014afw}. 
%We recomputed the soft function up to N$^3$LO based on the soft-virtual Higgs cross-section at N$^3$LO of ref.~%\cite{Anastasiou:2014vaa}, and we confirm the result of ref.~\cite{Bonvini:2014tea,Li:2014afw}. 
We note that if we choose the scales according to $\mu^2=\mu_h^2=\mu_s^2=\mu_t^2$, then $U(m_H^2,\mu^2,\mu_t^2,\mu_h,\mu_s^2)=1$, and eq.~\eqref{eq:resformula} corresponds to the fixed-order soft-virtual cross-section.
More precisely, if we expand the product of the soft and hard functions and the Wilson coefficient through order $a_s^n$, then we reproduce the  fixed-order soft-virtual cross-section at N$^n$LO up to terms that vanish in  the threshold limit.

Next, let us discuss the choice of the values of the scales $\mu_t$, $\mu_h$ and $\mu_s$. 
First, it is natural to choose the top-quark scale $\mu_t$ to be the top-quark mass, because this it is the only other mass scale in the Wilson coefficient. Similarly, it seems natural to choose the Higgs mass to be the hard scale entering the hard function $H$.
The cross-section depends on the hard function via its modulus squared, and the hard function depends on the Higgs boson mass via logarithms of the type $\log\left(\frac{-m_H^2}{\mu_h^2}\right)$. Choosing $\mu_h^2=m_H^2$, we have to analytically continue the logarithms, which then give rise terms proportional to $\pi^2$, 
\beq
\Big| \log\left(\frac{-m_H^2-i0}{\mu_h^2}\right)\Big|^2=\Big|\log\left(\frac{m_H^2}{\mu_h^2}\right)-i\pi\Big|^2=\log^2\left(\frac{m_H^2}{\mu_h^2}\right)+\pi^2.
\eeq
In ref.~\cite{Ahrens:2008qu} it was observed that at NLO the $\pi^2$ term is responsible for a large part of the perturbative corrections at this order. It was suggested to analytically continue the hard function to the space-like region by choosing $\mu_h^2=-m_H^2$. 
In this approach no $\pi^2$ is produced by the analytic continuation of the fixed-order hard coefficient, and the analytic continuation from the space-like to the time-like region is performed in the exponential of eq.~\eqref{eq:UFunc}. In the following we also adopt this procedure, which is sometimes referred to as \emph{$\pi^2$-resummation}, and we choose the hard scale as $\mu_h^2=-m_H^2$.

Finally, we have to make a suitable choice for the soft scale $\mu_s$. In ref.~\cite{Becher:2007ty,Ahrens:2008qu} two specific choices were outlined.
\begin{enumerate}
\item $\mu_I$: The value of $\mu_s$ where the contribution of the second-order coefficient of the soft function to the hadronic cross-section drops below 15\% of the leading order coefficient.
\item $\mu_{II}$: The value of $\mu_s$ that minimizes the contribution of the second-order coefficient of the soft function to the hadronic cross-section.
\end{enumerate}
Both of the above choices depend on $m_H$ and $\mu$, and following  ref.~\cite{Ahrens:2008nc,Becher:2007ty} we choose the average of both scales, 
\beq
\mu_s(\mu,m_H)=\frac{\mu_I(\mu,m_H)+\mu_{II}(\mu,m_H)}{2}\,.
\eeq

We have now all the ingredients to study the phenomenological implications of the SCET resummation. 
We have implemented the resummation formula, eq.~\eqref{eq:resformula}, into a {\tt C++} code and combined it with fixed-order cross-section through N$^3$LO. 
We write the full SCET-resummed cross-section as 
\beq
\label{eq:RGIXS}
\hat\sigma_{ij,\, EFT}^{SCET}=\hat\sigma_{ij,\, EFT}^{SCET,\text{thr}}-\hat\sigma_{ij,\, EFT}^{SCET,\text{thr}}\Big|_{\mu^2=\mu_h^2=\mu_s^2=\mu_t^2}+\hat\sigma_{ij,\, EFT}.
\eeq
The above formula matches the resummation to our fixed-order cross-section at N$^3$LO such as not to spoil our fixed-order accuracy, and resummation effects only start contributing from N$^4$LO.
In our implementation we followed closely the public code $\mathtt{RGHiggs}$~\cite{Ahrens:2008nc,Ahrens:2008qu,Ahrens:2010rs} and validated our results by comparison. 

Unfortunately, not all anomalous dimensions required for the evolution of the N$^3$LO cross-section are known at this point. We therefore truncate all anomalous dimensions at the maximally known order. 
Note that already at NNLO the unknown four-loop cusp anomalous dimension would be required. 
We checked that the numerical dependence of the result on the four-loop cusp anomalous dimension is small and insignificant for phenomenological purposes. 
\begin{figure}[!t]
\begin{center}
\includegraphics[width=0.8\textwidth]{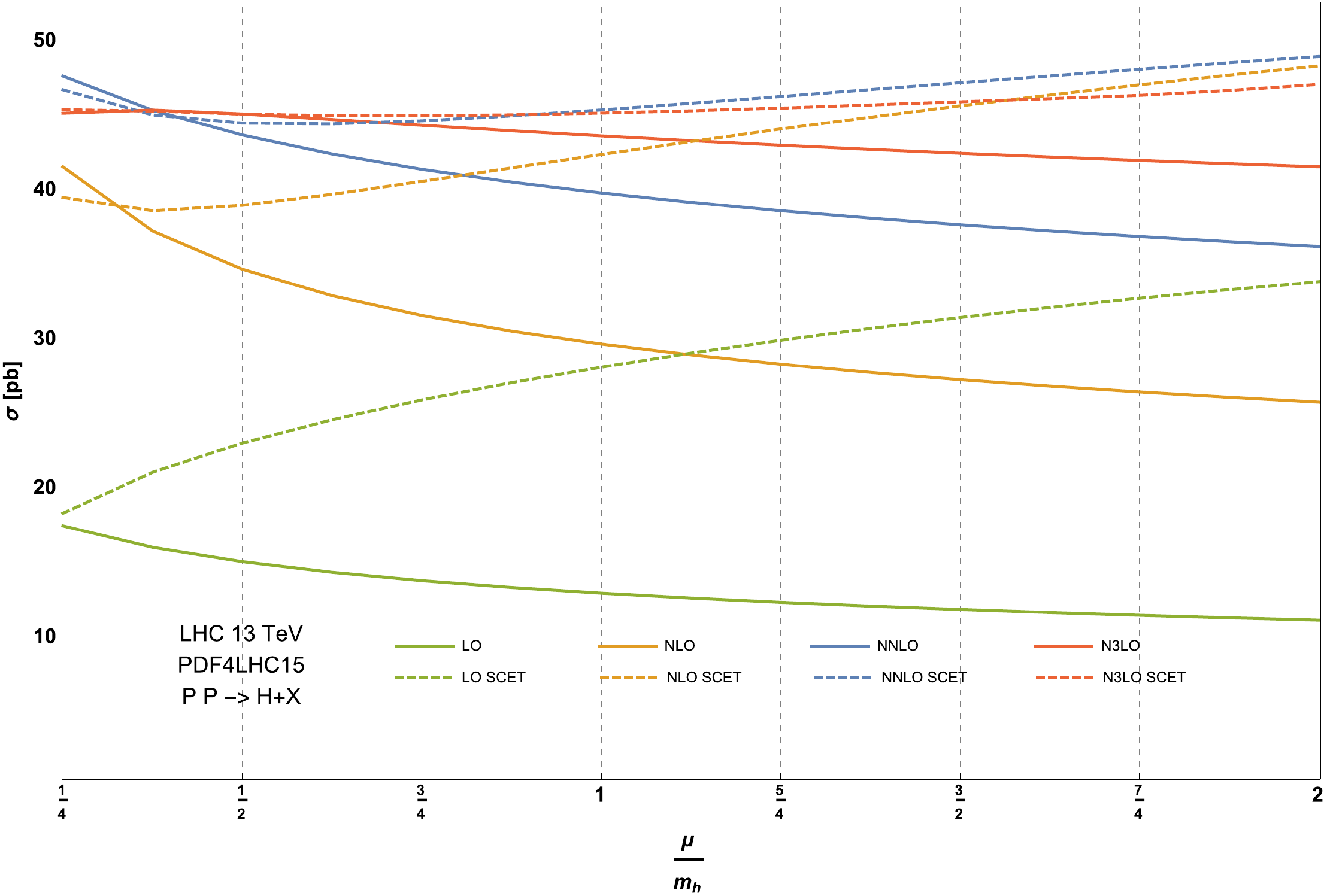}
\end{center}
\caption{\label{fig:SCETresum}The Higgs boson production cross-section computed for the LHC using Setup 2 at LO (green), NLO (orange), NNLO (blue), N$^3$LO (red). Solid lines correspond to fixed-order (FO) predictions and dashed lines to SCET predictions.}
\end{figure}

In Fig.~\ref{fig:SCETresum} we show the hadronic cross-section as a function of a common scale $\mu=\mu_R=\mu_F$. 
We observe that at lower orders there are significant differences between fixed-order and SCET-resummed cross-sections. 
At N$^3$LO, the scale dependence of the resummed cross-section is flat over a wide range of scales.
The dependence of the SCET-resummed cross-section on unphysical scales is reduced overall. This can be regarded as a means to find an optimal central value for our prediction.
Comparing fixed-order and SCET-resummed cross-section predictions at N$^3$LO we find perfect agreement for $\mu={m_H}/{2}$, which supports our preferred choice for the central scale. The upward bound of the uncertainty interval obtained by means of scale variation is comparable to the one obtained for the fixed-order cross-section. The lower bound of SCET-resummed cross-section scale variation interval is well contained within the fixed-order interval.

%, however, both methods show excellent agreement for the central scale $\mu=m_H/2$. 
%In general, we observe qualitative agreement between the SCET resummation considered in this section and the N$^3$LO+N$^3$LL resummation in Mellin space considered in the previous section: 
%, and the resummed cross-section is contained in the fixed-order scale variation in the range $[m_H/4,m_H]$. 
%As noted in ref.~\cite{Ahrens:2008nc,Ahrens:2008qu} at lower orders the main differences arise mainly due to the $\pi^2$ resummation. 

To conclude the analysis, we also need to assess the stability of our result under a variation of the soft, hard and top scales. We do this by varying these scales independently. The top-quark scale $\mu_t$ and the hard scale $\mu_h$ are varied by a factor of two up and down around their respective central values, while the soft scale is varied in the interval $\mu_s\in [\mu_s(m_H/4,m_H),\mu_s(m_H,m_H)]$. The effect of the variation of the hard, soft and top-quark scales is of the order of $\pm 0.1\%$ as noted already in ref.~\cite{Ahrens:2008nc}. 
%The result of the combined variation of all scales (including the perturbative scale $\mu$ for the central scale choice $\mu=m_H/2$), added linearly,  is shown in Tab.~\ref{tab:SCETTable}. 
As the derived uncertainty intervals and the central values of the SCET-resummed and fixed-order cross-sections are in very good agreement, we will not consider the SCET-resummed cross-section in subsequent chapters.

%\begin{table}
%\begin{center}
%\begin{tabular}{ c | c  c  c  c  c }
%\hline\hline
% &  $\delta_{\mu_h}$ [\%] & $\delta_{\mu_s}$ [\%] & $\delta_{\mu_t}$ [\%] & $\delta_{\mu}$ [\%] & $\sigma [pb]+\delta_{\text{all}}$ [\%]\\
% \hline
%%$ \sigma^{SCET}(\mu=m_h)$ & ${}^{0.05}_{-0.03}$ &  ${}^{0.05}_{-0.06}$   &  ${}^{0.007}_{-0.04} $   & ${}^{1.5}_{-0.26}$ & $44.3+{}^{1.6}_{0.4}$ \\
%%\hline
%$ \sigma^{SCET}(\mu=m_h/2)$ &  ${}^{0.08}_{-0.07} $   &  ${}^{0.04}_{-0.11} $   &  ${}^{0.04}_{-0.09} $ & ${}^{0.65}_{-0.25}$ & $45.08+{}^{0.81}_{-0.52} $ \\
% \hline
% $\sigma^{FO}(\mu=m_h/2)$ &     &    &  & ${}^{0.63}_{-3.31} $ & $45.08+{}^{0.63}_{-3.31} $ \\
% \hline\hline
%\end{tabular}
% \caption{
% \label{tab:SCETTable} 
% Prediction for the hadronic Higgs production cross-section using the SCET-resummed and fixed-order (FO) order partonic cross-sections and changes of the cross-sections under variation of the soft, hard, top-quark and perturbative scale.}
%\end{center}
%\end{table}

Let us conclude this section by commenting on the validity of using the $\pi^2$-resummation to predict constant terms at higher orders. Indeed, the exponentiation of the $\pi^2$ terms makes a prediction for terms proportional to powers of $\pi^2$, and it is of course interesting to see if these terms capture the bulk of the hard corrections not only at NLO, but also beyond. In particular, we can compare the numerical size of the constant term at N$^3$LO predicted by the exponentiation of $\pi^2$ to the exact soft-virtual cross-section at N$^3$LO of ref.~\cite{Anastasiou:2014vaa}. Since we are interested in fixed-order predictions, we start from eq.~\eqref{eq:resformula} and we choose the scales according to $\mu=\mu_t=\mu_s=m_H$ and $\mu_h^2=-m_H^2$. Note that choosing $\mu_h^2<0$ amounts to exponentiating $\pi^2$ terms to all orders.
Next, let us assume that we know the hard and soft functions to some order in perturbation theory, say through $\mathcal{O}(a_s^{n})$, and all anomalous dimensions governing the evolution equations to one order higher than required to obtain a result that is correct through order $n$. If we expand the SCET-resummed cross-section in perturbation theory, then we will reproduce the exact soft-virtual cross-section through $\mathcal{O}(a_s^{n})$. 
By expanding the SCET-resummed cross-section to one order higher we obtain a prediction of terms proportional to powers of $\pi^2$.
We want to assess the quality of this prediction by comparing it to the known values of soft-virtual cross-section at low orders. 
For example, before the coefficient of $\delta(1-z)$ at N$^3$LO was computed, all plus-distribution terms of the soft-virtual cross-section at N$^3$LO were already known~\cite{Moch:2005ky,Laenen:2005uz}. We could thus have made a prediction for the coefficient of $\delta(1-z)$ at N$^3$LO based on $\pi^2$ resummation. 
If we denote by $C^{(n)}_{\delta}$ the coefficient of the distribution $\delta(1-z)$ in the partonic soft-virtual cross-section accurate through $\ord(a_s^n)$, and where the term proportional to $a_s^{n+1}$ was obtained from the exponentiation of $\pi^2$, we obtain the following sequence of predictions:
\beq\bsp
C^{(0)}_\delta &\,= 1+14.80 \,a_s +\mathcal{O}\left(a_s^2\right)\,,\\
C^{(1)}_\delta &\,= 1+9.87\,a_s +45.35 \,a_s^2+\mathcal{O}\left(a_s^3\right)\,,\\
C^{(2)}_\delta &\,= 1+9.87 \,a_s +13.61\,a_s^2-554.79 \,a_s^3+\mathcal{O}\left(a_s^4\right)\,,\\
C^{(3)}_\delta &\,= 1+9.87 \,a_s+13.61 \,a_s^2+1124.31 \,a_s^3+\mathcal{O}\left(a_s^4\right)\,,
\esp\eeq
with $a_s\equiv a_s(m_H^2)$. In the previous expressions the Wilson coefficient was set to unity and the number of colours and light flavours are $N_c=3$ and $N_f=5$ respectively, and we truncated all numerical results after two digits.
We observe that, in the scenario where only the LO cross-section is known, we are able to predict the order of magnitude of the NLO correction, and this prediction would indeed suggest large corrections at NLO. At higher orders, however, the quality of the predictions deteriorates, and in $C_\delta^{(2)}$ even the prediction of the sign of the N$^3$LO correction is wrong. Even if we include the coefficients of the other distributions contributing to the soft-virtual cross-sections, we observe a similar unsatisfactory pattern. 
We conclude that $\pi^2$ terms originating from the systematic exponentiation of the analytic continuation of the hard function constitute only one source of large perturbative corrections to the Higgs boson cross-section, and hence on its own this procedure of predicting higher orders does not provide reliable estimates of the missing dominant corrections.

%To summarize, in this section we have studied  various ways of assessing the size of the missing higher-order terms at N$^4$LO and beyond. In particular, we have looked at higher-order terms generated by interference of quantum corrections to the Wilson coefficient and the coefficient functions $\eta_{ij}^{(n)}$ in eq.~\eqref{eq:sigmaeff}, as well as by studying different way of performing a resummation of logarithmically-enhanced terms in the threshold limit $z\to1$. While all these prescriptions to perform the threshold resummation are formally equivalent in the strict threshold limit, they differ by power-suppressed terms. While formally subleading, these terms can have sizeable phenomenological impact. We observe, however, that in all cases these higher-order corrections are negligible for the scale $\mu=m_H/2$, and they are contained in the scale variation at N$^3$LO for $\mu\in[m_H/4,m_H]$. We thus conclude that, in agreement with our expectation in Section~\ref{sec:EFT}, scale variation provides a reliable estimate of missing higher-order terms at N$^3$LO. For this reason, we do not consider resummation effects any further and we henceforth only consider fixed-order results.

% !TEX root = paper.tex

\section{Quark-mass effects}
\label{section:top_quark_mass_effects}

So far we have only considered QCD corrections to the effective theory where the top quark is infinitely heavy. In this section we discuss effects that are not captured by the effective theory, but that can still give rise to sizeable contributions. In particular, we discuss the inclusion of quark-mass effects from top, bottom and charm quarks, to the extent that these corrections are available in the literature.
We start by discussing the effect of quark masses at LO and NLO, where it is possible to obtain exact results including all quark-mass effects. In order to stress the importance of including these effects, we remind that the cross-section changes by $+6.3\%$ already at LO if the exact top-mass dependence is taken into account. The exact mass dependence of the cross-section is also known at NLO~\cite{Graudenz:1992pv,Djouadi:1991tka,Spira:1995rr, Harlander:2005rq,Aglietti:2006tp, Bonciani:2007ex, Anastasiou:2006hc, Anastasiou:2009kn, Aglietti:2004nj}, and we can thus include all effects from top, bottom and charm quarks up to that order.
The value of the cross-section through NLO as we add quark-mass effects for the parameters of Setup 1 (cf. Tab.~\ref{tab:setup1}) is summarized in Tab.~\ref{tab:quark_effects}. 
Beyond NLO finite quark-mass effects are in general unknown, and they can at best be included in an approximate fashion.

\begin{table}[!htb]
%\tiny\setlength{\tabcolsep}{5pt}
%
%    table 1
%
\begin{minipage}{\linewidth}
\centering

\caption{Quark-mass effects for the parameters of Setup 1.}
\label{tab:quark_effects}

\medskip

\begin{tabular}{cc|cc}
    \toprule
    
    $\sigma_{EFT}^{LO}$	&  15.05 pb 			&    		$\sigma_{EFT}^{NLO}$ & 34.66 pb               \\
    $R_{LO}\,\sigma_{EFT}^{LO}$	&  16.00 pb &    $R_{LO}\,\sigma_{EFT}^{NLO}$ & 36.84 pb                 \\
    $\sigma_{ex;t}^{LO}$	&  16.00 pb &    $\sigma_{ex;t}^{NLO}$ & 36.60 pb                 \\
    $\sigma_{ex;t+b}^{LO}$	&  14.94 pb &    $\sigma_{ex;t+b}^{NLO}$ & 34.96 pb                 \\
    $\sigma_{ex;t+b+c}^{LO}$	&  14.83 pb &    $\sigma_{ex;t+b+c}^{NLO}$ & 34.77 pb                 \\

    \bottomrule
\end{tabular}
\end{minipage}
\end{table}

Let us start by analyzing finite top-mass effects.
The exact NLO cross-section is approximated well by rescaling the EFT cross-section at NLO by the leading-order ratio $R_{LO}$ defined in eq.~\eqref{eq:KLO}. 
For example, within Setup 1 we have $R_{LO}=1.063$, and we see from Tab.~\ref{tab:quark_effects} that the rescaled NLO cross-section in the effective theory, $R_{LO}\,\sigma_{EFT}^{NLO}$, reproduces the NLO cross-section $\sigma_{ex;t}^{NLO}$ with full top-mass dependence within $0.65\%$. Because of this, it has become standard to multiply the EFT cross-section at NNLO by $R_{LO}$, and we follow this prescription also for the N$^3$LO coefficient. 

In addition to this rescaling, in ref.~\cite{Harlander:2009mq, Pak:2009dg, Harlander:2009bw, Harlander:2009my} top-mass corrections at NNLO were computed as an expansion in $m_H/m_t$, after factorizing the exact LO cross-section. We include these corrections into our prediction via the term $\delta_{t}\hat{\sigma}_{ij,EFT}^{NNLO}$ in eq.~\eqref{eq:master}. In particular, we include the contribution from the subleading $1/m_t$ terms for the numerically significant $gg$ and $qg$ channels~\cite{Harlander:2009my}. The $gg$ channel increases the rescaled EFT cross-section at NNLO by roughly $+0.8\%$, while the $qg$ channel leads to a negative contribution of $-0.1\%$, so that the total net effect is of the order of $+0.7\%$. Note that the small size of these effects corroborates the hypothesis that the cross-section in the effective theory rescaled by $R_{LO}$ gives a very good approximation of the exact result. 

Despite the fact that the approximation is good, 
these contributions come with an uncertainty of their own: the $1/m_t$ expansion is in fact an expansion in ${s}/m_t^2$, and consequently it needs to be matched to the high-energy limit of the cross-section, known to leading logarithmic accuracy from $k_t$-factorization. The high-energy limit corresponds to the contribution from small values of $z$ to the convolution integral in eq.~\eqref{eq:sigma}. Since this region is suppressed by the luminosity, a lack of knowledge of the precise matching term is not disastrous and induces an uncertainty of roughly $1\%$, which is of the order of magnitude of the net contribution. In conclusion, following the analysis of ref.~\cite{Harlander:2009my}, whose conclusions were confirmed by ref.~\cite{Pak:2009dg}, we assign an overall uncertainty of $1\%$ due to the unknown top-quark effects at NNLO.  

So far we have only discussed the effect of including top mass effects at NNLO. 
Despite their suppressed Yukawa couplings, the bottom and charm quarks also contribute to the Higgs cross-section, mainly through interference with the top quark. Indeed, we can easily see from Tab.~\ref{tab:quark_effects} that the inclusion of bottom-quark effects at LO and NLO leads to sizeable negative contributions to the cross-section, and hence it is not unreasonable to expect this trend to continue at NNLO. Unlike the case of the top quark, however, the contributions of the bottom and charm quarks at NNLO are entirely unknown. We estimate the uncertainty of the missing interference between the top and light quarks within the $\overline{ \rm MS }$ as:
\beq
\label{eq:tbc_uncertainty}
\delta(tbc)^{\overline{\rm MS}} = \pm\, \left| \,\frac{\delta\sigma_{ex;t}^{NLO}-\delta\sigma_{ex;t+b+c}^{NLO}}{\delta\sigma_{ex;t}^{NLO}}\,\right|\,
(R_{LO}\delta\sigma_{EFT}^{NNLO}+\delta_{t}\hat{\sigma}_{gg+qg,EFT}^{NNLO}) \simeq \pm 0.31\,\textrm{pb}\,,
\eeq
where 
\beq
\delta\sigma_{X}^{NLO} \equiv \sigma_{X}^{NLO}-\sigma_{X}^{LO}
{\rm~~and~~}
\delta\sigma_{X}^{NNLO} \equiv \sigma_{X}^{NNLO}-\sigma_{X}^{NLO}\,.
\eeq
With respect to the NNLO cross-section with the exact top effects described in the previous paragraph, this uncertainty is at the level of $0.6\%$, but it becomes slightly larger at lower energies. For example, at a 2 TeV proton-proton collider it increases to 1.1\%.

So far, we have assumed that all quark masses are given in the $\overline{\textrm{MS}}$-scheme. We now analyze how our predictions are affected if we use the on-shell (OS) scheme. In Tab.~\ref{tab:rs_dep_top} we summarize the values of the NLO cross-sections with the quark masses of Setup 1 ($\overline{\textrm{MS}}$) and Setup 2 (OS) for a common scale choice $\mu_F=\mu_R=m_H/2$.  Moreover, the ratio $R_{LO}$ as well as the Wilson coefficient multiplying the cross-section are functions of the top mass, and so they are affected by the choice of the renormalization scheme.

First, let us comment on the use of the OS-scheme for the top-quark mass on the Wilson coefficient. 
The analytic expression for the Wilson coefficient in the two schemes
is the same through NNLO but differs at N$^3$LO (see Appendix~\ref{app:WC_MSbar-OS}).  However, this difference is compensated 
by the different values of the 
top-quark mass in the two schemes and  the numerical value of the Wilson coefficient in the two schemes at \nnnlo agrees to better than a per mille
 (see penultimate line of Tab.~\ref{tab:rs_dep_top}). 
Next, let us turn to the scheme-dependence of $R_{LO}$.
For the top mass of Setup 1 ($\overline{\textrm{MS}}$), the value of this ratio is $R_{LO}=1.063$, while for the top mass of Setup 2 (OS), we find $R_{LO}=1.066$, i.e., the scheme dependence of the rescaled EFT prediction is at the level of $0.3\%$.

\begin{figure}[!t]
\begin{center}
\includegraphics[width=0.8\textwidth]{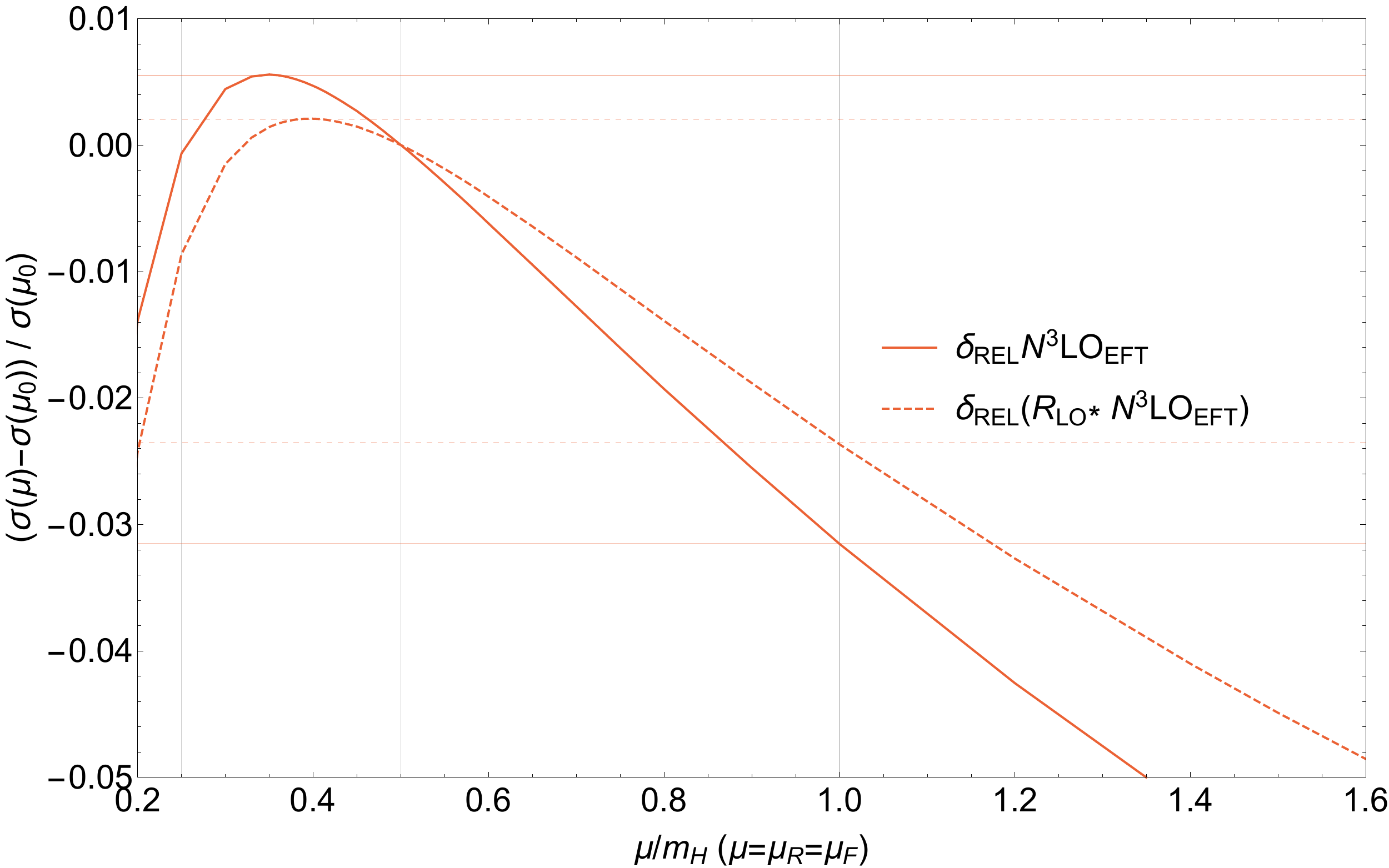}
\end{center}
\caption{
\label{fig:mu_eft_pure_vs_rescaled} 
The dependence of the cross-section on a common renormalization and 
 factorization scale $\mu = \mu_F = \mu_R$ in the EFT vs the EFT
 rescaled with the exact LO contribution in the $\overline{\textrm{MS}}$-scheme.}
\end{figure}

Since the top mass runs in the $\overline{\textrm{MS}}$-scheme, the LO cross-section
acquires its own scale dependence through the dependence of the
top mass on the renormalization scale. In Fig.~\ref{fig:mu_eft_pure_vs_rescaled} we
compare the two approximations as a function of the renormalization schale $\mu$ in the $\overline{\textrm{MS}}-$scheme.
We observe that the scale variation of the
rescaled-EFT cross-section is slightly smaller. The variation of the
rescaled N$^3$LO cross-section in the scale range $\mu\in[\frac{m_H}{4}, m_H]$ is $\pm 1.3\%$ (compared to $\pm 1.9\%$ in the pure EFT, cf. Section~\ref{sec:scale}). Note that in the OS-scheme the scale uncertainty is the same for the rescaled and pure EFT cross-sections, because the ratio $R_{LO}$ is a constant in this scheme.

%Finally, let us comment on the renormalization scheme dependence of the exact NLO cross-section when top, bottom and charm quark masses are included. 
%
%We quantify the dependence as the ratio
%\beq
%(\sigma_{ex;t+b+c}^{\overline{\textrm{MS}}}-\sigma_{ex;t+b+c}^{\textrm{OS}})/\sigma_{ex;t+b+c}^{\overline{\textrm{MS}}}\,.
%\eeq
%In Tab.~\ref{tab:rs_dep_top} we show the NLO cross-sections with the quark masses of setups 1 ($\overline{\textrm{MS}}$) and 2 (OS) for a common scale choice $\mu=\mu_F=\mu_R=m_H/2 (m_H)$ {\bf CD: Is it correct that we use $\mu=m_H/2$?}. We observe that the cross-section varies by $0.1\%$ ($0.3\%$) with the top quark only, but by $1.3\%$ ($2.3\%$) when we include the bottom quark, and by $2.1\%$ ($3.3\%$) when we also include the charm quark.  The scheme dependence is, in fact, varying with the renormalization scale, from $0.9\%$ at $\mu_R=m_h/4$ to $3.3\%$ at $\mu_R=m_h$, mainly due to the running of the bottom, and to a lesser degree, of the charm quark mass.

\begin{table}[!t]
%\tiny\setlength{\tabcolsep}{5pt}
%
%    table 1
%
\begin{minipage}{\linewidth}
\centering

\caption{Dependence on the renormalization scheme  for the  quark masses of Setup 1 and Setup 2. The relative scheme dependence is defined as $\delta\sigma^{sc}=(\sigma^{\textrm{OS}} / \sigma^{\overline{\textrm{MS}}} -1 )\times 100\%$.}
\label{tab:rs_dep_top}

\medskip

\begin{tabular}{cc|c|c}
    \toprule
    &     $\overline{\textrm{MS}}$ &     OS    & $\delta\sigma^{sc}$\\  
      \midrule
  $\sigma_{ex;t}^{LO}$	&  16.00 pb &      16.04 pb               & 0.25\%  \\
  $\sigma_{ex;t+b}^{LO}$	&  14.94 pb &     14.24 pb     &       -4.8\%     \\
  $\sigma_{ex;t+b+c}^{LO}$	&  14.83 pb &     13.81  pb  & -6.9\%             \\
 \midrule
    $\sigma_{ex;t}^{NLO}$	&  36.60 pb &      36.63 pb               & 0.08\%  \\
    $\sigma_{ex;t+b}^{NLO}$	&  34.96 pb &     34.49 pb     &       -1.3\%     \\
    $\sigma_{ex;t+b+c}^{NLO}$	&  34.77 pb &     34.04  pb  &          -2.1\%     \\
     \midrule
      $\sigma_{EFT}^{NNLO}$	&  43.65 pb &   43.66  pb            & 0.02\%    \\
      $R_{LO}\,\sigma_{EFT}^{NNLO}$	&  46.39 pb &    46.53  pb &        0.3\%        \\
      $\sigma_{EFT}^{N^3LO}$	&  45.06 pb &    45.06  pb              & 0\% \\
      $R_{LO}\, \sigma_{EFT}^{N^3LO}$	&  47.88 pb &     48.03  pb &      0.3\%          \\
       \bottomrule
\end{tabular}
\end{minipage}
\end{table}  

%The LO dependence on the scheme is at the level of $8.1\%$, as seen at the first entry of table~\ref{tab:rs_dep_top}. Hence we expect that once the Higgs cross-section including light quark interference effects is known at NNLO, the scheme dependence will be further reduced to per cent level or below.   

The largest scheme dependence appears at LO and NLO due to the
non-negligible interference between top and light quarks (see Tab.~\ref{tab:rs_dep_top}). At LO, the results for the cross-section 
in the two schemes are in excellent agreement.
However, including bottom and charm quark loops gives rise to substantial 
differences, which at LO are as large as $-6.9\%$. 
While the difference between the two schemes is reduced to  $-2.1\%$ at NLO, it still remains larger
than the uncertainty estimate of eq.~\eqref{eq:tbc_uncertainty}. 

From Tab.~\ref{tab:scheme_Kfactors} it becomes evident that
the difference between the results in the two schemes originates from the
light-quark contributions. The first line of Tab.~\ref{tab:scheme_Kfactors} shows that, if we only include mass effects from the top quark
through NLO, then the results in both schemes are in perfect agreement. Fortunately, 
light-quark contributions are suppressed in the Standard Model in comparison to the
pure top-quark contributions.  
\begin{table}[!t]
%\tiny\setlength{\tabcolsep}{5pt}
%
%    table 1
%
\centering
\caption{NLO K-factors ($K={\sigma^{ NLO}}/{\sigma^{ LO}}$) in the $\overline{{\rm MS}}$ and OS schemes and the ratio of the cross-sections
($R_{\rm scheme} = \sigma^{\overline{\rm MS}}/{\sigma^{\rm OS}}$) at LO and NLO, 
for various quark flavor combinations in the loops.}
\label{tab:scheme_Kfactors}
\medskip
\begin{tabular}{c|c|c|c|c}
    \toprule
  &  $K_{\overline{\rm MS}}$ &    $K_{\rm OS}$    & $ R_{\rm scheme}^{ LO}$ & $ R_{\rm scheme}^{ NLO}$  \\  
      \midrule
 $\sigma_{t}$  &  2.288& 2.284& 0.998& 0.999  \\
 $\sigma_{b}$& 2.39 & 1.58 & 0.22& 0.33  \\
 $\sigma_{c}$& 2.58 & 1.38 & 0.05& 0.09 \\
 $\sigma_{t+b}$& 2.34& 2.42& 1.05& 1.01 \\
 $\sigma_{t+c}$& 2.29& 2.32& 1.02& 1.01  \\
 $\sigma_{b+c}$& 2.41& 1.55& 0.18& 0.28  \\
 $\sigma_{t+b+c}$& 2.35& 2.47& 1.07& 1.02  \\
$\sigma_{t+b+c}-\sigma_{t}$& 1.56& 1.16& 0.53& 0.71 \\
\bottomrule
\end{tabular}
\end{table}  
Indeed, if we set the top-quark Yukawa coupling to zero and only include contributions from bottom and charm quarks
(see third line from the
bottom of Tab.~\ref{tab:scheme_Kfactors}), we observe that the
NLO cross-section in the $\overline{\rm MS}$ scheme is only about a third compared to its value in the OS-scheme. 
Similarly, the value of the
cross-section changes by an order of
magnitude between the two renormalization schemes if only the charm-quark loop is included and both bottom and top-quark Yukawa couplings are set to zero. 
The very large size of the differences between the two schemes shed some doubt on how well we control the perturbative corrections due to light-quark masses even at NLO. Nonetheless, it is at least encouraging that the scheme
dependence of the contributions from bottom and charm quarks alone is significantly smaller at NLO than
at LO. Moreover, since the cross-section is dominated by the top quark, 
the overall scheme dependence due to the light quarks is significantly
reduced 
when the top-quark Yukawa coupling is set to its physical coupling. 

For our purposes, the most interesting contribution is
$\sigma_{t+b+c}-\sigma_t$ in the last line of
Tab.~\ref{tab:scheme_Kfactors}, 
which is the difference between the exact cross-section and the cross-section when all
Yukawa couplings, except for the Yukawa coupling of the top quark, are set to zero.   
This part  of the cross-section is only known through NLO and 
is not captured (at least not in any direct or trustworthy way)
by existing NNLO computations\footnote{For first steps
  towards computing this contribution at NNLO we refer the reader to
  ref.~\cite{Mueller:2015lrx}.}. We observe that the NLO $K$-factor of this
contribution is smaller than the NLO $K$-factor of pure top-quark
contributions in the cross-section.  Therefore, we anticipate that the
estimate of the magnitude of the $\sigma_{t+b+c}-\sigma_t$ correction
at NNLO, based on the size of the top-only NNLO $K$-factor in 
eq.~\eqref{eq:tbc_uncertainty}, is a conservative estimate within the
$\overline{\rm MS}$-scheme. However, as we notice from the value of
$R_{\rm scheme}^{NLO}$, there  is a scheme dependence of $\sim 30
\%$ at NLO. Our preferred scheme is the $\overline{\rm MS}$-scheme
  due to the bad convergence of the perturbative series
  for the conversion from an $\overline{\rm MS}$ mass to a pole
  mass for the bottom and charm
  quarks~\cite{Marquard:2016vmy,Marquard:2015qpa}. 
To account for the difference with the OS scheme, we enlarge the uncertainty on 
$\sigma_{t+b+c} - \sigma_t$, as estimated via 
eq.~\eqref{eq:tbc_uncertainty}  within the  $\overline{\rm MS}$
scheme, by multiplying it with a factor of 1.3, %an appropriate  factor
%. At 13 TeV this factor is  1.3, i.e., we have
\begin{equation}
\delta(t,b,c) = 1.3  \, \delta(t,b,c)^{\overline{\rm MS}}\,.
\end{equation}
%but it increases with decreasing collider energy due to the larger NNLO to NLO 
%$K$-factors.  For example, at 2 TeV center of mass energy the rescaling factor is 1.x.  
%{\bf [CD: There is an `x' at the end of the previous paragraph.]}

Let us conclude this section by commenting on the amount by which the cross-section changes when the values of the quark masses used as input vary from those of Setup 1. As argued in the previous section, the dependence on the rescaled EFT cross-section on the top-quark mass is extremely mild. We will therefore focus in this section on the exact QCD corrections (including the light quarks) through NLO, and we study the variation of the cross-section when the quark masses are varied following the internal note of the HXSWG~\cite{Denner:2047636}, which either conforms to the PDG recommendation or is more conservative (see Tab.~\ref{tab:parametric_uncertainties}).
We see that the parametric uncertainties are entirely negligible, at the level of $0.1\%$ or below.  Finally, the parametric uncertainty on the ration $R_{LO}$ does not exceed $0.1\%$. For this reason, we will not consider parametric uncertainties on quark masses any further.

\begin{table}[!htb]
%\tiny
\scriptsize\setlength{\tabcolsep}{5pt}
%
%    table 1
%
\centering

\caption{Parametric uncertainties on quark masses.}
\label{tab:parametric_uncertainties}

\medskip

\begin{tabular}{c|cc|c|cc|c|cc}
    \toprule
    \multicolumn{3}{c|}{Top quark} & \multicolumn{3}{c|}{Bottom quark} & \multicolumn{3}{c}{Charm quark}\\
    \midrule
     $\delta m_t=1\textrm{ GeV}$ & $\sigma_{ex;t+b+c}^{NLO}$	&  34.77& 
     $\delta m_b= 0.03 \textrm{ GeV}$ & $\sigma_{ex;t+b+c}^{NLO}$	&  34.77&
     $\delta m_c = 0.026 \textrm{ GeV}$ & $\sigma_{ex;t+b+c}^{NLO}$	&  34.77 
        \\  
      \midrule
    $m_t+\delta m_t$&$\sigma_{ex;t+b+c}^{NLO}$ & 34.74   &
    $m_b+\delta m_b$&$\sigma_{ex;t+b+c}^{NLO}$ & 34.76  &
    $m_c+\delta m_c$&$\sigma_{ex;t+b+c}^{NLO}$ & 34.76             
     \\
    $m_t-\delta m_t$&$\sigma_{ex;t+b+c}^{NLO}$ & 34.80       &
    $m_b-\delta m_b$&$\sigma_{ex;t+b+c}^{NLO}$ & 34.79         &
    $m_c-\delta m_c$&$\sigma_{ex;t+b+c}^{NLO}$ & 34.78           \\
       \bottomrule
\end{tabular}
\end{table}

% !TEX root = paper.tex

\section{Electroweak corrections}
\label{sec:EW}

\begin{figure}[!t]
\begin{center}
\includegraphics[width=0.8\textwidth]{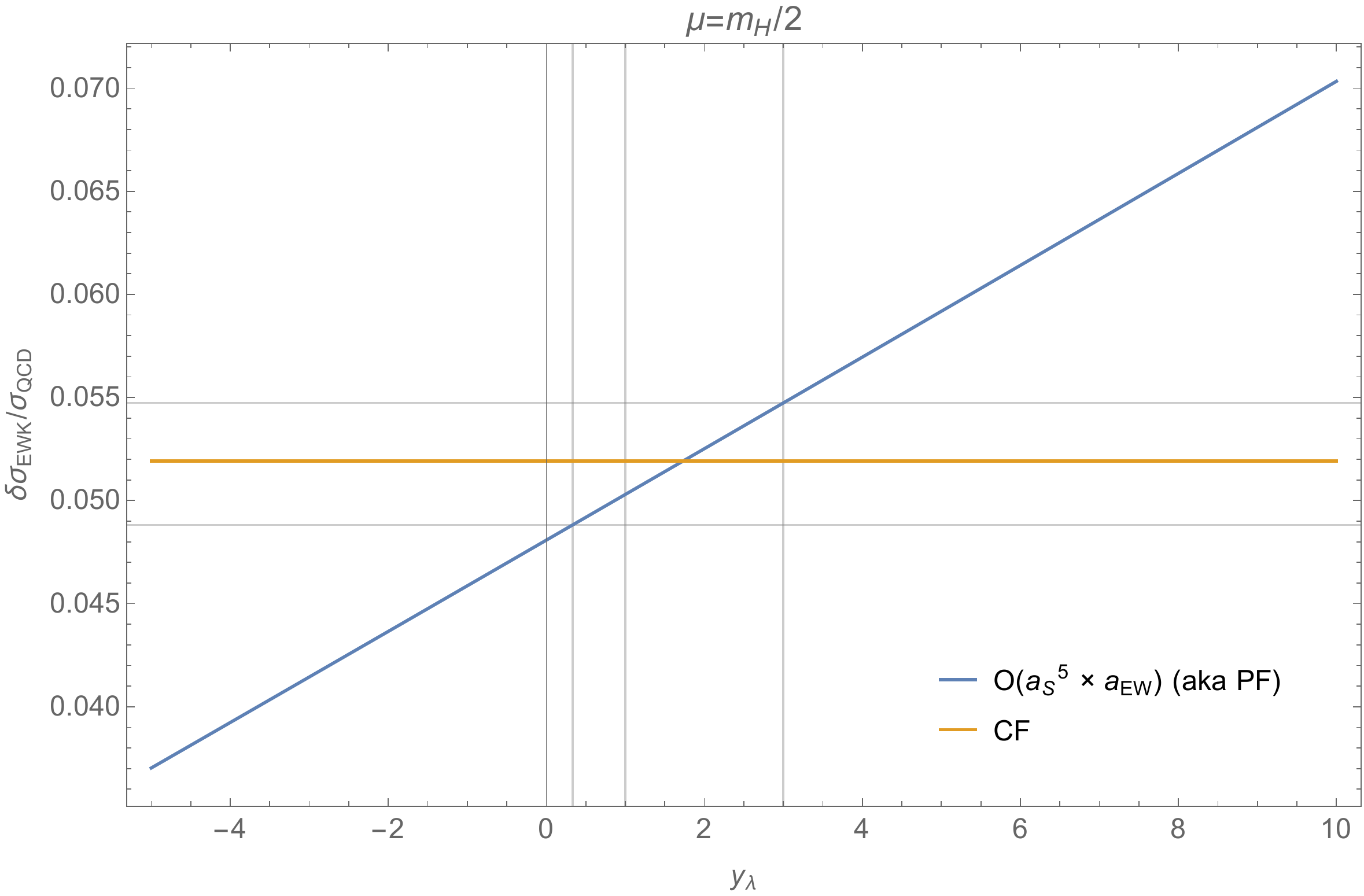}
\end{center}
\caption[EW corrections dependence on $C_{1w}$.]{
\label{fig:ew_corrections_study}
Relative EW corrections as a function of the parameter $y_\lambda$ defined in eq.~\eqref{eq:y_lamda}. Within the range $y_\lambda\in [1/3,3]$ the EW corrections are modified by $-0.2\%$ to $+0.4\% $. The EW correction under the assumption of complete factorization (CF) lies in the middle of the variation range.
}
\end{figure}

So far we have only considered higher-order QCD corrections to the gluon fusion cross-section. However, in order to obtain precise predictions for the Higgs cross-section also electroweak (EW) corrections need to be taken into account.
The EW corrections to the LO gluon fusion cross-section have been computed in ref.~\cite{Aglietti:2004nj,Actis:2008ug,Actis:2008ts}. For a Higgs mass of $m_H=125$ GeV, they increase the LO cross-section by $5.2\%$, and we take these corrections into account in our cross-section prediction.

Given the large size of the NLO QCD corrections to the Higgs cross-section, we may expect that also the EW corrections to the NLO QCD cross-section cannot be neglected. Unfortunately, these so-called mixed QCD-EW corrections are at present unknown. The contribution from light quarks, which at ${\cal O}(a_{EW}a_s^2)$ is the dominant one accounting for \mbox{$\sim95\%$} of the total EW corrections at that order, was computed in ref.~\cite{Anastasiou:2008tj} within an effective field theory approach where the $W$ and $Z$ bosons are assumed heavier than the Higgs boson and have been integrated out. This has the effect of introducing EW corrections to the Wilson coefficient describing the effective coupling of the Higgs boson to the gluons in eq.~\eqref{eq:L_eff},
\beq
\label{eq:EFTewk}
C \equiv C_{QCD} + \lambda_{EW}\, (1+C_{1w}\,a_s + C_{2w}\,a_s^2 +\ldots)\,,
\eeq 
where $C_{QCD}$ encodes the pure QCD corrections to the Wilson coefficient, $\lambda_{EW}$ denotes the EW corrections to the LO cross-section of ref.~\cite{Actis:2008ug} and $C_{1w}$ are the mixed QCD-EW corrections in the EFT approach of ref.~\cite{Anastasiou:2008tj}. The value of the coefficient $C_{1w}$ is~\cite{Anastasiou:2008tj}
\beq\label{eq:C1w}
C_{1w}=\frac{7}{6}\,.
\eeq
Adopting the modification of the Wilson coefficient also for higher orders in $a_s$ leads to a total correction of $5.0\%$.
%,  independently of the value of the coefficient $C_{2w}$ to below per mille level for reasonable values of $C_{2w}$
 We stress that the numerical effect of this correction is very similar to that of the `complete factorization' approach to include EW corrections of ref.~\cite{Actis:2008ug}, which lead to an increase of the NLO cross-section by $5.1\%$.
  
The effective theory method for the mixed QCD-EW corrections 
is of course not entirely satisfactory, because the computation of the EW Wilson coefficient assumes the validity of the $m_H/m_V$ expansion, $V=W,Z$ while clearly $m_H>m_V$. 
We thus need to carefully assess the uncertainty on the mixed QCD-EW corrections due to the EFT approximation. 
In the region $m_H > m_V$, we expect threshold effects to be important and one should not expect that a naive application of the  EFT can give a reliable value for the cross-section. However, in eq.~\eqref{eq:EFTewk} the EFT is only used to predict 
the relative size of QCD radiative corrections with respect to the leading 
order electroweak corrections. This can only vary mildly above and below threshold. For phenomenological purposes, we expect that the rescaling with the exact $\lambda_{EW}$ in  eq.~\eqref{eq:EFTewk} captures the bulk of threshold effects at 
all perturbative orders. To quantify the remaining uncertainty in this approach, we allow the coefficient $C_{1w}$ to vary by a factor of $3$ around its central value in eq.~\eqref{eq:C1w}. We do this by introducing a rescaling factor $y_{\lambda}$ by
\beq\label{eq:y_lamda}
\lambda_{EW}\, (1+C_{1w}\,a_s +\ldots) \to \lambda_{EW} \,(1+y_\lambda\, C_{1w}\,a_s +\ldots)\,.
\eeq
Varying $y_{\lambda}$ in the range $[1/3,3]$, we see that the cross-section varies by $-0.2\%$ to $+0.4\% $. We summarize the dependence of the cross-section on $y_{\lambda}$ in Fig.~\ref{fig:ew_corrections_study}. Note that the result obtained by assuming complete factorization of EW and QCD corrections (marked by `CF' in Fig.~\ref{fig:ew_corrections_study}) lies in the middle of the variation range, slightly higher than the $y_\lambda=1$ prediction.  Finally, we stress that the choice of the range is largely arbitrary of course. It is worth noting, however, that in order to reach uncertainties of the order of $1\%$, one needs to enlarge the range to $y_\lambda\in[-3,6]$.
% which is to our eyes overly conservative. 

An alternative way to assess the uncertainty on the mixed QCD-EW corrections is to note that the factorization of the EW corrections is exact in the soft and collinear limits of the NLO phase space. The hard contribution, however, might be badly captured. At NLO in QCD, the hard contribution amounts to $\sim 40\%$ of the $\ord(a_s^3)$ contribution to the cross-section, where we define the \emph{hard contribution} as the NLO cross-section minus its soft-virtual contribution, i.e., the NLO contribution that does not arise from the universal exponentiation of soft gluon radiation (see Section~\ref{sec:MHO}). In the notation of Section~\ref{sec:EFT} the \emph{hard contribution} is defined as the convolution of the parton-level quantity
\beq
\frac{\hat{\sigma}_{ij}^{(1),\textrm{hard}}}{z}\equiv\frac{\pi |C_0|^2}{8V}\,a_s^3\,\eta^{(1),\textrm{reg}}_{ij}(z)
\eeq
with the PDFs, which
receive contributions from the $gg$, $qg$ and $q\bar{q}$ initial state channels.
The mixed QCD-EW corrections  are $3.2\%$ of the total cross-section. Even if the 
 uncertainty of the  factorization ansatz is taken to be as large as the entire hard contribution, we will obtain
an estimate of the uncertainty equal to $0.4\times 3.2\%=1.3\%$ with respect to the total cross-section. 
  
An alternative way to define the \emph{hard contribution} is to look at the real emission cross-section regulated by a subtraction term in the FKS scheme~\cite{Frixione:1995ms}. We could then exclude the contribution of the integrated subtraction term, which is proportional to the Born matrix element, and hence of soft-collinear nature. We would then estimate the \emph{hard contribution} as $\sim 10\%$ of the $\ord(a_s^3)$ contribution to the cross-section, which would lead to an uncertainty equal to $0.1\times 3.2\%=0.32\%$.
  
 We note that the different estimates of the uncertainty range from $0.2\%$ to $1.3\%$.
We therefore assign, conservatively, an uncertainty of $1\%$ due to mixed QCD-EW corrections for 
LHC energies. This uncertainty decreases for smaller collider energies as the soft contributions become more 
important and the factorization ansatz becomes more accurate. For example,  at a 2 TeV proton-proton collider  
the most conservative estimate of the  uncertainty is $0.8\%$.

\section{PDF comparison}
\label{sec:pdfs}
So far we have only discussed perturbative higher-order corrections to the partonic cross-sections. 
The full hadronic cross-section is then obtained by convoluting the partonic coefficient functions
by the parton distribution functions.
In the last few years significant progress has been made towards
the improvement of the PDF fits, also  through the inclusion of new 
data from collider and fixed-target experiments. 
We refer to the analysis in the latest PDF4LHC working group paper~\cite{Butterworth:2015oua}
for a review of the updated sets 
ABM12~\cite{Alekhin:2013nda}, CT14~\cite{Dulat:2015mca}, JR14~\cite{Jimenez-Delgado:2014twa}, 
MMHT2014~\cite{Harland-Lang:2014zoa}, NNPDF3.0~\cite{Ball:2014uwa} and 
HERAPDF2.0~\cite{Abramowicz:2015mha}, which are available through NNLO, 
as well as the NLO set CJ12~\cite{Owens:2012bv}.
In this Section, we will compare the predictions from various pdf sets using 
Setup 1 and the partonic cross-sections derived in the rescaled EFT 
through \nnnlo for a factorisation and renormalisation scale $\mu = {m_H}/2$. 
 
%Comparison of the predictions from different sets in specific benchmark 
%scenarios have allowed a better understanding of the discrepancies 
%that were previously observed, and lead an improvement in the 
%techniques employed by the PDF4LHC group for the combination of these 
%sets\footnote{CT14, MMHT2014 and NNPDF3.0.}
%and the estimate of the final uncertainties. Furthermore, 
The three sets that enter the PDF4LHC fit (CT14, MMHT14 and NNPDF3.0)
and HERAPDF2.0, are provided 
at the same value of the strong coupling constant as the global
PDF4LHC15 combination~\cite{Butterworth:2015oua},
\beq
\alpha_s(m^2_Z) = 0.118 \,.
\eeq
This value is consistent with the PDG average~\cite{Agashe:2014kda}. 
%For all the sets, the $\alpha_s$ uncertainty is calculated over the range 
%given by the PDF4LHC group, 
%\beq
%	\alpha_s(m^2_Z) = 0.1180 \pm 0.0015 \;,
%\eeq
%as
\begin{figure}[tb]
\begin{center}
\includegraphics[width=0.9\textwidth]{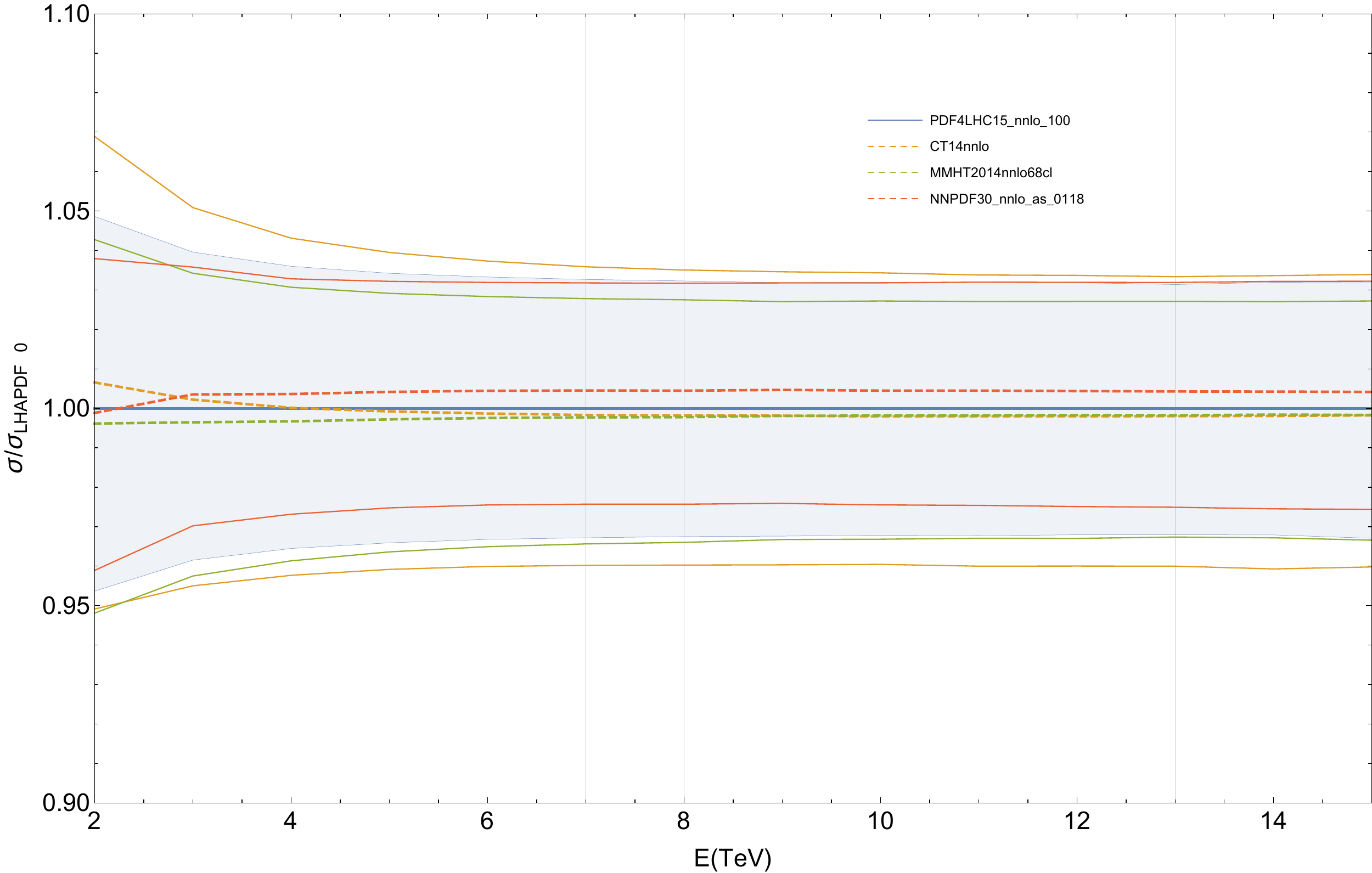}
\end{center}
\caption{
Higgs production cross-section and the relative PDF$+\alpha_s$ uncertainty 
at 68\% C.L. using the CT14, MMHT2014 and NNPDF3.0 sets, normalized by the 
central value obtained with the PDF4LHC15 combination.}
\label{fig:pdf_comparison} 
\end{figure}
 %
 %
%% \beq
%%	\delta(\alpha_s)=\frac{\sigma(\alpha_s=0.1195)-\sigma(\alpha_s=0.1165)}{2} \;.
%%\eeq

In Fig.~\ref{fig:pdf_comparison} we compare the 68\% C.L. predictions 
from CT14, MMHT2014 and NNPDF3.0 with those from the PDF4LHC15 combination. 
For comparison purposes, in this section we combine (potentially asymmetric) PDF and $\alpha_s$ uncertainties in quadrature\footnote{We note that the probabilistic interpretation of such an uncertainty combination in terms of confidence level intervals is not straightforward, when the individual uncertainties are not symmetric~\cite{Lai:2010nw}. 
%We circumvent this issue by following the recommendation of the PDF4LHC working group~\cite{Butterworth:2015oua} when it comes to uncertainty recommendations in Sec.~\ref{sec:recommendation}.
For a detailed discussion of the (PDF+$\alpha_s$) uncertainty entering our final recommendation for the value of the cross-section, see Section~\ref{sec:recommendation}.
},
\beq
\delta_\pm( PDF + \alpha_s ) = \sqrt{\delta_\pm(PDF)^2 + \delta_\pm(\alpha_s)^2 }\,.
\eeq
%Following  the recommendations of the PDF4LHC working group, 
%the PDF$+\alpha_s$ uncertainty is calculated as~\cite{Butterworth:2015oua} 
%\beq
%	\delta(\textrm{PDF}+\alpha_s) = \sqrt{\delta(\textrm{PDF})^2+\delta(\alpha_s)^2} \,.
%	\eeq
%This prescription, which we follow in this paper, yields symmetric error bands for the $\alpha_s$ uncertainty. 
%around the central value for $\alpha_s=0.118$. 
%We therefore follow this prescription.

From Fig.~\ref{fig:pdf_comparison}, we observe that the predictions obtained from the 
three sets that enter the PDF4LHC15 combination lie well 
within 1\% of each other over the 
whole range of center-of-mass energies from 2 to 15 TeV. 
In particular, MMHT2014 and NNPDF3.0 agree at the per mille level. 
The combined 
PDF$+\alpha_s$ uncertainty is at the level of $3-4\%$ for LHC 
energies, and it captures very well the small differences in the predictions 
among the different sets.

%%%We summarize the results in table 
%%%\begin{table}[tb]
%%%\begin{tabular}{c ccc ccccccccccccccc}
%%%$\sqrt{S}$ {TeV} & & 2 & & 7 & 8 & 13 & 14\\
%%%& $\sigma^0$ & $\sigma^{+}_{pdf+\alpha_s}$ &$\sigma^{-}_{pdf+\alpha_s}$\\
%%%$\sigma_{\rm \;PDF4LHC}$ (pb) & 1.10 & 1.13 & 1.06 & 16.66 & 21.17 & 47.89 & 53.88 \\
%%%$\frac{\sigma_{{\rm \;CT14}}}{\sigma^0_{{\rm \;PDF4LHC}}}$ & 1.007  & 0.998 & 0.998 & 0.998 & 0.998  \\
%%%$\frac{\sigma_{{\rm \;MMHT2014}}}{\sigma^0_{{\rm \;PDF4LHC}}}$ & 0.996 & 0.998 & 0.998 & 0.998 & 0.998  \\
%%%$\frac{\sigma_{{\rm \;NNPDF3.0}}}{\sigma^0_{{\rm \;PDF4LHC}}}$ & 0.999 & 1.005 & 1.004 & 1.004  & 1.004 \\
%%%\end{tabular}
%%%\caption{some}
%%%\label{tab:ratios_all_pdf4lhc}
%%%\end{table}
\begin{figure}[tb]
\begin{center}
\includegraphics[width=0.9\textwidth]{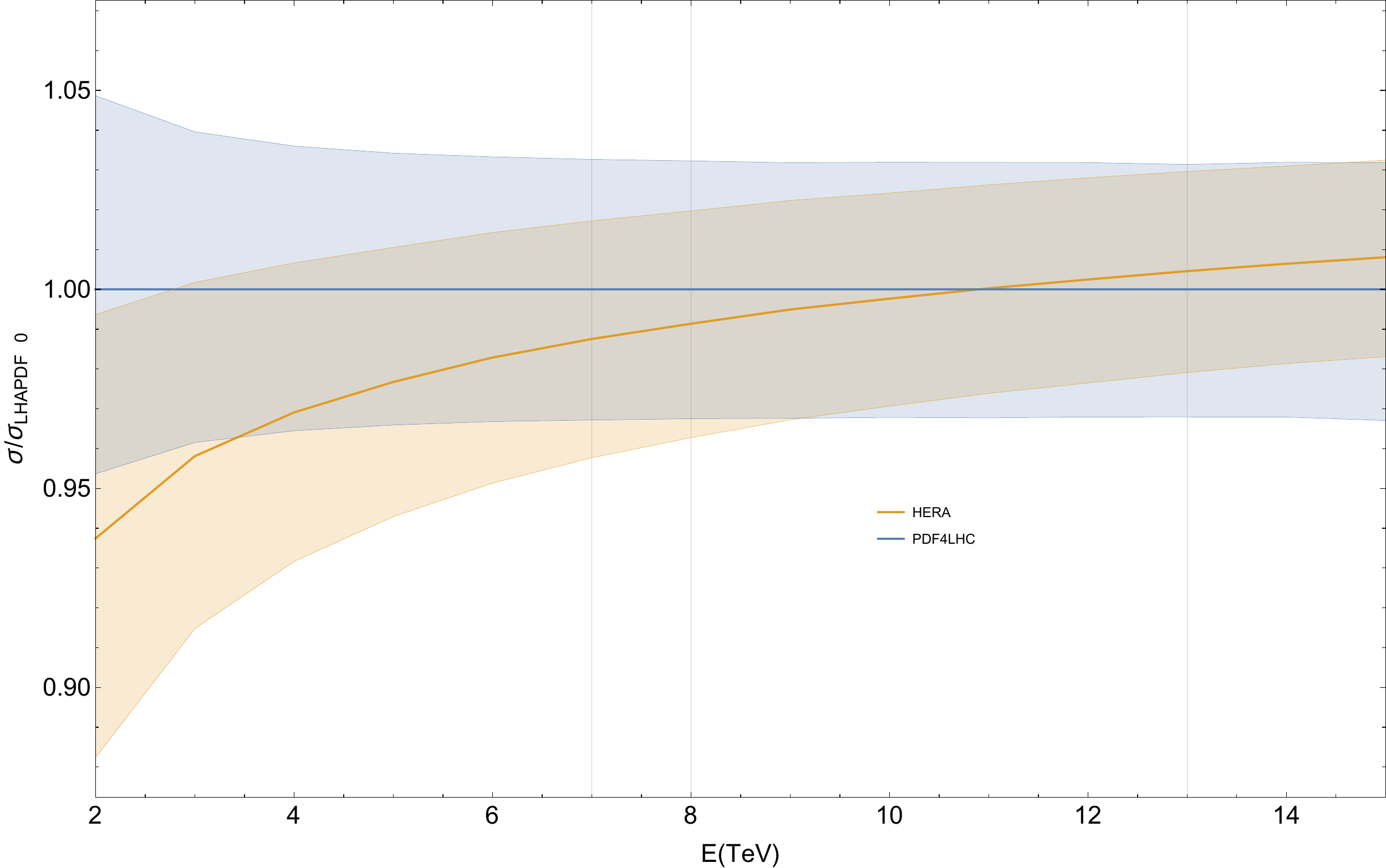}
\end{center}
\caption{
Higgs production cross-section and 68\%~C.L. PDF$+\alpha_s$ uncertainty 
from the HERAPDF2.0 fit, normalized by the 
central value obtained with the PDF4LHC combination.}
\label{fig:pdf_HERA} 
\end{figure}
Good agreement with the PDF4LHC15 predictions is also obtained for LHC energies 
using the HERAPDF2.0 set 
(Fig.~\ref{fig:pdf_HERA}). HERAPDF2.0 does not enter the 
PDF4LHC fit, but is given at the same central value of $\alpha_s$. 
However, these PDFs give a cross-section that is about 6\% lower at Tevatron 
energies, and  increase above the PDF4LHC15 predictions at higher 
center-of-mass energies.

\begin{figure}[!t]
\begin{center}
\includegraphics[width=0.9\textwidth]{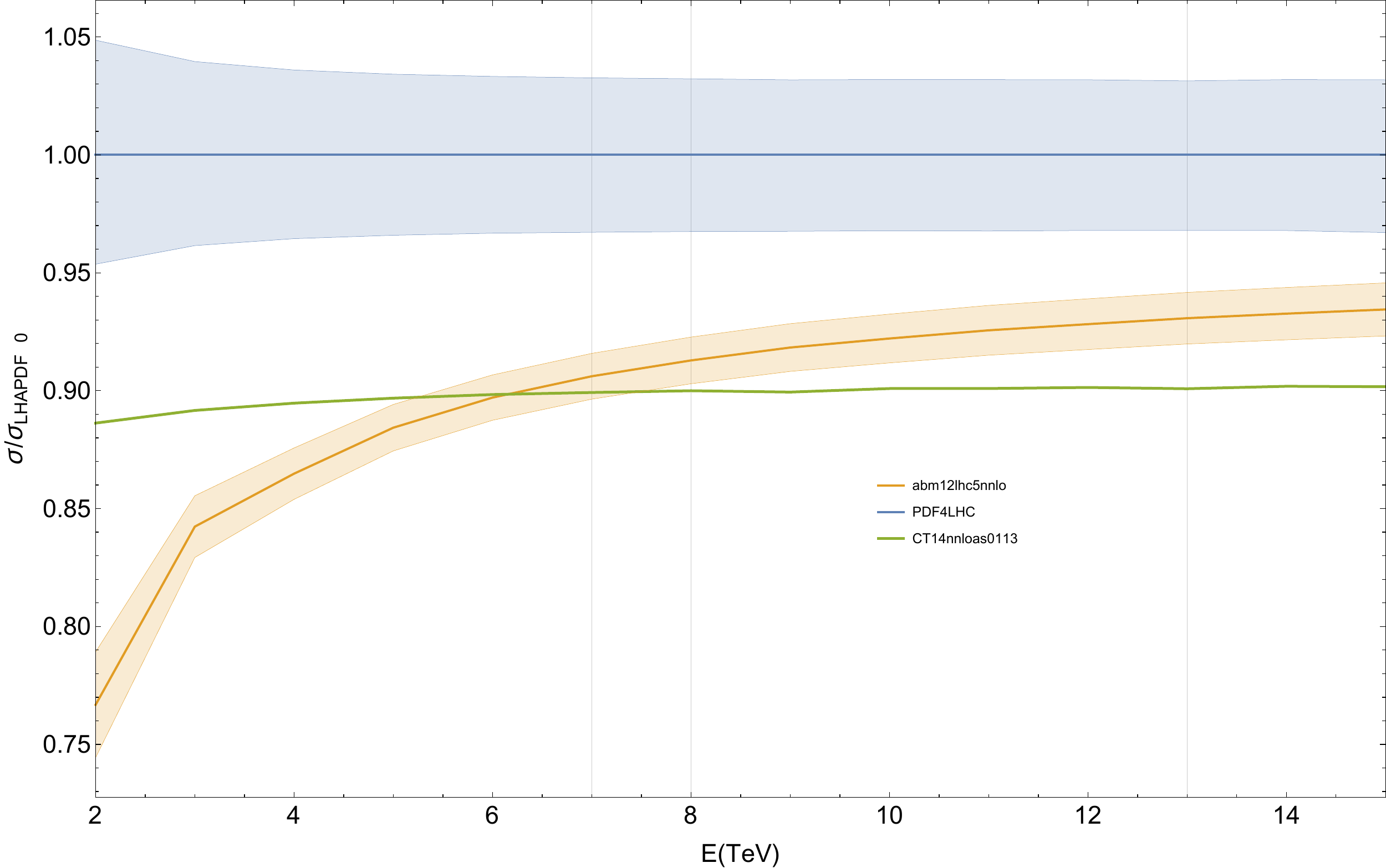}
\end{center}
\caption{
Higgs production cross-section and 68\%~C.L. PDF$+\alpha_s$ uncertainty 
from the ABM12 fit and from the CT14 set computed at $\alpha_s = \alpha_s^{ABM}$, normalized by the 
central value obtained with the PDF4LHC combination.}
\label{fig:pdf_ABM} 
\end{figure}

The situation is very different for the ABM12 set, 
which uses a lower central value of the strong coupling constant
\beq
\alpha_s^{ABM} = 0.1132 \pm 0.0011 \,.
\eeq
This value is the result of the ABM fit.
As one can see from Fig.~\ref{fig:pdf_ABM}, the ABM12 set gives a prediction that 
is about 23\% lower than the one from PDF4LHC15 at Tevatron energies, and 
$9-7\%$ lower at LHC energies. The PDF$+\alpha_s$ error is 1.2\%, which 
does not account for this discrepancy. We note here that the variation range for $\alpha_s$ used for the PDF$+\alpha_s$ variation in the ABM12 set  is determined by the fitting procedure and is slightly smaller than the range suggested by the PDF4LHC recommendation~\cite{Butterworth:2015oua}.

To understand how much of this difference comes from the choice of a different value 
of the strong coupling constant, we plot in Fig.~\ref{fig:pdf_ABM} the 
prediction from CT14 at the same value of $\alpha_s$ as the one obtained 
by ABM12. At $\alpha_s=0.118$ the predictions from CT14 are in
very good agreement with those from PDF4LHC15~(Fig.~\ref{fig:pdf_comparison}). 
At a lower value of $\alpha_s$, CT14 gives a cross-section that is about 
10\% smaller than the result at $\alpha_s=0.118$ (12\% at Tevatron energies). 
The dependence on the center-of-mass energy appears to be much milder than 
the one exhibited by ABM12. However, the PDF$+\alpha_s$ uncertainty might 
improve the agreement between the two sets. Unfortunately, only one error set
for CT14 at $\alpha_s=0.113$ is available, and we cannot assess this uncertainty.

\section{Recommendation for the LHC}
\label{sec:recommendation}
In previous sections we have considered various effects that contribute to the gluon-fusion Higgs production cross-section at higher orders. In this section we combine all these effects, and as a result we are able to present the most precise prediction for the gluon-fusion cross-section available to date. In particular (for the Setup 1 of Tab.~\ref{tab:setup1})
 for a Higgs boson with a mass $m_H = 125$ GeV, the
cross-section at the LHC with a center-of-mass energy of 13 TeV is
\begin{equation}
\label{eq:finalresult_13TeV}
\boxed{
\sigma = 48.58\,{\rm pb} {}^{+2.22\, {\rm pb}\, (+4.56\%)}_{-3.27\, {\rm pb}\, (-6.72\%)} \mbox{ (theory)} 
\pm 1.56 \,{\rm pb}\, (3.20\%)  \mbox{ (PDF+$\alpha_s$)} \,.
}
\end{equation}
Equation~\eqref{eq:finalresult_13TeV} is one of the main results of our work. In the following, we will analyze it in some detail. 

Let us start by commenting on the central value of the
prediction~\eqref{eq:finalresult_13TeV}. Since
eq.~\eqref{eq:finalresult_13TeV} is the combination of all the effects
considered in previous sections, it is interesting to see how the
final prediction is built up from the different contributions. The
breakdown of the different effects is:
\beq\label{eq:central_value_breakdown}
\begin{array}{rrrl}
48.58\, \textrm{pb} =&  16.00 \, {\rm pb}  & \quad (+32.9\%) & \qquad (\textrm{LO, rEFT})   \\ 
                              & \,+\,  20.84 \,{\rm pb} &  \quad
                                (+42.9\%) & \qquad (\textrm{NLO, rEFT} )   \\
                              &\,-\, \phantom{a}2.05\, {\rm pb}  & \quad (-4.2\%)
                                                              &\qquad  ((t, b, c)\textrm{, exact NLO})  \\
                              & \, + \,\phantom{a} 9.56 \, {\rm pb} & \quad (+19.7\%) & \qquad (\textrm{NNLO, rEFT}) \\
                             &\,+\,\phantom{a} 0.34\, {\rm pb} &  \quad (+0.2\%) &  \qquad (\textrm{NNLO, }1/m_t)  \\
                             &\,+ \,\phantom{a} 2.40\, {\rm pb}  & \quad (+4.9\%) & \qquad
                                                         (\textrm{EW, QCD-EW}) \\
                              &\,+\,\phantom{a}  1.49\, {\rm pb}  & \quad (+3.1\%)
                                                              & \qquad (\textrm{N$^3$LO, rEFT})   
\end{array}
\eeq
where we denote  by rEFT the contributions in the large-$m_t$ limit,
rescaled by the ratio $R_{LO}$ of the exact LO 
cross-section by the cross-section in the EFT (see Section~\ref{section:top_quark_mass_effects}). 
All the numbers in eq.~\eqref{eq:central_value_breakdown} have been
obtained by setting the renormalization and factorization scales equal
to $m_H/2$ and using the same set of parton densities at all
perturbative orders. 
Specifically, the first line, (LO, rEFT), is the cross-section at LO
taking into account only the top quark. The second line, (NLO, rEFT)
are the NLO corrections to the LO cross-section in the rescaled EFT, and the third line,
(($t$, $b$, $c$), exact NLO), is the correction that needs to be added to the first two lines
in order to obtain the exact QCD cross-section through NLO, including the full 
dependence on top, bottom and charm quark masses. 
The fourth and fifth lines contain the NNLO QCD corrections
to the NLO cross-section in the  rescaled EFT: 
(NNLO, rEFT) denotes the NNLO corrections in the EFT rescaled by $R_{LO}$,
and (NNLO, $1/m_t$) contains subleading corrections in the top mass at NNLO
computed as an expansion in $1/m_t$. The sixth line, 
(EW, QCD-EW), contains the two-loop electroweak corrections, computed exactly, and
three-loop mixed QCD-electroweak corrections, computed in an effective theory approach. 
The last line, (N$^3$LO, rEFT),  is the main addition of our work and contains the N$^3$LO corrections to the NNLO rEFT cross-section,
rescaled by $R_{LO}$.
Resummation effects, within the resummation frameworks studied in Section~\ref{sec:MHO}, contribute at the per mille level for our choice of the central scale, $\mu = m_H/2$, and are therefore neglected.   

Next, let us analyze the uncertainties quoted in our cross-section prediction. We present our result in eq.~\eqref{eq:finalresult_13TeV} with two uncertainties which we describe in the following. 
The first uncertainty in eq.~\eqref{eq:finalresult_13TeV} is the theory uncertainty related to
missing corrections in the perturbative description of the
cross-section. Just like for the central value, it is interesting to look at the breakdown of how the different effects build up the final number. Collecting all the uncertainties described in previous sections, we find the following components:
\begin{center}
\begin{tabular}{cccccc}
\toprule
\begin{tabular}{c} $\delta$(scale) \end{tabular} &
\begin{tabular}{c} $\delta$(trunc) \end{tabular} &
\begin{tabular}{c} $\delta$(PDF-TH) \end{tabular} &
\begin{tabular}{c} $\delta$(EW) \end{tabular} & 
\begin{tabular}{c} $\delta(t,b,c)$  \end{tabular} & 
\begin{tabular}{c} $\delta(1/m_t)$ \end{tabular}\\ \midrule
${}^{+0.10\textrm{ pb}}_{-1.15\textrm{ pb}} $ & $\pm$0.18 pb & $\pm$0.56 pb & $\pm$0.49 pb& $\pm$0.40 pb& $\pm$0.49 pb
\\ \midrule
${}^{+0.21\%}_{-2.37\%}$ & $\pm 0.37\%$ & $\pm 1.16\%$ & $\pm 1\%$ & $\pm 0.83\%$ 
& $\pm 1\%$ \\
 \bottomrule
\end{tabular}
\end{center}
In the previous table, $\delta$(scale) and $\delta$(trunc) denote the scale and truncation uncertainties on the rEFT cross-section, and $\delta$(PDF-TH) denotes the uncertainty on the cross-section prediction due to our ignorance of N$^3$LO parton densities, cf. Section~\ref{sec:EFT}. $\delta$(EW),  $\delta(t,b,c)$ and $\delta(1/m_t)$ denote the uncertainties on the cross-section due to missing quark-mass effects at NNLO and mixed QCD-EW corrections. The first uncertainty in eq.~\eqref{eq:finalresult_13TeV} is then obtained by adding linearly all these effects. The parametric uncertainty due to the mass values of the
top, bottom and charm quarks is at the per mille level, and hence completely negligible.
We note that including into our
prediction resummation effects in the schemes that we have studied
in Section~\ref{sec:MHO} would lead to a very small scale variation, which we believe unrealistic
and which we do not expect to capture the uncertainty due to
missing higher-order corrections at N$^4$LO and beyond. Based on this observation, as
well as on the fact that the definition of the resummation scheme may suffer from large ambiguities, 
we prefer a prudent approach and we adopt to adhere to fixed-order
perturbation theory as an estimator of remaining theoretical uncertainty from QCD.   

The second uncertainty in eq.~\eqref{eq:finalresult_13TeV} is the PDF+$\alpha_s$ uncertainty due to the determination of  the parton distribution functions and the strong coupling constant, 
following the PDF4LHC recommendation.
When studying the correlations with other uncertainties in Monte-Carlo simulations,
it is often necessary to separate the
PDF and $\alpha_s$ uncertainties:
\begin{center}
\begin{tabular}{ccc}
\toprule
\begin{tabular}{c} $\delta$(PDF) \end{tabular} &
\begin{tabular}{c} $\delta$($\alpha_s$) \end{tabular} \\ \midrule
$\pm 0.90$ pb & $ {}^{+1.27{\rm pb}}_{-1.25{\rm pb}}$\\ \midrule
$\pm 1.86\%$ & ${}^{+2.61\%}_{-2.58\%}$  \\ \bottomrule
\end{tabular}
\end{center}
Since the $\delta(\alpha_s)$ error is asymmetric, in the combination presented in eq.~\eqref{eq:finalresult_13TeV}
we conservatively add in quadrature the largest of the two errors to the PDF error.

As pointed out in Section~\ref{sec:pdfs}, the PDF4LHC  uncertainty estimate
quoted above does not cover the cross-section value as predicted by
the ABM12 set of parton distribution functions. For comparison we
quote here the corresponding
cross-section value and PDF+$\alpha_s$ uncertainty with the ABM12 
set\footnote{We use the {\tt abm11\_5\_as\_nlo} and {\tt abm11\_5\_as\_nnlo} 
set to estimate the $\delta$(PDF-TH): these sets are fits with a fixed value of $\alpha_s$ which allows us to compare NLO and NNLO grids for the same $\alpha_s$ value.  Using this prescription $\delta$(PDF-TH)$=1.1\%$ very similar to the corresponding uncertainty for  the  set.}:
\begin{equation}
\label{eq:finalresult_13TeV_ABM}
\sigma_{\rm ABM12} = 45.07\,{\rm pb} {}^{+2.00\, {\rm pb}\, (+4.43\%)}_{-2.88\, {\rm pb}\, (-6.39\%)} \mbox{ (theory)} 
\pm 0.52 \,{\rm pb}\, (1.17\%)  \mbox{ (PDF+$\alpha_s$)}\,. 
\end{equation}

The significantly lower central
value is mostly due to the smaller value of $\alpha_s$, which 
however is also smaller than the world average.  

It is also interesting to compare our prediction~\eqref{eq:finalresult_13TeV} to the value one would have obtained without the knowledge of the N$^3$LO corrections in the rEFT. We find 
\begin{equation}\label{eq:final_NNLO}
\sigma^{NNLO} =  47.02\textrm{ pb } {}^{+5.13 \textrm{ pb }(10.9\%)}_{-5.17 \textrm{ pb }(11.0\%)}  \mbox{ (theory)}  {}^{+1.48 \textrm{ pb }}_{-1.46 \textrm{ pb }}  {}^{(3.14\%)}_{(3.11\%)} \mbox{ (PDF+$\alpha_s$)} \,.
\end{equation}
The central value in eq.~\eqref{eq:final_NNLO} is obtained by summing all terms
in eq.~\eqref{eq:central_value_breakdown} except for the term in the
last line.  Moreover, we do not include the uncertainties $\delta$(PDF-TH)
and $\delta$(trunc) from
 missing higher orders in the extraction of the parton densities 
and from the truncation of the threshold expansion (because the NNLO cross-sections are
known in a closed analytic form). The scale variation uncertainty
$\delta$(scale) at NNLO is approximately five times larger than at
N$^3$LO.  This explains the reduction by a factor of two in the total
$\delta$(theory) uncertainty by including the N$^3$LO corrections
presented in this publication. We stress at this point that uncertainties on the 
NNLO cross-section have been investigated by different groups in the past, yielding a variety of uncertainty estimates at NNLO~\cite{Dittmaier:2011ti,Bonvini:2014jma,deFlorian:2014vta,Ahrens:2010rs,Anastasiou:2011pi,Anastasiou:2012hx,Baglio:2010ae,Baglio:2010um}. Here we adopt exactly the same prescription to estimate the uncertainty at NNLO and at N$^3$LO, and we do not only rely on scale variation for the NNLO uncertainty estimate, as was often done in the past.

Finally, we have also studied how our predictions change as we vary the center-of-mass energy and the value of the Higgs mass. Our predictions for different
values of the proton-proton collision energy and a Higgs mass of
$m_{H}=125$ GeV are summarized in Tab.~\ref{tab:xs_relerr}. 
In comparison to the official recommendation of the  LHC Higgs Cross-section
Working Group earlier than our work~\cite{Heinemeyer:2013tqa}, 
our results have a larger central value by about $11\%$. The
difference can be attributed to the choice of optimal renormalization
and factorization scale, the effect of the \nnnlo corrections, the different sets of parton distribution
functions and value of $\alpha_s$ as well as smaller differences due
to the treatment of finite quark-mass effects.  In comparison to the
earlier recommendation from some of the authors in
ref.~\cite{Anastasiou:2012hx}, our result has a central value which is
higher by $3.5\%$. The difference can be attributed to the effect of
the \nnnlo corrections, the different sets of parton distribution
functions and value of $\alpha_s$ as well as smaller differences due
to the treatment of finite quark-mass effects.  

Additional cross-section predictions for a variety of collider energies and Higgs boson masses can be found in Appendix~\ref{app:predictions}.

\begin{table}[!t]
\normalsize\setlength{\tabcolsep}{2pt}
\begin{center}
\begin{tabular}{rcrcrcrcr}
\toprule
\multicolumn{1}{c}{$E_{CM}$}& \phantom{a}&
\multicolumn{1}{c}{$\sigma$}& \phantom{a}&
\multicolumn{1}{c}{$\delta(\mbox{theory})$}& \phantom{a}&
\multicolumn{1}{c}{$\delta(\textrm{PDF})$}& \phantom{a}&
\multicolumn{1}{c}{$\delta(\alpha_s)$} 
\\ \midrule
2 TeV && 1.10 pb&& ${}^{+0.04\textrm{pb}}_{-0.09\textrm{pb}}$(${}^{+4.06\%}_{-7.88\%}$)&& $\pm$ 0.03 pb ($\pm$ 3.17\%) && ${}^{+0.04\textrm{pb}}_{-0.04\textrm{pb}}$(${}^{+3.36\%}_{-3.69\%}$)  \\ \midrule 
7 TeV&& 16.85 pb&& ${}^{+0.74\textrm{pb}}_{-1.17\textrm{pb}}$(${}^{+4.41\%}_{-6.96\%}$)&& $\pm$ 0.32 pb ($\pm$ 1.89\%) && ${}^{+0.45\textrm{pb}}_{-0.45\textrm{pb}}$(${}^{+2.67\%}_{-2.66\%}$)  \\ \midrule 
8 TeV&& 21.42 pb&& ${}^{+0.95\textrm{pb}}_{-1.48\textrm{pb}}$(${}^{+4.43\%}_{-6.90\%}$)&& $\pm$ 0.40 pb ($\pm$ 1.87\%) && ${}^{+0.57\textrm{pb}}_{-0.56\textrm{pb}}$(${}^{+2.65\%}_{-2.62\%}$)  \\ \midrule
13 TeV&& 48.58 pb&& ${}^{+2.22\textrm{pb}}_{-3.27\textrm{pb}}$(${}^{+4.56\%}_{-6.72\%}$)&& $\pm$ 0.90 pb ($\pm$ 1.86\%) && ${}^{+1.27\textrm{pb}}_{-1.25\textrm{pb}}$(${}^{+2.61\%}_{-2.58\%}$) \\ \midrule
14 TeV&& 54.67 pb&& ${}^{+2.51\textrm{ pb}}_{-3.67\textrm{ pb}}$ (${}^{+4.58\%}_{-6.71\%}$) && $\pm$1.02 pb ($\pm$ 1.86\%) && ${}^{+1.43\textrm{pb}}_{-1.41\textrm{pb}}$(${}^{+2.61\%}_{-2.59\%}$)  \\ \bottomrule
\end{tabular}
\end{center}
\caption{Gluon-fusion Higgs cross-section at a proton-proton collider 
for various values of the collision energy.}
\label{tab:xs_relerr}
\end{table} 

\begin{table}[!t]
\normalsize\setlength{\tabcolsep}{2pt}
\begin{center}
\begin{tabular}{rcrcccccc}
\toprule
\multicolumn{1}{c}{$E_{CM}$}& \phantom{a}&
\multicolumn{1}{c}{$\sigma$}& \phantom{a}&
\multicolumn{1}{c}{$\delta(\mbox{theory})$}& \phantom{a}&
\multicolumn{1}{c}{$\delta(\textrm{PDF}+\alpha_s)$}& \phantom{a}&
\\ \midrule
7 TeV && 15.13 pb && ${}^{+7.1\%}_{-7.8\%}$  && ${}^{+7.6\%}_{-7.1\%}$   \\ \midrule 
8 TeV && 19.27 pb && ${}^{+7.2\%}_{-7.8\%}$  && ${}^{+7.5\%}_{-6.9\%}$   \\ \bottomrule 
\end{tabular}
\end{center}
\caption{Earlier recommendation for the gluon-fusion Higgs cross-section at a proton-proton collider 
by the Higgs Cross-Section Working Group~\cite{Heinemeyer:2013tqa}.}
\label{tab:xs_wg}
\end{table}  

\begin{table}[!t]
\normalsize\setlength{\tabcolsep}{2pt}
\begin{center}
\begin{tabular}{rcrcccccc}
\toprule
\multicolumn{1}{c}{$E_{CM}$}& \phantom{a}&
\multicolumn{1}{c}{$\sigma$}& \phantom{a}&
\multicolumn{1}{c}{$\delta(\mbox{theory})$}& \phantom{a}&
\multicolumn{1}{c}{$\delta(\textrm{PDF}+\alpha_s)$}& \phantom{a}&
\\ \midrule
8 TeV && 20.69 pb &&  ${}^{+8.37\%}_{-9.26\%}$ &&  ${}^{+7.79\%}_{-7.53\%}$ \\ \bottomrule 
\end{tabular}
\end{center}
\caption{Earlier recommendation for the gluon-fusion Higgs cross-section at a proton-proton collider 
by some of the authors in ref.~\cite{Anastasiou:2012hx}.}
\label{tab:xs_ihixs}
\end{table}

%\input{table_sigmas_7Tev_HXSWG.tex}

%%% Local Variables: 
%%% mode: latex
%%% TeX-master: "paper"
%%% End: 

%\section{Predictions for a generic scalar}
%\label{sec:BSM}
%\input{BSM.tex}

\section{Conclusion}
\label{sec:conclusion}

In this paper we have presented the most precise prediction for the
Higgs boson gluon-fusion cross-section at the LHC. In order to achieve this task, we have combined 
all known higher-order effects from QCD, EW and quark-mass corrections. 
The main component that made our computation possible was the recent computation of the N$^3$LO
correction to the cross-section in an effective field
theory where the top quark was integrated out.  
In an appendix we present analytic expressions for the
partonic subchannels of the N$^3$LO partonic cross-sections which have 
not been presented elsewhere in the literature,  
in the form of a series expansion around the threshold limit.  

The N$^3$LO corrections moderately increase ($\sim 3\%$) 
the cross-section for renormalization and factorization scales equal
to  $m_H/2$. In addition, they notably stabilize the scale variation,
reducing it almost by a factor of five compared to NNLO. The N$^3$LO scale-variation band 
is included entirely within the NNLO scale-variation band for scales
in the interval $[{m_H}/{4}, m_H ]$. Moreover, we have found 
good evidence that the N$^3$LO scale variation captures the effects of missing higher 
perturbative orders in the EFT. We base this conclusion on the following observations:
First, we observed that expanding in
$\alpha_s$ separately the Wilson coefficient and matrix-element
factors in the cross-section gives results consistent with expanding
directly their product through N$^3$LO. Second, 
a traditional threshold resummation in Mellin space up to N$^3$LL did not contribute 
significantly to the cross-section beyond N$^3$LO in the range of scales 
$\mu \in [{m_H}/{4}, m_H ]$. 
Although  the effects of threshold resummation are in general sensitive to
ambiguities due to subleading terms beyond the soft limit, we found
that within our preferred range of scales, several variants of the exponentiation formula
gave very similar phenomenological results, which are always consistent with
fixed-order perturbation theory.  Finally, a soft-gluon and
$\pi^2$-resummation using the SCET formalism also gave consistent
results with fixed-order perturbation theory at N$^3$LO. While 
ambiguities in subleading soft terms limit the use of soft-gluon
resummation as an estimator of higher-order effects, and while it is of
course possible that some variant of resummation may  
yield larger corrections, it is encouraging that this does not happen
for the mainstream prescriptions studied here. 

Besides studying the effect of QCD corrections in the EFT at high orders, 
we also investigated the cross-section in the EFT after inclusion of exact LO and
NLO QCD corrections in the full Standard Model theory (with finite
top, bottom and charm quark masses)  and $1/m_t$ corrections at NNLO.  
We also included known two-loop electroweak corrections and an
estimate of three-loop mixed QCD-EW corrections into our final prediction. 

No prediction for the cross-section would be complete without estimating the residual uncertainties 
that may affect our result.
We have identified several sources of theoretical uncertainties, 
namely, the truncation of the threshold
expansion, the QCD scale variation, missing higher-order
corrections in the extraction of parton densities, missing finite
quark-mass effects beyond NLO and missing mixed QCD-EW 
corrections. After adding all these uncertainties linearly, we obtain a residual theoretical uncertainty of about $5-6\%$. 
We have also studied the sensitivity of the cross-section on the
choice of parton distribution functions. The CT14, MSTW and NNPDF sets
are in good agreement among themselves, and have been combined together according to the
PDF4LHC recommendation. They yield a combined uncertainty due to both
$\alpha_s$ and parton densitites of the order of $\sim 3.5\%$. The
PDF4LHC sets give cross-section values that are in good agreement
with the cross-section as computed with HERAPDF sets. 
However, the ABM12 set of parton densities yields results which are
significantly lower and outside the quoted range of uncertainty.

We expect that
further progress can be made in order to improve even more the
precision of our computation.  A forthcoming computation of the N$^3$LO 
cross-section in the EFT in a closed analytic form will remove the
truncation uncertainty. Future computations of the NNLO QCD
cross-section in the full Standard Model (including finite top, bottom and charm
masses) and a complete computation of three-loop mixed QCD-EW
corrections will remove further significant sources of uncertainties.  
Progress in the determination of parton densities, with more precise
LHC data  and more precise computations of cross-sections used in the
extraction of parton densities, will be crucial to corroborate the ${\rm
PDF}+\alpha_s$ uncertainty and to resolve discrepancies due to systematic
effects.  

To conclude, we have presented the predictions for the Higgs boson cross-section in gluon fusion,
based on very high orders in perturbation theory. In this way, we have obtained 
 the most precise prediction of the Higgs boson production
cross-section at the LHC to date.  We are looking forward to comparisons of our
results with precise measurements of the Higgs boson cross-section at
the LHC in the future.

\section*{Acknowledgements}
We thank Nigel Glover, Gavin Salam and Michael Spira for useful discussions and
suggestions and Thomas Becher and Li Lin Yang for their support
regarding the SCET resummation code {\tt RGHiggs}.
We thank Robert Harlander for providing us with analytic results for the subleading $1/m_t$ contributions at NNLO and Tobias Neumann  for providing us with a private version of the {\tt ggh@nnlo} program and helping us with numerical comparisons. We also thank Sven Moch for suggestions and clarifications with respect to the ABM PDF sets.  
This research was supported by the Swiss National Science Foundation (SNF) under 
contracts 200021-143781 and  200020-162487 and by the European Commission 
through the ERC grants ``IterQCD'', ``HEPGAME'' (320651), ``HICCUP'', ``MathAm'' and ``MC@NNLO'' (340983).

\bibliographystyle{JHEP.bst}
\bibliography{higgsrefs}

\appendix

\section{Analytic expression for the Wilson coefficient}
\label{app:WC_MSbar-OS}

In the effective theory (i.e., for $N_f$ light flavors), with the top quark decoupled from 
the running of the strong coupling constant,
the $\overline{\textrm{MS}}$-scheme Wilson coefficient 
 reads~\cite{Chetyrkin:1997un,Schroder:2005hy}
\begin{eqnarray}
C^{\overline{\textrm{MS}}} & = & - \frac{a_s}{3 v} \Bigg\{ 1 +a_s\,\frac{11}{4} 
+ a_s^2 \left[\frac{2777}{288} - \frac{19}{16} \log\left(\frac{m_t^2}{\mu^2}\right) -
 N_f\left(\frac{67}{96}+\frac{1}{3}\log\left(\frac{m_t^2}{\mu^2}\right)\right)\right] 
 \nonumber \\
  &&+ a_s^3\Bigg[-\left(\frac{6865}{31104} 
  + \frac{77}{1728} \log\left(\frac{m_t^2}{\mu^2}\right) 
  + \frac{1}{18}\log^2\left(\frac{m_t^2}{\mu^2}\right)\right) N_f^2 \\
  && + \left(\frac{23}{32} \log^2\left(\frac{m_t^2}{\mu^2}\right) - 
  \frac{55}{54} \log\left(\frac{m_t^2}{\mu^2}\right)+\frac{40291}{20736} 
 - \frac{110779}{13824} \zeta_3 \right) N_f \nonumber\\
  &&  -\frac{2892659}{41472}+\frac{897943}{9216}\zeta_3 
  + \frac{209}{64} \log^2\left(\frac{m_t^2}{\mu^2}\right)
  - \frac{1733}{288}\log\left(\frac{m_t^2}{\mu^2}\right)\Bigg] +\ord(a_s^4) \Bigg\} \, .\nonumber
\end{eqnarray}
The analogous result in the on-shell scheme can be derived combining
the OS decoupling constant $\zeta_g$ and the OS Wilson coefficient with 
$\alpha_s$ running in the full 
theory, that one can find in the literature (see, for 
example, ref.~\cite{Chetyrkin:1997un,Schroder:2005hy}). The result is
\begin{eqnarray}
C^{\textrm{OS}} & = & - \frac{a_s}{3 v} \Bigg\{ 1 +a_s\,\frac{11}{4} 
+ a_s^2 \left[\frac{2777}{288} - \frac{19}{16} \log\left(\frac{m_t^2}{\mu^2}\right) -
 N_f\left(\frac{67}{96}+\frac{1}{3}\log\left(\frac{m_t^2}{\mu^2}\right)\right)\right] 
 \nonumber \\
  &&+ a_s^3\Bigg[-\left(\frac{6865}{31104} 
  + \frac{77}{1728} \log\left(\frac{m_t^2}{\mu^2}\right) 
  + \frac{1}{18}\log^2\left(\frac{m_t^2}{\mu^2}\right)\right) N_f^2 \\
  && + \left(\frac{23}{32} \log^2\left(\frac{m_t^2}{\mu^2}\right) - 
  \frac{91}{54} \log\left(\frac{m_t^2}{\mu^2}\right)+\frac{58723}{20736} 
 - \frac{110779}{13824} \zeta_3 \right) N_f \nonumber\\
  &&  -\frac{2761331}{41472}+\frac{897943}{9216}\zeta_3 
  + \frac{209}{64} \log^2\left(\frac{m_t^2}{\mu^2}\right)
  - \frac{2417}{288}\log\left(\frac{m_t^2}{\mu^2}\right)\Bigg] +\ord(a_s^4) \Bigg\} \nonumber\, .
\end{eqnarray}

\section{Numerical implementation of the Mellin inversion}
\label{app:inverse_Mellin}
In this section we describe our numerical implementation of the inverse Mellin transform,
\beq
\sigma_{gg}(\tau) =  \int_{c-i\infty}^{c+i\infty}\frac{\dd N}{2\pi i}\,\tau^{1-N}\,f_g^2(N)\,\hat{\sigma}_{gg}(N)\,,
\eeq
where $f_g(N)$ are the Mellin moments of the gluon density and $\hat{\sigma}_{gg}(N)$ the (resummed) partonic cross-section. The integration contour is the straight vertical line $\textrm{Re}(N) = c$, chosen according to the \emph{minimal prescription}~\cite{Catani:1996yz}, i.e., all the poles in $\hat{\sigma}_{gg}(N)$ lie to the left of the contour, except for the Landau pole in Mellin space, which lies to the right of the contour. The position of the Landau pole in Mellin space is given by
\beq
N_L \equiv\,\exp\frac{1}{2\beta_0\,a_s(\mu_R^2)}\,.
\eeq
We parametrize the integration contour as $N=c+it$, and we obtain
\beq\label{eq:IMellin}
\sigma_{gg}(\tau) =  \tau^{-c}\, \int_0^\infty\frac{\dd t}{\pi}\,\textrm{Re}\left[\tau^{-it}\,f_g^2(c+it)\,\hat{\sigma}_{gg}(c+it)\right]\,.
\eeq
In order to evaluate the integral, we need to know the Mellin moments of the gluon density for complex values of the Mellin variable $N$. To our knowledge, the public PDF sets do not provide grids which allow one to immediately obtain the Mellin moments of the PDFs, and so we need to use our own method to perform the inverse Mellin transform in eq.~\eqref{eq:IMellin}. This method is described in the remainder of this appendix.

We start by truncating the integral~\eqref{eq:IMellin} at some large value $t_{\textrm{max}}$, and we approximate the integral over the range $[0,t_{\textrm{max}}]$ by a Gauss-Legendre quadrature of order $l$,
\beq\bsp\label{eq:Mellin_GL}
\sigma_{gg}(\tau) 
&\,\simeq  \frac{\tau^{-c}\,t_{\textrm{max}}}{2\pi}\, \sum_{k=1}^l\,w_k^{(l)}\,\textrm{Re}\left[\tau^{-it^{(l)}_k}\,f_g^2\Big(c+it^{(l)}_k\Big)\,\hat{\sigma}_{gg}\Big(c+it^{(l)}_k\Big)\right]\,,
\esp\eeq
where $t^{(l)}_k = \frac{t_{\textrm{max}}}{2}\Big(1+u_k^{(l)}\Big)$, and $u_k^{(l)}$ are the zeroes of the $l$-th Legendre polynomial $P_l(x)$. The Gauss-Legendre weights are given by
\beq
w_k^{(l)} = \frac{a_l}{a_{l-1}}\,\frac{1}{P'_{l}\Big(u_k^{(l)}\Big)\,P_{l-1}\Big(u_k^{(l)}\Big)} \int_{-1}^1\dd x\,P_{l-1}^2(x) 
=\frac{2}{l\,P'_{l}\Big(u_k^{(l)}\Big)\,P_{l-1}\Big(u_k^{(l)}\Big)}\,,
\eeq
where $a_l$ denotes the coefficient of $x^l$ in $P_l(x)$.
The advantage of eq.~\eqref{eq:Mellin_GL} is that we only need to know the Mellin moments of the gluon density on the finite set of points $N_k^{(l)} \equiv c + \frac{it_{\textrm{max}}}{2}\Big(1+u_k^{(l)}\Big)$. We can evaluate these Mellin moments numerically once and for all (for a given PDF set and factorization scale) and store them in a grid,
\beq\label{eq:f_g_N}
f_g\Big(N_k^{(l)}\Big) = \int_0^1\dd x\,x^{N_k^{(l)}-1}\,f_g(x)\,.
\eeq
Note that the integral~\eqref{eq:f_g_N} is numerically convergent for $\textrm{Re}\Big(N_k^{(l)}\Big)>0$.

Equation~\eqref{eq:Mellin_GL} is our master formula for the computation of inverse Mellin transforms. We have generated grids $f_g\Big(N_k^{(l)}\Big)$ for various choices of PDF sets and factorization scales, making it straightforward to compute eq.~\eqref{eq:Mellin_GL} for any of these choices. Let us make some comments about the master formula~\eqref{eq:Mellin_GL}. First, we see that the right-hand side of eq.~\eqref{eq:Mellin_GL} depends on three free parameters: the real part $c$ of the integration contour, the truncation $t_{\textrm{max}}$ and the order $l$ of the Gauss-Legendre quadrature. While the inverse Mellin transform must obviously be independent of these parameters, they may introduce some systematic uncertainties. In our implementation we choose $c=2.5$, and we checked that the value of the integral remains unchanged under small deformations of this value. Next, it is easy to check that the Mellin moments are highly suppressed for $\textrm{Im}(N)\gg1$. In our implementation we choose $t_{\textrm{max}} = 125$, and we checked that the contribution to the integral from the range $[100,125]$ is completely negligible. Note that this implies that the bulk of the value of the integral comes from the region where $\textrm{Im}(N)$ is small. We therefore partition the range $[0,t_{\textrm{max}}]$ into subregions of increasing length, and in every subregion we approximate the integral by a Gauss-Legendre quadrature of order $l=20$. In this way we make sure that the sum in eq.~\eqref{eq:Mellin_GL} receives mostly contributions from points where $\textrm{Im}(N)$ is small.

Finally, let us briefly comment on the choice of the straight-line contour in the inverse Mellin transform. Indeed, we could deform the contour such as to maximize the convergence of the numerical integration. In particular, in ref.~\cite{Vogt:2004ns} it was argued that the inverse Mellin transform converges faster if the contour is chosen as $N=c+t\,\exp(i\phi)$, $\pi/2<\phi<\pi$. In this case, however, we have $\textrm{Re}(N) = c+t\,\cos\phi$, and so $\textrm{Re}(N)<0$ for large enough $t$, which contradicts the convergence criterion for the integral~\eqref{eq:f_g_N}. Hence, as we need to perform the integral~\eqref{eq:f_g_N} numerically in our approach, we cannot choose the optimized integration contour of ref.~\cite{Vogt:2004ns}. We note, however, that since it is sufficient to generate the grids $f_g\Big(N_k^{(l)}\Big)$ for a sufficiently large number of points once and for all, speed is not an issue and we do not loose anything by choosing $\phi=\pi/2$.

\newpage

\section{Numerical values for the coefficients of the threshold expansion}
\label{app:expansioncoefficients_numeric}

%%%%%%%%%%%%%%%%%%%%%%%%%%%%%%%%%%%%%%%%%%%%%
%%%%%%%%%%%%%%%%%%%%%%%%%%%%%%%%%%%%%%%%%%%%%

\subsection{The $gg$ channel}
%the gg 3,2
\begin{gather}
\begin{aligned}
\eta_{gg}^{(3,2),\textrm{reg}} &= -11089.328+1520.0814 \bar{z}+8805.7669 \bar{z}^2-12506.932 \bar{z}^3\\
&-440.32959 \bar{z}^4+1232.0873 \bar{z}^5+1646.4249 \bar{z}^6+1781.8637 \bar{z}^7\\
&+1835.6555 \bar{z}^8+1861.3612 \bar{z}^9+1876.6428 \bar{z}^{10}+1888.2649 \bar{z}^{11}\\
&+1899.1749 \bar{z}^{12}+1910.7995 \bar{z}^{13}+1923.8791 \bar{z}^{14}+1938.8053 \bar{z}^{15}\\
&+1955.7742 \bar{z}^{16}+1974.8643 \bar{z}^{17}+1996.0810 \bar{z}^{18}+2019.3836 \bar{z}^{19}\\
&+2044.7025 \bar{z}^{20}+2071.9510 \bar{z}^{21}+2101.0331 \bar{z}^{22}+2131.8486 \bar{z}^{23}\\
&+2164.2968 \bar{z}^{24}+2198.2785 \bar{z}^{25}+2233.6976 \bar{z}^{26}+2270.4621 \bar{z}^{27}\\
&+2308.4845 \bar{z}^{28}+2347.6819 \bar{z}^{29}+2387.9764 \bar{z}^{30}+2429.2946 \bar{z}^{31}\\
&+2471.5678 \bar{z}^{32}+2514.7317 \bar{z}^{33}+2558.7261 \bar{z}^{34}+2603.4947 \bar{z}^{35}\\
&+2648.9850 \bar{z}^{36}+2695.1477 \bar{z}^{37}
\end{aligned}
\end{gather}
%the gg 3,1
\begin{gather}
\begin{aligned}
\eta_{gg}^{(3,1),\textrm{reg}} &= 15738.441-13580.184 \bar{z}+1757.5646 \bar{z}^2+16078.884 \bar{z}^3\\
&+82.947070 \bar{z}^4+222.78697 \bar{z}^5+947.71319 \bar{z}^6+1490.0998 \bar{z}^7\\
&+1869.9658 \bar{z}^8+2145.3018 \bar{z}^9+2354.6608 \bar{z}^{10}+2520.8158 \bar{z}^{11}\\
&+2657.1437 \bar{z}^{12}+2771.7331 \bar{z}^{13}+2869.6991 \bar{z}^{14}+2954.4505 \bar{z}^{15}\\
&+3028.3834 \bar{z}^{16}+3093.2654 \bar{z}^{17}+3150.4554 \bar{z}^{18}+3201.0314 \bar{z}^{19}\\
&+3245.8702 \bar{z}^{20}+3285.6978 \bar{z}^{21}+3321.1237 \bar{z}^{22}+3352.6649 \bar{z}^{23}\\
&+3380.7639 \bar{z}^{24}+3405.8019 \bar{z}^{25}+3428.1091 \bar{z}^{26}+3447.9734 \bar{z}^{27}\\
&+3465.6466 \bar{z}^{28}+3481.3499 \bar{z}^{29}+3495.2787 \bar{z}^{30}+3507.6057 \bar{z}^{31}\\
&+3518.4844 \bar{z}^{32}+3528.0516 \bar{z}^{33}+3536.4294 \bar{z}^{34}+3543.7272 \bar{z}^{35}\\
&+3550.0434 \bar{z}^{36}+3555.4664 \bar{z}^{37}
\end{aligned}
\end{gather}
%the gg 3,0
\begin{gather}
\begin{aligned}
\eta_{gg}^{(3,0),\textrm{reg}} &= -5872.5889+13334.440 \bar{z}-8488.6090 \bar{z}^2-4281.1568 \bar{z}^3\\
&+2157.5052 \bar{z}^4+907.63249 \bar{z}^5+234.32211 \bar{z}^6-49.179428 \bar{z}^7\\
&-157.42872 \bar{z}^8-187.57931 \bar{z}^9-182.18174 \bar{z}^{10}-160.17000 \bar{z}^{11}\\
&-130.14932 \bar{z}^{12}-96.114987 \bar{z}^{13}-59.980602 \bar{z}^{14}-22.710016 \bar{z}^{15}\\
&+15.172227 \bar{z}^{16}+53.350662 \bar{z}^{17}+91.615033 \bar{z}^{18}+129.81315 \bar{z}^{19}\\
&+167.82893 \bar{z}^{20}+205.57154 \bar{z}^{21}+242.96939 \bar{z}^{22}+279.96631 \bar{z}^{23}\\
&+316.51876 \bar{z}^{24}+352.59361 \bar{z}^{25}+388.16630 \bar{z}^{26}+423.21933 \bar{z}^{27}\\
&+457.74098 \bar{z}^{28}+491.72418 \bar{z}^{29}+525.16567 \bar{z}^{30}+558.06524 \bar{z}^{31}\\
&+590.42510 \bar{z}^{32}+622.24942 \bar{z}^{33}+653.54391 \bar{z}^{34}+684.31546 \bar{z}^{35}\\
&+714.57190 \bar{z}^{36}+744.32176 \bar{z}^{37}
\end{aligned}
\end{gather}

%%%%%%%%%%%%%%%%%%%%%%%%%%%%%%%%%%%%%%%%%%%%%
%%%%%%%%%%%%%%%%%%%%%%%%%%%%%%%%%%%%%%%%%%%%%

\subsection{The $qg$ channel}

\begin{gather}
\begin{aligned}
\eta_{qg}^{(3,2),\textrm{reg}} &= 513.56298-754.78793 \bar{z}-280.97494 \bar{z}^2-2.0101406 \bar{z}^3\\
&+503.52967 \bar{z}^4+627.89991 \bar{z}^5+691.45552 \bar{z}^6+733.60753 \bar{z}^7\\
&+765.14788 \bar{z}^8+790.66308 \bar{z}^9+812.57547 \bar{z}^{10}+832.30620 \bar{z}^{11}\\
&+850.73481 \bar{z}^{12}+868.42184 \bar{z}^{13}+885.73010 \bar{z}^{14}+902.89588 \bar{z}^{15}\\
&+920.07262 \bar{z}^{16}+937.35866 \bar{z}^{17}+954.81528 \bar{z}^{18}+972.47867 \bar{z}^{19}\\
&+990.36794 \bar{z}^{20}+1008.4906 \bar{z}^{21}+1026.8464 \bar{z}^{22}+1045.4298 \bar{z}^{23}\\
&+1064.2318 \bar{z}^{24}+1083.2414 \bar{z}^{25}+1102.4464 \bar{z}^{26}+1121.8338 \bar{z}^{27}\\
&+1141.3904 \bar{z}^{28}+1161.1034 \bar{z}^{29}+1180.9600 \bar{z}^{30}+1200.9479 \bar{z}^{31}\\
&+1221.0555 \bar{z}^{32}+1241.2716 \bar{z}^{33}+1261.5856 \bar{z}^{34}+1281.9875 \bar{z}^{35}\\
&+1302.4680 \bar{z}^{36}+1323.0182 \bar{z}^{37}
\end{aligned}
\end{gather}
%the qg 3,1
\begin{gather}
\begin{aligned}
\eta_{qg}^{(3,1),\textrm{reg}} &= -313.98523+807.28021 \bar{z}+673.01632 \bar{z}^2+424.92437 \bar{z}^3\\
&-94.523260 \bar{z}^4-16.197667 \bar{z}^5+53.689920 \bar{z}^6+107.82115 \bar{z}^7\\
&+152.20191 \bar{z}^8+190.11227 \bar{z}^9+223.24799 \bar{z}^{10}+252.59416 \bar{z}^{11}\\
&+278.80517 \bar{z}^{12}+302.36320 \bar{z}^{13}+323.64795 \bar{z}^{14}+342.97017 \bar{z}^{15}\\
&+360.58960 \bar{z}^{16}+376.72599 \bar{z}^{17}+391.56667 \bar{z}^{18}+405.27209 \bar{z}^{19}\\
&+417.98023 \bar{z}^{20}+429.81014 \bar{z}^{21}+440.86488 \bar{z}^{22}+451.23389 \bar{z}^{23}\\
&+460.99506 \bar{z}^{24}+470.21638 \bar{z}^{25}+478.95737 \bar{z}^{26}+487.27030 \bar{z}^{27}\\
&+495.20115 \bar{z}^{28}+502.79050 \bar{z}^{29}+510.07423 \bar{z}^{30}+517.08417 \bar{z}^{31}\\
&+523.84857 \bar{z}^{32}+530.39259 \bar{z}^{33}+536.73868 \bar{z}^{34}+542.90692 \bar{z}^{35}\\
&+548.91525 \bar{z}^{36}+554.77980 \bar{z}^{37}
\end{aligned}
\end{gather}
%the qg 3,0
\begin{gather}
\begin{aligned}
\eta_{qg}^{(3,0),\textrm{reg}} &= 204.62079+94.711709 \bar{z}-336.52127 \bar{z}^2+51.214999 \bar{z}^3\\
&+240.58379 \bar{z}^4+132.45353 \bar{z}^5+96.832530 \bar{z}^6+88.263488 \bar{z}^7\\
&+90.475716 \bar{z}^8+97.701845 \bar{z}^9+107.59956 \bar{z}^{10}+119.06337 \bar{z}^{11}\\
&+131.49609 \bar{z}^{12}+144.53662 \bar{z}^{13}+157.94768 \bar{z}^{14}+171.56432 \bar{z}^{15}\\
&+185.26783 \bar{z}^{16}+198.97100 \bar{z}^{17}+212.60919 \bar{z}^{18}+226.13430 \bar{z}^{19}\\
&+239.51072 \bar{z}^{20}+252.71229 \bar{z}^{21}+265.72011 \bar{z}^{22}+278.52088 \bar{z}^{23}\\
&+291.10563 \bar{z}^{24}+303.46873 \bar{z}^{25}+315.60716 \bar{z}^{26}+327.51992 \bar{z}^{27}\\
&+339.20753 \bar{z}^{28}+350.67174 \bar{z}^{29}+361.91520 \bar{z}^{30}+372.94122 \bar{z}^{31}\\
&+383.75366 \bar{z}^{32}+394.35673 \bar{z}^{33}+404.75490 \bar{z}^{34}+414.95284 \bar{z}^{35}\\
&+424.95529 \bar{z}^{36}+434.76706 \bar{z}^{37}
\end{aligned}
\end{gather}

%%%%%%%%%%%%%%%%%%%%%%%%%%%%%%%%%%%%%%%%%%%%%
%%%%%%%%%%%%%%%%%%%%%%%%%%%%%%%%%%%%%%%%%%%%%

\subsection{The $q\bar{q}$ channel}

\begin{gather}
\begin{aligned}
\eta_{q\bar{q}}^{(3,2),\textrm{reg}} &= 52.489897 \bar{z}+121.14225 \bar{z}^2+546.26186 \bar{z}^3\\
&+430.10665 \bar{z}^4+395.20262 \bar{z}^5+377.03244 \bar{z}^6+365.05682 \bar{z}^7\\
&+356.30539 \bar{z}^8+349.64832 \bar{z}^9+344.54422 \bar{z}^{10}+340.68027 \bar{z}^{11}\\
&+337.84848 \bar{z}^{12}+335.89655 \bar{z}^{13}+334.70587 \bar{z}^{14}+334.18036 \bar{z}^{15}\\
&+334.24025 \bar{z}^{16}+334.81815 \bar{z}^{17}+335.85649 \bar{z}^{18}+337.30562 \bar{z}^{19}\\
&+339.12245 \bar{z}^{20}+341.26935 \bar{z}^{21}+343.71330 \bar{z}^{22}+346.42520 \bar{z}^{23}\\
&+349.37931 \bar{z}^{24}+352.55277 \bar{z}^{25}+355.92522 \bar{z}^{26}+359.47847 \bar{z}^{27}\\
&+363.19620 \bar{z}^{28}+367.06378 \bar{z}^{29}+371.06801 \bar{z}^{30}+375.19698 \bar{z}^{31}\\
&+379.43991 \bar{z}^{32}+383.78703 \bar{z}^{33}+388.22945 \bar{z}^{34}+392.75910 \bar{z}^{35}\\
&+397.36861 \bar{z}^{36}+402.05123 \bar{z}^{37}
\end{aligned}
\end{gather}
%the qqb 3,1
\begin{gather}
\begin{aligned}
\eta_{q\bar{q}}^{(3,1),\textrm{reg}} &= -13.561787 \bar{z}-122.83887 \bar{z}^2-747.63122 \bar{z}^3\\
&-396.29959 \bar{z}^4-305.88934 \bar{z}^5-259.42707 \bar{z}^6-228.03650 \bar{z}^7\\
&-204.06989 \bar{z}^8-184.61437 \bar{z}^9-168.25305 \bar{z}^{10}-154.17060 \bar{z}^{11}\\
&-141.84193 \bar{z}^{12}-130.90258 \bar{z}^{13}-121.08653 \bar{z}^{14}-112.19267 \bar{z}^{15}\\
&-104.06504 \bar{z}^{16}-96.580303 \bar{z}^{17}-89.639462 \bar{z}^{18}-83.162027 \bar{z}^{19}\\
&-77.081876 \bar{z}^{20}-71.344195 \bar{z}^{21}-65.903187 \bar{z}^{22}-60.720315 \bar{z}^{23}\\
&-55.762951 \bar{z}^{24}-51.003310 \bar{z}^{25}-46.417609 \bar{z}^{26}-41.985393 \bar{z}^{27}\\
&-37.688995 \bar{z}^{28}-33.513090 \bar{z}^{29}-29.444339 \bar{z}^{30}-25.471089 \bar{z}^{31}\\
&-21.583127 \bar{z}^{32}-17.771478 \bar{z}^{33}-14.028228 \bar{z}^{34}-10.346383 \bar{z}^{35}\\
&-6.7197486 \bar{z}^{36}-3.1428213 \bar{z}^{37}
\end{aligned}
\end{gather}
%the qqb 3,0
\begin{gather}
\begin{aligned}
\eta_{q\bar{q}}^{(3,0),\textrm{reg}} &= -37.707516 \bar{z}+64.836867 \bar{z}^2+370.64251 \bar{z}^3\\
&+99.620940 \bar{z}^4+86.612810 \bar{z}^5+95.425837 \bar{z}^6+108.12237 \bar{z}^7\\
&+121.73122 \bar{z}^8+135.50571 \bar{z}^9+149.18659 \bar{z}^{10}+162.65901 \bar{z}^{11}\\
&+175.86438 \bar{z}^{12}+188.77159 \bar{z}^{13}+201.36522 \bar{z}^{14}+213.63958 \bar{z}^{15}\\
&+225.59530 \bar{z}^{16}+237.23713 \bar{z}^{17}+248.57249 \bar{z}^{18}+259.61045 \bar{z}^{19}\\
&+270.36110 \bar{z}^{20}+280.83500 \bar{z}^{21}+291.04293 \bar{z}^{22}+300.99558 \bar{z}^{23}\\
&+310.70347 \bar{z}^{24}+320.17681 \bar{z}^{25}+329.42545 \bar{z}^{26}+338.45884 \bar{z}^{27}\\
&+347.28601 \bar{z}^{28}+355.91555 \bar{z}^{29}+364.35564 \bar{z}^{30}+372.61407 \bar{z}^{31}\\
&+380.69820 \bar{z}^{32}+388.61503 \bar{z}^{33}+396.37117 \bar{z}^{34}+403.97291 \bar{z}^{35}\\
&+411.42620 \bar{z}^{36}+418.73668 \bar{z}^{37}
\end{aligned}
\end{gather}

%%%%%%%%%%%%%%%%%%%%%%%%%%%%%%%%%%%%%%%%%%%%%
%%%%%%%%%%%%%%%%%%%%%%%%%%%%%%%%%%%%%%%%%%%%%

\subsection{The $qq$ channel}

\begin{gather}
\begin{aligned}
\eta_{qq}^{(3,2),\textrm{reg}} &= 52.489897 \bar{z}+115.88299 \bar{z}^2+206.89141 \bar{z}^3\\
&+237.16727 \bar{z}^4+253.85312 \bar{z}^5+264.50690 \bar{z}^6+271.88762 \bar{z}^7\\
&+277.47724 \bar{z}^8+282.11036 \bar{z}^9+286.26594 \bar{z}^{10}+290.22209 \bar{z}^{11}\\
&+294.14093 \bar{z}^{12}+298.11608 \bar{z}^{13}+302.20004 \bar{z}^{14}+306.42029 \bar{z}^{15}\\
&+310.78904 \bar{z}^{16}+315.30914 \bar{z}^{17}+319.97778 \bar{z}^{18}+324.78884 \bar{z}^{19}\\
&+329.73434 \bar{z}^{20}+334.80540 \bar{z}^{21}+339.99280 \bar{z}^{22}+345.28737 \bar{z}^{23}\\
&+350.68023 \bar{z}^{24}+356.16287 \bar{z}^{25}+361.72729 \bar{z}^{26}+367.36598 \bar{z}^{27}\\
&+373.07194 \bar{z}^{28}+378.83869 \bar{z}^{29}+384.66021 \bar{z}^{30}+390.53096 \bar{z}^{31}\\
&+396.44582 \bar{z}^{32}+402.40008 \bar{z}^{33}+408.38938 \bar{z}^{34}+414.40972 \bar{z}^{35}\\
&+420.45742 \bar{z}^{36}+426.52908 \bar{z}^{37}
\end{aligned}
\end{gather}
%the qq 3,1
\begin{gather}
\begin{aligned}
\eta_{qq}^{(3,1),\textrm{reg}} &= -13.561787 \bar{z}-100.44381 \bar{z}^2-197.02897 \bar{z}^3\\
&-201.49505 \bar{z}^4-196.70233 \bar{z}^5-189.72948 \bar{z}^6-181.90181 \bar{z}^7\\
&-174.01305 \bar{z}^8-166.44104 \bar{z}^9-159.32993 \bar{z}^{10}-152.70888 \bar{z}^{11}\\
&-146.55489 \bar{z}^{12}-140.82408 \bar{z}^{13}-135.46673 \bar{z}^{14}-130.43420 \bar{z}^{15}\\
&-125.68188 \bar{z}^{16}-121.17016 \bar{z}^{17}-116.86451 \bar{z}^{18}-112.73513 \bar{z}^{19}\\
&-108.75638 \bar{z}^{20}-104.90636 \bar{z}^{21}-101.16628 \bar{z}^{22}-97.520078 \bar{z}^{23}\\
&-93.954009 \bar{z}^{24}-90.456274 \bar{z}^{25}-87.016749 \bar{z}^{26}-83.626728 \bar{z}^{27}\\
&-80.278716 \bar{z}^{28}-76.966251 \bar{z}^{29}-73.683754 \bar{z}^{30}-70.426402 \bar{z}^{31}\\
&-67.190021 \bar{z}^{32}-63.970996 \bar{z}^{33}-60.766195 \bar{z}^{34}-57.572901 \bar{z}^{35}\\
&-54.388756 \bar{z}^{36}-51.211716 \bar{z}^{37}
\end{aligned}
\end{gather}
%the qq 3,0
\begin{gather}
\begin{aligned}
\eta_{qq}^{(3,0),\textrm{reg}} &= -39.014783 \bar{z}+16.214979 \bar{z}^2+49.524960 \bar{z}^3\\
&+45.647897 \bar{z}^4+49.192648 \bar{z}^5+56.534430 \bar{z}^6+65.488703 \bar{z}^7\\
&+75.308404 \bar{z}^8+85.593580 \bar{z}^9+96.089042 \bar{z}^{10}+106.62551 \bar{z}^{11}\\
&+117.09068 \bar{z}^{12}+127.41095 \bar{z}^{13}+137.53925 \bar{z}^{14}+147.44661 \bar{z}^{15}\\
&+157.11652 \bar{z}^{16}+166.54094 \bar{z}^{17}+175.71767 \bar{z}^{18}+184.64838 \bar{z}^{19}\\
&+193.33742 \bar{z}^{20}+201.79079 \bar{z}^{21}+210.01561 \bar{z}^{22}+218.01959 \bar{z}^{23}\\
&+225.81076 \bar{z}^{24}+233.39727 \bar{z}^{25}+240.78718 \bar{z}^{26}+247.98841 \bar{z}^{27}\\
&+255.00867 \bar{z}^{28}+261.85538 \bar{z}^{29}+268.53568 \bar{z}^{30}+275.05640 \bar{z}^{31}\\
&+281.42407 \bar{z}^{32}+287.64490 \bar{z}^{33}+293.72482 \bar{z}^{34}+299.66944 \bar{z}^{35}\\
&+305.48412 \bar{z}^{36}+311.17392 \bar{z}^{37}
\end{aligned}
\end{gather}

%%%%%%%%%%%%%%%%%%%%%%%%%%%%%%%%%%%%%%%%%%%%%
%%%%%%%%%%%%%%%%%%%%%%%%%%%%%%%%%%%%%%%%%%%%%

\subsection{The $qq^\prime$ channel}

\begin{gather}
\begin{aligned}
\eta_{qq'}^{(3,2),\textrm{reg}} &= 52.489897 \bar{z}+115.95707 \bar{z}^2+207.09717 \bar{z}^3\\
&+237.47076 \bar{z}^4+254.23192 \bar{z}^5+264.94538 \bar{z}^6+272.37440 \bar{z}^7\\
&+278.00388 \bar{z}^8+282.67045 \bar{z}^9+286.85449 \bar{z}^{10}+290.83519 \bar{z}^{11}\\
&+294.77542 \bar{z}^{12}+298.76938 \bar{z}^{13}+302.87003 \bar{z}^{14}+307.10519 \bar{z}^{15}\\
&+311.48736 \bar{z}^{16}+316.01959 \bar{z}^{17}+320.69926 \bar{z}^{18}+325.52041 \bar{z}^{19}\\
&+330.47516 \bar{z}^{20}+335.55474 \bar{z}^{21}+340.75002 \bar{z}^{22}+346.05189 \bar{z}^{23}\\
&+351.45154 \bar{z}^{24}+356.94052 \bar{z}^{25}+362.51087 \bar{z}^{26}+368.15511 \bar{z}^{27}\\
&+373.86629 \bar{z}^{28}+379.63795 \bar{z}^{29}+385.46411 \bar{z}^{30}+391.33924 \bar{z}^{31}\\
&+397.25825 \bar{z}^{32}+403.21643 \bar{z}^{33}+409.20946 \bar{z}^{34}+415.23335 \bar{z}^{35}\\
&+421.28443 \bar{z}^{36}+427.35932 \bar{z}^{37}
\end{aligned}
\end{gather}
%the qQ 3,1
\begin{gather}
\begin{aligned}
\eta_{qq'}^{(3,1),\textrm{reg}} &= -13.561787 \bar{z}-101.23393 \bar{z}^2-199.27314 \bar{z}^3\\
&-204.58988 \bar{z}^4-200.32378 \bar{z}^5-193.67683 \bar{z}^6-186.04539 \bar{z}^7\\
&-178.26735 \bar{z}^8-170.74845 \bar{z}^9-163.65084 \bar{z}^{10}-157.01563 \bar{z}^{11}\\
&-150.82793 \bar{z}^{12}-145.04949 \bar{z}^{13}-139.63457 \bar{z}^{14}-134.53740 \bar{z}^{15}\\
&-129.71545 \bar{z}^{16}-125.13066 \bar{z}^{17}-120.74964 \bar{z}^{18}-116.54344 \bar{z}^{19}\\
&-112.48710 \bar{z}^{20}-108.55918 \bar{z}^{21}-104.74128 \bar{z}^{22}-101.01763 \bar{z}^{23}\\
&-97.374684 \bar{z}^{24}-93.800809 \bar{z}^{25}-90.286003 \bar{z}^{26}-86.821649 \bar{z}^{27}\\
&-83.400317 \bar{z}^{28}-80.015589 \bar{z}^{29}-76.661913 \bar{z}^{30}-73.334486 \bar{z}^{31}\\
&-70.029139 \bar{z}^{32}-66.742261 \bar{z}^{33}-63.470710 \bar{z}^{34}-60.211761 \bar{z}^{35}\\
&-56.963043 \bar{z}^{36}-53.722495 \bar{z}^{37}
\end{aligned}
\end{gather}
%the qQ 3,0
\begin{gather}
\begin{aligned}
\eta_{qq'}^{(3,0),\textrm{reg}} &= -38.124370 \bar{z}+21.925696 \bar{z}^2+62.593745 \bar{z}^3\\
&+62.740689 \bar{z}^4+68.779415 \bar{z}^5+77.692571 \bar{z}^6+87.639674 \bar{z}^7\\
&+98.079383 \bar{z}^8+108.73738 \bar{z}^9+119.43753 \bar{z}^{10}+130.06186 \bar{z}^{11}\\
&+140.53231 \bar{z}^{12}+150.79875 \bar{z}^{13}+160.83049 \bar{z}^{14}+170.61028 \bar{z}^{15}\\
&+180.13007 \bar{z}^{16}+189.38806 \bar{z}^{17}+198.38664 \bar{z}^{18}+207.13098 \bar{z}^{19}\\
&+215.62804 \bar{z}^{20}+223.88584 \bar{z}^{21}+231.91300 \bar{z}^{22}+239.71843 \bar{z}^{23}\\
&+247.31108 \bar{z}^{24}+254.69977 \bar{z}^{25}+261.89310 \bar{z}^{26}+268.89942 \bar{z}^{27}\\
&+275.72674 \bar{z}^{28}+282.38271 \bar{z}^{29}+288.87465 \bar{z}^{30}+295.20952 \bar{z}^{31}\\
&+301.39392 \bar{z}^{32}+307.43414 \bar{z}^{33}+313.33611 \bar{z}^{34}+319.10548 \bar{z}^{35}\\
&+324.74760 \bar{z}^{36}+330.26751 \bar{z}^{37}
\end{aligned}
\end{gather}

\section{Color and flavor number dependence  for the first coefficients of the threshold expansion}
\label{app:expansioncoefficients_analytic}

In this appendix we present analytic result for the first few coefficients in the threshold expansion for each partonic channel. We use the notation
\begin{equation}
\label{eq:analytic_n3lo}
\eta_{ij}^{(3,m),{\rm reg}} = \sum_{n=0}^\infty  \sum_{a=-3}^3 \sum_{b=0}^2 N_c^a
N_f^b {\bar z}^n  {\cal C}_{ij}[m,a,b,n],
\end{equation}
The non-zero coefficients ${\cal C}_{ij}[m,a,b,n]$ for $m=0,1,2$ and $n \leq 5$ are given in the remainder of this appendix.

\subsection{The $gg$ channel}
\begingroup
\addtolength{\jot}{1em}
{\small
\begin{gather*}
\mathcal{C}_{gg}[2,3,0,0]=-\frac{2147 \zeta _2}{12}-181 \zeta _3+\frac{2711}{27}\\
\mathcal{C}_{gg}[2,3,0,1]=\frac{22645 \zeta _2}{48}+362 \zeta _3-\frac{363355}{288}\\
\mathcal{C}_{gg}[2,3,0,2]=\frac{2006159}{2592}-\frac{30767 \zeta _2}{144}\\
\mathcal{C}_{gg}[2,3,0,3]=\frac{7441 \zeta _2}{36}+181 \zeta _3-\frac{11532781}{10368}\\
\mathcal{C}_{gg}[2,3,0,4]=\frac{11263 \zeta _2}{90}+181 \zeta _3-\frac{609287813}{1296000}\\
\mathcal{C}_{gg}[2,3,0,5]=\frac{779 \zeta _2}{9}+181 \zeta _3-\frac{849910693}{2592000}\\
\mathcal{C}_{gg}[2,2,1,0]=\frac{545 \zeta _2}{48}-\frac{4139}{216}\\
\mathcal{C}_{gg}[2,2,1,1]=\frac{192943}{1728}-\frac{629 \zeta _2}{24}\\
\mathcal{C}_{gg}[2,2,1,2]=\frac{\zeta _2}{72}-\frac{315721}{5184}\\
\mathcal{C}_{gg}[2,2,1,3]=\frac{285745}{3456}-\frac{1069 \zeta _2}{72}\\
\mathcal{C}_{gg}[2,2,1,4]=\frac{15200689}{345600}-\frac{1069 \zeta _2}{72}\\
\mathcal{C}_{gg}[2,2,1,5]=\frac{174604379}{5184000}-\frac{2155 \zeta _2}{144}\\
\mathcal{C}_{gg}[2,1,2,0]=\frac{59}{108}\\
\mathcal{C}_{gg}[2,1,2,1]=-\frac{1571}{864}\\
\mathcal{C}_{gg}[2,1,2,2]=\frac{145}{216}\\
\mathcal{C}_{gg}[2,1,2,3]=-\frac{2693}{2592}\\
\mathcal{C}_{gg}[2,1,2,4]=-\frac{19133}{25920}\\
\mathcal{C}_{gg}[2,1,2,5]=-\frac{31681}{51840}\\
\mathcal{C}_{gg}[2,0,1,0]=\frac{1}{4}\\
\mathcal{C}_{gg}[2,0,1,1]=\frac{119 \zeta _2}{24}-\frac{1193}{54}\\
\mathcal{C}_{gg}[2,0,1,2]=\frac{5221}{288}\\
\mathcal{C}_{gg}[2,0,1,3]=\frac{173 \zeta _2}{36}-\frac{40231}{1728}\\
\mathcal{C}_{gg}[2,0,1,4]=\frac{173 \zeta _2}{36}-\frac{32005}{3456}\\
\mathcal{C}_{gg}[2,0,1,5]=\frac{359 \zeta _2}{72}-\frac{3976679}{518400}\\
\mathcal{C}_{gg}[2,-1,2,1]=\frac{199}{432}\\
\mathcal{C}_{gg}[2,-1,2,2]=-\frac{19}{72}\\
\mathcal{C}_{gg}[2,-1,2,3]=\frac{355}{864}\\
\mathcal{C}_{gg}[2,-1,2,4]=\frac{17}{64}\\
\mathcal{C}_{gg}[2,-1,2,5]=\frac{11201}{51840}\\
\mathcal{C}_{gg}[2,-2,1,1]=\frac{475}{192}-\frac{23 \zeta _2}{24}\\
\mathcal{C}_{gg}[2,-2,1,2]=-\frac{275}{96}\\
\mathcal{C}_{gg}[2,-2,1,3]=\frac{5053}{1152}-\frac{29 \zeta _2}{36}\\
\mathcal{C}_{gg}[2,-2,1,4]=\frac{26765}{13824}-\frac{29 \zeta _2}{36}\\
\mathcal{C}_{gg}[2,-2,1,5]=\frac{1660559}{1036800}-\frac{37 \zeta _2}{45}\\
\mathcal{C}_{gg}[1,3,0,0]=\frac{2375 \zeta _2}{18}+362 \zeta _3+77 \zeta _4-\frac{9547}{108}\\
\mathcal{C}_{gg}[1,3,0,1]=-\frac{142441 \zeta _2}{144}-\frac{11093 \zeta _3}{12}-154 \zeta _4+\frac{1077125}{432}\\
\mathcal{C}_{gg}[1,3,0,2]=\frac{26317 \zeta _2}{48}+\frac{16115 \zeta _3}{36}-\frac{645199}{432}\\
\mathcal{C}_{gg}[1,3,0,3]=-\frac{66785 \zeta _2}{96}-\frac{3613 \zeta _3}{9}-77 \zeta _4+\frac{25909463}{10368}\\
\mathcal{C}_{gg}[1,3,0,4]=-\frac{513361 \zeta _2}{1440}-\frac{20779 \zeta _3}{90}-77 \zeta _4+\frac{5210522741}{5184000}\\
\mathcal{C}_{gg}[1,3,0,5]=-\frac{5660203 \zeta _2}{21600}-\frac{27031 \zeta _3}{180}-77 \zeta _4+\frac{115009800821}{155520000}\\
\mathcal{C}_{gg}[1,2,1,0]=-\frac{1813 \zeta _2}{72}-\frac{223 \zeta _3}{12}+\frac{8071}{324}\\
\mathcal{C}_{gg}[1,2,1,1]=\frac{673 \zeta _2}{9}+\frac{349 \zeta _3}{8}-\frac{608693}{2592}\\
\mathcal{C}_{gg}[1,2,1,2]=-\frac{1601 \zeta _2}{48}-\frac{\zeta _3}{6}+\frac{692437}{5184}\\
\mathcal{C}_{gg}[1,2,1,3]=\frac{4127 \zeta _2}{108}+\frac{599 \zeta _3}{24}-\frac{741821}{3456}\\
\mathcal{C}_{gg}[1,2,1,4]=\frac{101351 \zeta _2}{4320}+\frac{599 \zeta _3}{24}-\frac{263548441}{2592000}\\
\mathcal{C}_{gg}[1,2,1,5]=\frac{36493 \zeta _2}{2160}+\frac{18137 \zeta _3}{720}-\frac{515585813}{6480000}\\
\mathcal{C}_{gg}[1,1,2,0]=\frac{4 \zeta _2}{9}-\frac{163}{324}\\
\mathcal{C}_{gg}[1,1,2,1]=\frac{2195}{648}-\frac{11 \zeta _2}{12}\\
\mathcal{C}_{gg}[1,1,2,2]=-\frac{4109}{1728}\\
\mathcal{C}_{gg}[1,1,2,3]=\frac{1619}{576}-\frac{17 \zeta _2}{36}\\
\mathcal{C}_{gg}[1,1,2,4]=\frac{720019}{518400}-\frac{17 \zeta _2}{36}\\
\mathcal{C}_{gg}[1,1,2,5]=\frac{130223}{129600}-\frac{511 \zeta _2}{1080}\\
\mathcal{C}_{gg}[1,0,1,0]=\frac{\zeta _2}{24}+3 \zeta _3-\frac{17}{4}\\
\mathcal{C}_{gg}[1,0,1,1]=-\frac{239 \zeta _2}{18}-\frac{365 \zeta _3}{24}+\frac{142381}{2592}\\
\mathcal{C}_{gg}[1,0,1,2]=\frac{147 \zeta _2}{16}-\frac{20647}{432}\\
\mathcal{C}_{gg}[1,0,1,3]=-\frac{5221 \zeta _2}{432}-\frac{889 \zeta _3}{72}+\frac{102361}{1728}\\
\mathcal{C}_{gg}[1,0,1,4]=-\frac{5129 \zeta _2}{864}-\frac{889 \zeta _3}{72}+\frac{14772497}{829440}\\
\mathcal{C}_{gg}[1,0,1,5]=-\frac{5363 \zeta _2}{1440}-\frac{9137 \zeta _3}{720}+\frac{313545373}{20736000}\\
\mathcal{C}_{gg}[1,-1,2,1]=\frac{\zeta _2}{36}-\frac{181}{162}\\
\mathcal{C}_{gg}[1,-1,2,2]=\frac{1187}{864}\\
\mathcal{C}_{gg}[1,-1,2,3]=\frac{\zeta _2}{36}-\frac{725}{648}\\
\mathcal{C}_{gg}[1,-1,2,4]=\frac{\zeta _2}{36}-\frac{599}{2304}\\
\mathcal{C}_{gg}[1,-1,2,5]=\frac{31 \zeta _2}{1080}-\frac{136769}{1296000}\\
\mathcal{C}_{gg}[1,-2,1,1]=\frac{71 \zeta _2}{24}+\frac{23 \zeta _3}{12}-\frac{245}{48}\\
\mathcal{C}_{gg}[1,-2,1,2]=\frac{47}{8}-\frac{109 \zeta _2}{48}\\
\mathcal{C}_{gg}[1,-2,1,3]=\frac{1129 \zeta _2}{432}+\frac{14 \zeta _3}{9}-\frac{117785}{10368}\\
\mathcal{C}_{gg}[1,-2,1,4]=\frac{1313 \zeta _2}{864}+\frac{14 \zeta _3}{9}-\frac{655181}{165888}\\
\mathcal{C}_{gg}[1,-2,1,5]=\frac{25019 \zeta _2}{21600}+\frac{19 \zeta _3}{12}-\frac{1041496027}{311040000}\\
\mathcal{C}_{gg}[0,3,0,0]=\frac{725 \zeta _3 \zeta _2}{6}-\frac{11183 \zeta _2}{162}-\frac{32849 \zeta _3}{216}-\frac{821 \zeta _4}{12}-186 \zeta _5+\frac{834419}{23328}\\
\mathcal{C}_{gg}[0,3,0,1]=-\frac{725}{3} \zeta _3 \zeta _2+\frac{2578495 \zeta _2}{2592}+\frac{54373 \zeta _3}{54}+\frac{9287 \zeta _4}{48}+372 \zeta _5-\frac{112071959}{46656}\\
\mathcal{C}_{gg}[0,3,0,2]=-\frac{1587065 \zeta _2}{2592}-\frac{12167 \zeta _3}{24}-\frac{14905 \zeta _4}{144}+\frac{136933337}{93312}\\
\mathcal{C}_{gg}[0,3,0,3]=-\frac{725}{6} \zeta _3 \zeta _2+\frac{17524253 \zeta _2}{20736}+\frac{315079 \zeta _3}{432}+\frac{22799 \zeta _4}{288}+186 \zeta _5-\frac{7695352049}{2985984}\\
\mathcal{C}_{gg}[0,3,0,4]=-\frac{725}{6} \zeta _3 \zeta _2+\frac{225231577 \zeta _2}{648000}+\frac{173857 \zeta _3}{432}+\frac{58783 \zeta _4}{1440}+186 \zeta _5-\frac{4501027226621}{4665600000}\\
\mathcal{C}_{gg}[0,3,0,5]=-\frac{725}{6} \zeta _3 \zeta _2+\frac{22725557 \zeta _2}{96000}+\frac{378617 \zeta _3}{1200}+\frac{6739 \zeta _4}{288}+186 \zeta _5-\frac{6555187542491}{9331200000}\\
\mathcal{C}_{gg}[0,2,1,0]=\frac{4579 \zeta _2}{324}+\frac{1789 \zeta _3}{72}+\frac{19 \zeta _4}{8}-\frac{527831}{46656}\\
\mathcal{C}_{gg}[0,2,1,1]=-\frac{60211 \zeta _2}{648}-\frac{670 \zeta _3}{9}-\frac{569 \zeta _4}{96}+\frac{9673753}{46656}\\
\mathcal{C}_{gg}[0,2,1,2]=\frac{111257 \zeta _2}{2592}+\frac{1333 \zeta _3}{48}+\frac{17 \zeta _4}{36}-\frac{26275573}{186624}\\
\mathcal{C}_{gg}[0,2,1,3]=-\frac{42613 \zeta _2}{648}-\frac{16411 \zeta _3}{432}-\frac{955 \zeta _4}{288}+\frac{57063737}{248832}\\
\mathcal{C}_{gg}[0,2,1,4]=-\frac{4880261 \zeta _2}{129600}-\frac{28189 \zeta _3}{1080}-\frac{955 \zeta _4}{288}+\frac{9425777309}{103680000}\\
\mathcal{C}_{gg}[0,2,1,5]=-\frac{1879457 \zeta _2}{64800}-\frac{907349 \zeta _3}{43200}-\frac{9799 \zeta _4}{2880}+\frac{32259119399}{466560000}\\
\mathcal{C}_{gg}[0,1,2,0]=-\frac{19 \zeta _2}{36}-\frac{5 \zeta _3}{27}+\frac{49}{729}\\
\mathcal{C}_{gg}[0,1,2,1]=\frac{677 \zeta _2}{432}+\frac{10 \zeta _3}{27}-\frac{59731}{23328}\\
\mathcal{C}_{gg}[0,1,2,2]=\frac{36443}{15552}-\frac{109 \zeta _2}{216}\\
\mathcal{C}_{gg}[0,1,2,3]=\frac{137 \zeta _2}{162}+\frac{5 \zeta _3}{27}-\frac{79931}{23328}\\
\mathcal{C}_{gg}[0,1,2,4]=\frac{1043 \zeta _2}{1620}+\frac{5 \zeta _3}{27}-\frac{66874081}{46656000}\\
\mathcal{C}_{gg}[0,1,2,5]=\frac{35627 \zeta _2}{64800}+\frac{5 \zeta _3}{27}-\frac{6687083}{6220800}\\
\mathcal{C}_{gg}[0,0,1,0]=-\frac{5 \zeta _2}{24}-\frac{149 \zeta _3}{72}-\frac{\zeta _4}{4}+\frac{5065}{1728}\\
\mathcal{C}_{gg}[0,0,1,1]=\frac{3689 \zeta _2}{216}+\frac{2273 \zeta _3}{144}+\frac{5 \zeta _4}{3}-\frac{401911}{7776}\\
\mathcal{C}_{gg}[0,0,1,2]=-\frac{3859 \zeta _2}{288}-\frac{1199 \zeta _3}{96}+\frac{578489}{10368}\\
\mathcal{C}_{gg}[0,0,1,3]=\frac{52769 \zeta _2}{2592}+\frac{2485 \zeta _3}{216}+\frac{179 \zeta _4}{96}-\frac{25339}{384}\\
\mathcal{C}_{gg}[0,0,1,4]=\frac{190591 \zeta _2}{20736}+\frac{36403 \zeta _3}{8640}+\frac{179 \zeta _4}{96}-\frac{705541369}{49766400}\\
\mathcal{C}_{gg}[0,0,1,5]=\frac{19819151 \zeta _2}{2592000}+\frac{31817 \zeta _3}{21600}+\frac{1801 \zeta _4}{960}-\frac{244316987519}{18662400000}\\
\mathcal{C}_{gg}[0,-1,2,1]=\frac{583}{486}-\frac{7 \zeta _2}{27}\\
\mathcal{C}_{gg}[0,-1,2,2]=\frac{7 \zeta _2}{72}-\frac{9235}{5184}\\
\mathcal{C}_{gg}[0,-1,2,3]=\frac{374}{243}-\frac{29 \zeta _2}{108}\\
\mathcal{C}_{gg}[0,-1,2,4]=\frac{53797}{248832}-\frac{2 \zeta _2}{9}\\
\mathcal{C}_{gg}[0,-1,2,5]=\frac{37539559}{233280000}-\frac{13387 \zeta _2}{64800}\\
\mathcal{C}_{gg}[0,-2,1,1]=-\frac{47 \zeta _2}{16}-\frac{43 \zeta _3}{16}+\frac{23 \zeta _4}{96}+\frac{61}{12}\\
\mathcal{C}_{gg}[0,-2,1,2]=\frac{117 \zeta _2}{32}+\frac{67 \zeta _3}{32}-\frac{4673}{768}\\
\mathcal{C}_{gg}[0,-2,1,3]=-\frac{12257 \zeta _2}{2592}-\frac{53 \zeta _3}{27}-\frac{5 \zeta _4}{144}+\frac{221}{18}\\
\mathcal{C}_{gg}[0,-2,1,4]=-\frac{11377 \zeta _2}{5184}-\frac{1631 \zeta _3}{1728}-\frac{5 \zeta _4}{144}+\frac{6529079}{1990656}\\
\mathcal{C}_{gg}[0,-2,1,5]=-\frac{2375053 \zeta _2}{1296000}-\frac{1691 \zeta _3}{2880}-\frac{13 \zeta _4}{240}+\frac{54978128417}{18662400000}
\end{gather*}
}
\endgroup
%%%%%%%%%%%%%%%%%%%%%%%%%%%%%%%%%%%%%%%%%%%%%
%%%%%%%%%%%%%%%%%%%%%%%%%%%%%%%%%%%%%%%%%%%%%

\subsection{The $qg$ channel}
\begingroup
\addtolength{\jot}{1em}
{\small \begin{gather*}
  \mathcal{C}_{qg}[2,3,0,0]=\frac{1729 \zeta _2}{576}+\frac{1687 \zeta _3}{96}-\frac{120073}{41472}\\
  \mathcal{C}_{qg}[2,3,0,1]=\frac{4577 \zeta _2}{288}+\frac{1687 \zeta _3}{96}-\frac{3446137}{41472}\\
  \mathcal{C}_{qg}[2,3,0,2]=\frac{7249 \zeta _2}{288}+\frac{1687 \zeta _3}{48}-\frac{2118601}{20736}\\
  \mathcal{C}_{qg}[2,3,0,3]=\frac{1613 \zeta _2}{96}+\frac{1687 \zeta _3}{48}-\frac{1498555}{20736}\\
  \mathcal{C}_{qg}[2,3,0,4]=\frac{9919 \zeta _2}{1152}+\frac{1687 \zeta _3}{48}-\frac{22757717}{663552}\\
  \mathcal{C}_{qg}[2,3,0,5]=\frac{15487 \zeta _2}{5760}+\frac{1687 \zeta _3}{48}-\frac{1473963793}{82944000}\\
  \mathcal{C}_{qg}[2,2,1,0]=\frac{6427}{10368}-\frac{185 \zeta _2}{288}\\
  \mathcal{C}_{qg}[2,2,1,1]=\frac{33127}{10368}-\frac{185 \zeta _2}{288}\\
  \mathcal{C}_{qg}[2,2,1,2]=\frac{116677}{20736}-\frac{185 \zeta _2}{144}\\
  \mathcal{C}_{qg}[2,2,1,3]=\frac{69853}{20736}-\frac{185 \zeta _2}{144}\\
  \mathcal{C}_{qg}[2,2,1,4]=\frac{117373}{82944}-\frac{185 \zeta _2}{144}\\
  \mathcal{C}_{qg}[2,2,1,5]=\frac{860453}{2073600}-\frac{185 \zeta _2}{144}\\
  \mathcal{C}_{qg}[2,1,2,0]=-\frac{11}{432}\\
  \mathcal{C}_{qg}[2,1,2,1]=-\frac{5}{432}\\
  \mathcal{C}_{qg}[2,1,2,2]=-\frac{25}{432}\\
  \mathcal{C}_{qg}[2,1,2,3]=-\frac{13}{432}\\
  \mathcal{C}_{qg}[2,1,2,4]=-\frac{25}{1728}\\
  \mathcal{C}_{qg}[2,1,2,5]=-\frac{29}{8640}\\
  \mathcal{C}_{qg}[2,1,0,0]=-\frac{1589 \zeta _2}{288}-\frac{4241 \zeta _3}{192}+\frac{46025}{13824}\\
  \mathcal{C}_{qg}[2,1,0,1]=-\frac{11845 \zeta _2}{576}-\frac{4241 \zeta _3}{192}+\frac{1439677}{13824}\\
  \mathcal{C}_{qg}[2,1,0,2]=-\frac{2665 \zeta _2}{72}-\frac{4241 \zeta _3}{96}+\frac{1986127}{13824}\\
  \mathcal{C}_{qg}[2,1,0,3]=-\frac{7019 \zeta _2}{288}-\frac{4241 \zeta _3}{96}+\frac{4076855}{41472}\\
  \mathcal{C}_{qg}[2,1,0,4]=-\frac{1903 \zeta _2}{144}-\frac{4241 \zeta _3}{96}+\frac{35022785}{663552}\\
  \mathcal{C}_{qg}[2,1,0,5]=-\frac{15013 \zeta _2}{2880}-\frac{4241 \zeta _3}{96}+\frac{302835701}{9216000}\\
  \mathcal{C}_{qg}[2,0,1,0]=\frac{59 \zeta _2}{72}-\frac{215}{288}\\
  \mathcal{C}_{qg}[2,0,1,1]=\frac{59 \zeta _2}{72}-\frac{6229}{1728}\\
  \mathcal{C}_{qg}[2,0,1,2]=\frac{59 \zeta _2}{36}-\frac{329}{48}\\
  \mathcal{C}_{qg}[2,0,1,3]=\frac{59 \zeta _2}{36}-\frac{21001}{5184}\\
  \mathcal{C}_{qg}[2,0,1,4]=\frac{59 \zeta _2}{36}-\frac{69071}{41472}\\
  \mathcal{C}_{qg}[2,0,1,5]=\frac{59 \zeta _2}{36}-\frac{516223}{1036800}\\
  \mathcal{C}_{qg}[2,-1,2,0]=\frac{11}{432}\\
  \mathcal{C}_{qg}[2,-1,2,1]=\frac{5}{432}\\
  \mathcal{C}_{qg}[2,-1,2,2]=\frac{25}{432}\\
  \mathcal{C}_{qg}[2,-1,2,3]=\frac{13}{432}\\
  \mathcal{C}_{qg}[2,-1,2,4]=\frac{25}{1728}\\
  \mathcal{C}_{qg}[2,-1,2,5]=\frac{29}{8640}\\
  \mathcal{C}_{qg}[2,-1,0,0]=\frac{541 \zeta _2}{192}+\frac{485 \zeta _3}{96}-\frac{14087}{41472}\\
  \mathcal{C}_{qg}[2,-1,0,1]=\frac{979 \zeta _2}{192}+\frac{485 \zeta _3}{96}-\frac{966503}{41472}\\
  \mathcal{C}_{qg}[2,-1,0,2]=\frac{2597 \zeta _2}{192}+\frac{485 \zeta _3}{48}-\frac{237743}{5184}\\
  \mathcal{C}_{qg}[2,-1,0,3]=\frac{4883 \zeta _2}{576}+\frac{485 \zeta _3}{48}-\frac{7263}{256}\\
  \mathcal{C}_{qg}[2,-1,0,4]=\frac{737 \zeta _2}{144}+\frac{485 \zeta _3}{48}-\frac{4504849}{221184}\\
  \mathcal{C}_{qg}[2,-1,0,5]=\frac{997 \zeta _2}{360}+\frac{485 \zeta _3}{48}-\frac{1382797423}{82944000}\\
  \mathcal{C}_{qg}[2,-2,1,0]=\frac{1313}{10368}-\frac{17 \zeta _2}{96}\\
  \mathcal{C}_{qg}[2,-2,1,1]=\frac{4247}{10368}-\frac{17 \zeta _2}{96}\\
  \mathcal{C}_{qg}[2,-2,1,2]=\frac{25451}{20736}-\frac{17 \zeta _2}{48}\\
  \mathcal{C}_{qg}[2,-2,1,3]=\frac{4717}{6912}-\frac{17 \zeta _2}{48}\\
  \mathcal{C}_{qg}[2,-2,1,4]=\frac{6923}{27648}-\frac{17 \zeta _2}{48}\\
  \mathcal{C}_{qg}[2,-2,1,5]=\frac{57331}{691200}-\frac{17 \zeta _2}{48}\\
  \mathcal{C}_{qg}[2,-3,0,0]=-\frac{29 \zeta _2}{96}-\frac{103 \zeta _3}{192}-\frac{145}{1536}\\
  \mathcal{C}_{qg}[2,-3,0,1]=-\frac{41 \zeta _2}{96}-\frac{103 \zeta _3}{192}+\frac{3467}{1536}\\
  \mathcal{C}_{qg}[2,-3,0,2]=-\frac{323 \zeta _2}{192}-\frac{103 \zeta _3}{96}+\frac{6695}{1536}\\
  \mathcal{C}_{qg}[2,-3,0,3]=-\frac{523 \zeta _2}{576}-\frac{103 \zeta _3}{96}+\frac{32287}{13824}\\
  \mathcal{C}_{qg}[2,-3,0,4]=-\frac{197 \zeta _2}{384}-\frac{103 \zeta _3}{96}+\frac{46277}{24576}\\
  \mathcal{C}_{qg}[2,-3,0,5]=-\frac{157 \zeta _2}{640}-\frac{103 \zeta _3}{96}+\frac{131239907}{82944000}\\
  \mathcal{C}_{qg}[1,3,0,0]=-\frac{3755 \zeta _2}{1152}-\frac{283 \zeta _3}{72}-\frac{871 \zeta _4}{96}+\frac{1641013}{248832}\\
  \mathcal{C}_{qg}[1,3,0,1]=-\frac{60559 \zeta _2}{1152}-\frac{9739 \zeta _3}{288}-\frac{871 \zeta _4}{96}+\frac{45245365}{248832}\\
  \mathcal{C}_{qg}[1,3,0,2]=-\frac{7921 \zeta _2}{128}-\frac{30493 \zeta _3}{576}-\frac{871 \zeta _4}{48}+\frac{28811983}{124416}\\
  \mathcal{C}_{qg}[1,3,0,3]=-\frac{166477 \zeta _2}{3456}-\frac{2065 \zeta _3}{64}-\frac{871 \zeta _4}{48}+\frac{20750399}{124416}\\
  \mathcal{C}_{qg}[1,3,0,4]=-\frac{398975 \zeta _2}{13824}-\frac{7661 \zeta _3}{576}-\frac{871 \zeta _4}{48}+\frac{341816329}{3981312}\\
  \mathcal{C}_{qg}[1,3,0,5]=-\frac{6957287 \zeta _2}{345600}+\frac{1087 \zeta _3}{2880}-\frac{871 \zeta _4}{48}+\frac{143432872057}{2488320000}\\
  \mathcal{C}_{qg}[1,2,1,0]=\frac{155 \zeta _2}{288}+\frac{125 \zeta _3}{144}-\frac{157411}{62208}\\
  \mathcal{C}_{qg}[1,2,1,1]=\frac{361 \zeta _2}{288}+\frac{125 \zeta _3}{144}-\frac{592291}{62208}\\
  \mathcal{C}_{qg}[1,2,1,2]=\frac{205 \zeta _2}{96}+\frac{125 \zeta _3}{72}-\frac{1036031}{62208}\\
  \mathcal{C}_{qg}[1,2,1,3]=\frac{1073 \zeta _2}{864}+\frac{125 \zeta _3}{72}-\frac{687467}{62208}\\
  \mathcal{C}_{qg}[1,2,1,4]=\frac{551 \zeta _2}{1728}+\frac{125 \zeta _3}{72}-\frac{3241357}{497664}\\
  \mathcal{C}_{qg}[1,2,1,5]=-\frac{3101 \zeta _2}{8640}+\frac{125 \zeta _3}{72}-\frac{12023201}{2488320}\\
  \mathcal{C}_{qg}[1,1,2,0]=\frac{29}{432}\\
  \mathcal{C}_{qg}[1,1,2,1]=-\frac{11}{432}\\
  \mathcal{C}_{qg}[1,1,2,2]=\frac{59}{288}\\
  \mathcal{C}_{qg}[1,1,2,3]=\frac{61}{864}\\
  \mathcal{C}_{qg}[1,1,2,4]=\frac{47}{1728}\\
  \mathcal{C}_{qg}[1,1,2,5]=\frac{13}{1728}\\
  \mathcal{C}_{qg}[1,1,0,0]=\frac{20545 \zeta _2}{3456}+\frac{2297 \zeta _3}{288}+\frac{3787 \zeta _4}{384}-\frac{46859}{9216}\\
  \mathcal{C}_{qg}[1,1,0,1]=\frac{241543 \zeta _2}{3456}+\frac{11645 \zeta _3}{288}+\frac{3787 \zeta _4}{384}-\frac{2069755}{9216}\\
  \mathcal{C}_{qg}[1,1,0,2]=\frac{330233 \zeta _2}{3456}+\frac{20977 \zeta _3}{288}+\frac{3787 \zeta _4}{192}-\frac{489517}{1536}\\
  \mathcal{C}_{qg}[1,1,0,3]=\frac{241585 \zeta _2}{3456}+\frac{12959 \zeta _3}{288}+\frac{3787 \zeta _4}{192}-\frac{27336073}{124416}\\
  \mathcal{C}_{qg}[1,1,0,4]=\frac{202327 \zeta _2}{4608}+\frac{6185 \zeta _3}{288}+\frac{3787 \zeta _4}{192}-\frac{491187695}{3981312}\\
  \mathcal{C}_{qg}[1,1,0,5]=\frac{2218337 \zeta _2}{69120}+\frac{139 \zeta _3}{30}+\frac{3787 \zeta _4}{192}-\frac{24426241103}{276480000}\\
  \mathcal{C}_{qg}[1,0,1,0]=-\frac{473 \zeta _2}{432}-\frac{55 \zeta _3}{36}+\frac{9859}{3456}\\
  \mathcal{C}_{qg}[1,0,1,1]=-\frac{695 \zeta _2}{432}-\frac{55 \zeta _3}{36}+\frac{39623}{3456}\\
  \mathcal{C}_{qg}[1,0,1,2]=-\frac{689 \zeta _2}{216}-\frac{55 \zeta _3}{18}+\frac{5873}{288}\\
  \mathcal{C}_{qg}[1,0,1,3]=-\frac{137 \zeta _2}{72}-\frac{55 \zeta _3}{18}+\frac{209369}{15552}\\
  \mathcal{C}_{qg}[1,0,1,4]=-\frac{1015 \zeta _2}{1728}-\frac{55 \zeta _3}{18}+\frac{977155}{124416}\\
  \mathcal{C}_{qg}[1,0,1,5]=\frac{1103 \zeta _2}{2880}-\frac{55 \zeta _3}{18}+\frac{22420963}{3888000}\\
  \mathcal{C}_{qg}[1,-1,2,0]=-\frac{29}{432}\\
  \mathcal{C}_{qg}[1,-1,2,1]=\frac{11}{432}\\
  \mathcal{C}_{qg}[1,-1,2,2]=-\frac{59}{288}\\
  \mathcal{C}_{qg}[1,-1,2,3]=-\frac{61}{864}\\
  \mathcal{C}_{qg}[1,-1,2,4]=-\frac{47}{1728}\\
  \mathcal{C}_{qg}[1,-1,2,5]=-\frac{13}{1728}\\
  \mathcal{C}_{qg}[1,-1,0,0]=-\frac{7039 \zeta _2}{3456}-\frac{85 \zeta _3}{18}-\frac{53 \zeta _4}{96}-\frac{340909}{248832}\\
  \mathcal{C}_{qg}[1,-1,0,1]=-\frac{65383 \zeta _2}{3456}-\frac{2071 \zeta _3}{288}-\frac{53 \zeta _4}{96}+\frac{11607347}{248832}\\
  \mathcal{C}_{qg}[1,-1,0,2]=-\frac{130325 \zeta _2}{3456}-\frac{12637 \zeta _3}{576}-\frac{53 \zeta _4}{48}+\frac{11806439}{124416}\\
  \mathcal{C}_{qg}[1,-1,0,3]=-\frac{83435 \zeta _2}{3456}-\frac{7969 \zeta _3}{576}-\frac{53 \zeta _4}{48}+\frac{7098125}{124416}\\
  \mathcal{C}_{qg}[1,-1,0,4]=-\frac{235115 \zeta _2}{13824}-\frac{1267 \zeta _3}{144}-\frac{53 \zeta _4}{48}+\frac{163055503}{3981312}\\
  \mathcal{C}_{qg}[1,-1,0,5]=-\frac{4762843 \zeta _2}{345600}-\frac{191 \zeta _3}{36}-\frac{53 \zeta _4}{48}+\frac{27832701149}{829440000}\\
  \mathcal{C}_{qg}[1,-2,1,0]=\frac{481 \zeta _2}{864}+\frac{95 \zeta _3}{144}-\frac{20051}{62208}\\
  \mathcal{C}_{qg}[1,-2,1,1]=\frac{307 \zeta _2}{864}+\frac{95 \zeta _3}{144}-\frac{120923}{62208}\\
  \mathcal{C}_{qg}[1,-2,1,2]=\frac{911 \zeta _2}{864}+\frac{95 \zeta _3}{72}-\frac{232537}{62208}\\
  \mathcal{C}_{qg}[1,-2,1,3]=\frac{571 \zeta _2}{864}+\frac{95 \zeta _3}{72}-\frac{50003}{20736}\\
  \mathcal{C}_{qg}[1,-2,1,4]=\frac{29 \zeta _2}{108}+\frac{95 \zeta _3}{72}-\frac{222421}{165888}\\
  \mathcal{C}_{qg}[1,-2,1,5]=-\frac{13 \zeta _2}{540}+\frac{95 \zeta _3}{72}-\frac{58155383}{62208000}\\
  \mathcal{C}_{qg}[1,-3,0,0]=-\frac{83 \zeta _2}{128}+\frac{65 \zeta _3}{96}-\frac{91 \zeta _4}{384}-\frac{431}{3072}\\
  \mathcal{C}_{qg}[1,-3,0,1]=\frac{613 \zeta _2}{384}+\frac{55 \zeta _3}{96}-\frac{91 \zeta _4}{384}-\frac{3989}{1024}\\
  \mathcal{C}_{qg}[1,-3,0,2]=\frac{517 \zeta _2}{128}+\frac{49 \zeta _3}{24}-\frac{91 \zeta _4}{192}-\frac{11945}{1536}\\
  \mathcal{C}_{qg}[1,-3,0,3]=\frac{8327 \zeta _2}{3456}+\frac{53 \zeta _3}{48}-\frac{91 \zeta _4}{192}-\frac{56939}{13824}\\
  \mathcal{C}_{qg}[1,-3,0,4]=\frac{27109 \zeta _2}{13824}+\frac{359 \zeta _3}{576}-\frac{91 \zeta _4}{192}-\frac{4561379}{1327104}\\
  \mathcal{C}_{qg}[1,-3,0,5]=\frac{125689 \zeta _2}{69120}+\frac{283 \zeta _3}{960}-\frac{91 \zeta _4}{192}-\frac{7094805577}{2488320000}\\
  \mathcal{C}_{qg}[0,3,0,0]=-\frac{505}{48} \zeta _3 \zeta _2+\frac{3691 \zeta _2}{1296}+\frac{34117 \zeta _3}{3456}-\frac{649 \zeta _4}{2304}+\frac{1687 \zeta _5}{96}-\frac{1457441}{995328}\\
  \mathcal{C}_{qg}[0,3,0,1]=-\frac{505}{48} \zeta _3 \zeta _2+\frac{624575 \zeta _2}{10368}+\frac{54085 \zeta _3}{864}+\frac{17939 \zeta _4}{2304}+\frac{1687 \zeta _5}{96}-\frac{59565061}{331776}\\
  \mathcal{C}_{qg}[0,3,0,2]=-\frac{505}{24} \zeta _3 \zeta _2+\frac{168217 \zeta _2}{2592}+\frac{143947 \zeta _3}{1728}+\frac{2927 \zeta _4}{288}+\frac{1687 \zeta _5}{48}-\frac{118254245}{497664}\\
  \mathcal{C}_{qg}[0,3,0,3]=-\frac{505}{24} \zeta _3 \zeta _2+\frac{128537 \zeta _2}{2592}+\frac{76855 \zeta _3}{1152}+\frac{1285 \zeta _4}{288}+\frac{1687 \zeta _5}{48}-\frac{26723083}{165888}\\
  \mathcal{C}_{qg}[0,3,0,4]=-\frac{505}{24} \zeta _3 \zeta _2+\frac{1711207 \zeta _2}{82944}+\frac{167051 \zeta _3}{3456}-\frac{261 \zeta _4}{256}+\frac{1687 \zeta _5}{48}-\frac{1182915463}{15925248}\\
  \mathcal{C}_{qg}[0,3,0,5]=-\frac{505}{24} \zeta _3 \zeta _2+\frac{27596011 \zeta _2}{3456000}+\frac{7045877 \zeta _3}{172800}-\frac{57137 \zeta _4}{11520}+\frac{1687 \zeta _5}{48}-\frac{737446993121}{16588800000}\\
  \mathcal{C}_{qg}[0,2,1,0]=-\frac{139 \zeta _2}{324}-\frac{47 \zeta _3}{27}+\frac{193 \zeta _4}{576}+\frac{82171}{248832}\\
  \mathcal{C}_{qg}[0,2,1,1]=-\frac{7163 \zeta _2}{2592}-\frac{1585 \zeta _3}{864}+\frac{193 \zeta _4}{576}+\frac{2073437}{248832}\\
  \mathcal{C}_{qg}[0,2,1,2]=-\frac{529 \zeta _2}{162}-\frac{3401 \zeta _3}{864}+\frac{193 \zeta _4}{288}+\frac{1035035}{62208}\\
  \mathcal{C}_{qg}[0,2,1,3]=-\frac{3133 \zeta _2}{1296}-\frac{2801 \zeta _3}{864}+\frac{193 \zeta _4}{288}+\frac{795053}{93312}\\
  \mathcal{C}_{qg}[0,2,1,4]=-\frac{1151 \zeta _2}{1152}-\frac{2939 \zeta _3}{1152}+\frac{193 \zeta _4}{288}+\frac{18979949}{5971968}\\
  \mathcal{C}_{qg}[0,2,1,5]=-\frac{24941 \zeta _2}{86400}-\frac{35309 \zeta _3}{17280}+\frac{193 \zeta _4}{288}+\frac{4717058237}{3732480000}\\
  \mathcal{C}_{qg}[0,1,2,0]=-\frac{\zeta _3}{72}-\frac{125}{3888}\\
  \mathcal{C}_{qg}[0,1,2,1]=\frac{59}{1944}-\frac{\zeta _3}{72}\\
  \mathcal{C}_{qg}[0,1,2,2]=-\frac{\zeta _3}{36}-\frac{4573}{15552}\\
  \mathcal{C}_{qg}[0,1,2,3]=-\frac{\zeta _3}{36}-\frac{1051}{15552}\\
  \mathcal{C}_{qg}[0,1,2,4]=-\frac{\zeta _3}{36}-\frac{179}{7776}\\
  \mathcal{C}_{qg}[0,1,2,5]=-\frac{\zeta _3}{36}-\frac{139}{15552}\\
  \mathcal{C}_{qg}[0,1,0,0]=\frac{2807 \zeta _3 \zeta _2}{192}-\frac{5833 \zeta _2}{1296}-\frac{4001 \zeta _3}{432}-\frac{73 \zeta _4}{64}-\frac{1447 \zeta _5}{64}+\frac{53237}{995328}\\
  \mathcal{C}_{qg}[0,1,0,1]=\frac{2807 \zeta _3 \zeta _2}{192}-\frac{822677 \zeta _2}{10368}-\frac{259225 \zeta _3}{3456}-\frac{2081 \zeta _4}{256}-\frac{1447 \zeta _5}{64}+\frac{223748699}{995328}\\
  \mathcal{C}_{qg}[0,1,0,2]=\frac{2807 \zeta _3 \zeta _2}{96}-\frac{2146133 \zeta _2}{20736}-\frac{188917 \zeta _3}{1728}-\frac{5365 \zeta _4}{384}-\frac{1447 \zeta _5}{32}+\frac{162487697}{497664}\\
  \mathcal{C}_{qg}[0,1,0,3]=\frac{2807 \zeta _3 \zeta _2}{96}-\frac{1519481 \zeta _2}{20736}-\frac{2993 \zeta _3}{36}-\frac{371 \zeta _4}{48}-\frac{1447 \zeta _5}{32}+\frac{106728905}{497664}\\
  \mathcal{C}_{qg}[0,1,0,4]=\frac{2807 \zeta _3 \zeta _2}{96}-\frac{6084335 \zeta _2}{165888}-\frac{832939 \zeta _3}{13824}-\frac{21791 \zeta _4}{9216}-\frac{1447 \zeta _5}{32}+\frac{1784995513}{15925248}\\
  \mathcal{C}_{qg}[0,1,0,5]=\frac{2807 \zeta _3 \zeta _2}{96}-\frac{431450027 \zeta _2}{20736000}-\frac{17502863 \zeta _3}{345600}+\frac{1493 \zeta _4}{1024}-\frac{1447 \zeta _5}{32}+\frac{11203424776447}{149299200000}\\
  \mathcal{C}_{qg}[0,0,1,0]=\frac{229 \zeta _2}{324}+\frac{1723 \zeta _3}{864}-\frac{5 \zeta _4}{32}-\frac{17219}{124416}\\
  \mathcal{C}_{qg}[0,0,1,1]=\frac{8471 \zeta _2}{2592}+\frac{1789 \zeta _3}{864}-\frac{5 \zeta _4}{32}-\frac{1295837}{124416}\\
  \mathcal{C}_{qg}[0,0,1,2]=\frac{395 \zeta _2}{81}+\frac{4109 \zeta _3}{864}-\frac{5 \zeta _4}{16}-\frac{617845}{31104}\\
  \mathcal{C}_{qg}[0,0,1,3]=\frac{4427 \zeta _2}{1296}+\frac{3001 \zeta _3}{864}-\frac{5 \zeta _4}{16}-\frac{945307}{93312}\\
  \mathcal{C}_{qg}[0,0,1,4]=\frac{15587 \zeta _2}{10368}+\frac{2015 \zeta _3}{864}-\frac{5 \zeta _4}{16}-\frac{2590181}{746496}\\
  \mathcal{C}_{qg}[0,0,1,5]=\frac{153551 \zeta _2}{259200}+\frac{1291 \zeta _3}{864}-\frac{5 \zeta _4}{16}-\frac{386889707}{373248000}\\
  \mathcal{C}_{qg}[0,-1,2,0]=\frac{\zeta _3}{72}+\frac{125}{3888}\\
  \mathcal{C}_{qg}[0,-1,2,1]=\frac{\zeta _3}{72}-\frac{59}{1944}\\
  \mathcal{C}_{qg}[0,-1,2,2]=\frac{\zeta _3}{36}+\frac{4573}{15552}\\
  \mathcal{C}_{qg}[0,-1,2,3]=\frac{\zeta _3}{36}+\frac{1051}{15552}\\
  \mathcal{C}_{qg}[0,-1,2,4]=\frac{\zeta _3}{36}+\frac{179}{7776}\\
  \mathcal{C}_{qg}[0,-1,2,5]=\frac{\zeta _3}{36}+\frac{139}{15552}\\
  \mathcal{C}_{qg}[0,-1,0,0]=-\frac{55}{12} \zeta _3 \zeta _2+\frac{95 \zeta _2}{72}-\frac{463 \zeta _3}{1152}+\frac{2245 \zeta _4}{2304}+\frac{545 \zeta _5}{96}+\frac{422195}{331776}\\
  \mathcal{C}_{qg}[0,-1,0,1]=-\frac{55}{12} \zeta _3 \zeta _2+\frac{72847 \zeta _2}{3456}+\frac{1261 \zeta _3}{96}-\frac{35 \zeta _4}{2304}+\frac{545 \zeta _5}{96}-\frac{48336041}{995328}\\
  \mathcal{C}_{qg}[0,-1,0,2]=-\frac{55}{6} \zeta _3 \zeta _2+\frac{48605 \zeta _2}{1152}+\frac{517 \zeta _3}{18}+\frac{443 \zeta _4}{144}+\frac{545 \zeta _5}{48}-\frac{197233}{2048}\\
  \mathcal{C}_{qg}[0,-1,0,3]=-\frac{55}{6} \zeta _3 \zeta _2+\frac{266753 \zeta _2}{10368}+\frac{61867 \zeta _3}{3456}+\frac{799 \zeta _4}{288}+\frac{545 \zeta _5}{48}-\frac{85508497}{1492992}\\
  \mathcal{C}_{qg}[0,-1,0,4]=-\frac{55}{6} \zeta _3 \zeta _2+\frac{1464109 \zeta _2}{82944}+\frac{22661 \zeta _3}{1728}+\frac{6935 \zeta _4}{2304}+\frac{545 \zeta _5}{48}-\frac{1960413023}{47775744}\\
  \mathcal{C}_{qg}[0,-1,0,5]=-\frac{55}{6} \zeta _3 \zeta _2+\frac{29578963 \zeta _2}{2073600}+\frac{631069 \zeta _3}{57600}+\frac{36971 \zeta _4}{11520}+\frac{545 \zeta _5}{48}-\frac{4948133456027}{149299200000}\\
  \mathcal{C}_{qg}[0,-2,1,0]=-\frac{5 \zeta _2}{18}-\frac{73 \zeta _3}{288}-\frac{103 \zeta _4}{576}-\frac{15911}{82944}\\
  \mathcal{C}_{qg}[0,-2,1,1]=-\frac{109 \zeta _2}{216}-\frac{17 \zeta _3}{72}-\frac{103 \zeta _4}{576}+\frac{518237}{248832}\\
  \mathcal{C}_{qg}[0,-2,1,2]=-\frac{29 \zeta _2}{18}-\frac{59 \zeta _3}{72}-\frac{103 \zeta _4}{288}+\frac{22295}{6912}\\
  \mathcal{C}_{qg}[0,-2,1,3]=-\frac{647 \zeta _2}{648}-\frac{25 \zeta _3}{108}-\frac{103 \zeta _4}{288}+\frac{75127}{46656}\\
  \mathcal{C}_{qg}[0,-2,1,4]=-\frac{1307 \zeta _2}{2592}+\frac{757 \zeta _3}{3456}-\frac{103 \zeta _4}{288}+\frac{1741499}{5971968}\\
  \mathcal{C}_{qg}[0,-2,1,5]=-\frac{9841 \zeta _2}{32400}+\frac{3163 \zeta _3}{5760}-\frac{103 \zeta _4}{288}-\frac{282720389}{1244160000}\\
  \mathcal{C}_{qg}[0,-3,0,0]=\frac{31 \zeta _3 \zeta _2}{64}+\frac{\zeta _2}{3}-\frac{5 \zeta _3}{24}+\frac{43 \zeta _4}{96}-\frac{41 \zeta _5}{64}+\frac{1699}{12288}\\
  \mathcal{C}_{qg}[0,-3,0,1]=\frac{31 \zeta _3 \zeta _2}{64}-\frac{757 \zeta _2}{384}-\frac{93 \zeta _3}{128}+\frac{275 \zeta _4}{768}-\frac{41 \zeta _5}{64}+\frac{40525}{12288}\\
  \mathcal{C}_{qg}[0,-3,0,2]=\frac{31 \zeta _3 \zeta _2}{32}-\frac{2759 \zeta _2}{768}-\frac{259 \zeta _3}{96}+\frac{281 \zeta _4}{384}-\frac{41 \zeta _5}{32}+\frac{45607}{6144}\\
  \mathcal{C}_{qg}[0,-3,0,3]=\frac{31 \zeta _3 \zeta _2}{32}-\frac{14107 \zeta _2}{6912}-\frac{319 \zeta _3}{216}+\frac{71 \zeta _4}{144}-\frac{41 \zeta _5}{32}+\frac{5829529}{1492992}\\
  \mathcal{C}_{qg}[0,-3,0,4]=\frac{31 \zeta _3 \zeta _2}{32}-\frac{266297 \zeta _2}{165888}-\frac{16553 \zeta _3}{13824}+\frac{383 \zeta _4}{1024}-\frac{41 \zeta _5}{32}+\frac{154172873}{47775744}\\
  \mathcal{C}_{qg}[0,-3,0,5]=\frac{31 \zeta _3 \zeta _2}{32}-\frac{29915669 \zeta _2}{20736000}-\frac{75061 \zeta _3}{69120}+\frac{4493 \zeta _4}{15360}-\frac{41 \zeta _5}{32}+\frac{381731617669}{149299200000}
\end{gather*} }
\endgroup
%%%%%%%%%%%%%%%%%%%%%%%%%%%%%%%%%%%%%%%%%%%%%
%%%%%%%%%%%%%%%%%%%%%%%%%%%%%%%%%%%%%%%%%%%%%

\subsection{The $q\bar{q}$ channel}
\begingroup
\addtolength{\jot}{1em}
{\small }
\endgroup
%%%%%%%%%%%%%%%%%%%%%%%%%%%%%%%%%%%%%%%%%%%%%
%%%%%%%%%%%%%%%%%%%%%%%%%%%%%%%%%%%%%%%%%%%%%

\subsection{The $qq$ channel}
\begingroup
\addtolength{\jot}{1em}
{\small \begin{gather*}
\mathcal{C}_{\text{qq}}[2,3,0,1]=\frac{1985}{384}-\frac{43 \zeta _2}{32}\\
\mathcal{C}_{\text{qq}}[2,3,0,2]=\frac{2403}{256}-\frac{129 \zeta _2}{64}\\
\mathcal{C}_{\text{qq}}[2,3,0,3]=\frac{37955}{2304}-\frac{215 \zeta _2}{64}\\
\mathcal{C}_{\text{qq}}[2,3,0,4]=\frac{182495}{9216}-\frac{559 \zeta _2}{128}\\
\mathcal{C}_{\text{qq}}[2,3,0,5]=\frac{14956}{675}-\frac{10019 \zeta _2}{1920}\\
\mathcal{C}_{\text{qq}}[2,2,1,1]=-\frac{29}{96}\\
\mathcal{C}_{\text{qq}}[2,2,1,2]=-\frac{23}{64}\\
\mathcal{C}_{\text{qq}}[2,2,1,3]=-\frac{121}{192}\\
\mathcal{C}_{\text{qq}}[2,2,1,4]=-\frac{595}{768}\\
\mathcal{C}_{\text{qq}}[2,2,1,5]=-\frac{50621}{57600}\\
\mathcal{C}_{\text{qq}}[2,2,0,2]=-\frac{1}{64}\\
\mathcal{C}_{\text{qq}}[2,2,0,3]=-\frac{3}{64}\\
\mathcal{C}_{\text{qq}}[2,2,0,4]=-\frac{107}{1536}\\
\mathcal{C}_{\text{qq}}[2,2,0,5]=-\frac{671}{7680}\\
\mathcal{C}_{\text{qq}}[2,1,0,1]=\frac{115 \zeta _2}{32}-\frac{4525}{384}\\
\mathcal{C}_{\text{qq}}[2,1,0,2]=\frac{345 \zeta _2}{64}-\frac{5463}{256}\\
\mathcal{C}_{\text{qq}}[2,1,0,3]=\frac{575 \zeta _2}{64}-\frac{88427}{2304}\\
\mathcal{C}_{\text{qq}}[2,1,0,4]=\frac{1495 \zeta _2}{128}-\frac{425861}{9216}\\
\mathcal{C}_{\text{qq}}[2,1,0,5]=\frac{5359 \zeta _2}{384}-\frac{89556961}{1728000}\\
\mathcal{C}_{\text{qq}}[2,0,1,1]=\frac{29}{48}\\
\mathcal{C}_{\text{qq}}[2,0,1,2]=\frac{23}{32}\\
\mathcal{C}_{\text{qq}}[2,0,1,3]=\frac{121}{96}\\
\mathcal{C}_{\text{qq}}[2,0,1,4]=\frac{595}{384}\\
\mathcal{C}_{\text{qq}}[2,0,1,5]=\frac{50621}{28800}\\
\mathcal{C}_{\text{qq}}[2,0,0,2]=\frac{5}{64}\\
\mathcal{C}_{\text{qq}}[2,0,0,3]=\frac{49}{192}\\
\mathcal{C}_{\text{qq}}[2,0,0,4]=\frac{587}{1536}\\
\mathcal{C}_{\text{qq}}[2,0,0,5]=\frac{1233}{2560}\\
\mathcal{C}_{\text{qq}}[2,-1,0,1]=\frac{3095}{384}-\frac{101 \zeta _2}{32}\\
\mathcal{C}_{\text{qq}}[2,-1,0,2]=\frac{3717}{256}-\frac{303 \zeta _2}{64}\\
\mathcal{C}_{\text{qq}}[2,-1,0,3]=\frac{62989}{2304}-\frac{505 \zeta _2}{64}\\
\mathcal{C}_{\text{qq}}[2,-1,0,4]=\frac{304237}{9216}-\frac{1313 \zeta _2}{128}\\
\mathcal{C}_{\text{qq}}[2,-1,0,5]=\frac{32125921}{864000}-\frac{23533 \zeta _2}{1920}\\
\mathcal{C}_{\text{qq}}[2,-2,1,1]=-\frac{29}{96}\\
\mathcal{C}_{\text{qq}}[2,-2,1,2]=-\frac{23}{64}\\
\mathcal{C}_{\text{qq}}[2,-2,1,3]=-\frac{121}{192}\\
\mathcal{C}_{\text{qq}}[2,-2,1,4]=-\frac{595}{768}\\
\mathcal{C}_{\text{qq}}[2,-2,1,5]=-\frac{50621}{57600}\\
\mathcal{C}_{\text{qq}}[2,-2,0,2]=-\frac{7}{64}\\
\mathcal{C}_{\text{qq}}[2,-2,0,3]=-\frac{71}{192}\\
\mathcal{C}_{\text{qq}}[2,-2,0,4]=-\frac{853}{1536}\\
\mathcal{C}_{\text{qq}}[2,-2,0,5]=-\frac{359}{512}\\
\mathcal{C}_{\text{qq}}[2,-3,0,1]=\frac{29 \zeta _2}{32}-\frac{185}{128}\\
\mathcal{C}_{\text{qq}}[2,-3,0,2]=\frac{87 \zeta _2}{64}-\frac{657}{256}\\
\mathcal{C}_{\text{qq}}[2,-3,0,3]=\frac{145 \zeta _2}{64}-\frac{12517}{2304}\\
\mathcal{C}_{\text{qq}}[2,-3,0,4]=\frac{377 \zeta _2}{128}-\frac{60871}{9216}\\
\mathcal{C}_{\text{qq}}[2,-3,0,5]=\frac{6757 \zeta _2}{1920}-\frac{12982241}{1728000}\\
\mathcal{C}_{\text{qq}}[2,-4,0,2]=\frac{3}{64}\\
\mathcal{C}_{\text{qq}}[2,-4,0,3]=\frac{31}{192}\\
\mathcal{C}_{\text{qq}}[2,-4,0,4]=\frac{373}{1536}\\
\mathcal{C}_{\text{qq}}[2,-4,0,5]=\frac{2357}{7680}\\
\mathcal{C}_{\text{qq}}[1,3,0,1]=\frac{277 \zeta _2}{96}+\frac{27 \zeta _3}{8}-\frac{17327}{1728}\\
\mathcal{C}_{\text{qq}}[1,3,0,2]=\frac{149 \zeta _2}{32}+\frac{81 \zeta _3}{16}-\frac{44593}{2304}\\
\mathcal{C}_{\text{qq}}[1,3,0,3]=\frac{773 \zeta _2}{96}+\frac{135 \zeta _3}{16}-\frac{240269}{6912}\\
\mathcal{C}_{\text{qq}}[1,3,0,4]=\frac{7331 \zeta _2}{768}+\frac{351 \zeta _3}{32}-\frac{751631}{18432}\\
\mathcal{C}_{\text{qq}}[1,3,0,5]=\frac{66187 \zeta _2}{6400}+\frac{2097 \zeta _3}{160}-\frac{14463727}{324000}\\
\mathcal{C}_{\text{qq}}[1,2,1,1]=\frac{233}{432}-\frac{\zeta _2}{24}\\
\mathcal{C}_{\text{qq}}[1,2,1,2]=\frac{379}{576}-\frac{\zeta _2}{16}\\
\mathcal{C}_{\text{qq}}[1,2,1,3]=\frac{2359}{1728}-\frac{5 \zeta _2}{48}\\
\mathcal{C}_{\text{qq}}[1,2,1,4]=\frac{2447}{1536}-\frac{13 \zeta _2}{96}\\
\mathcal{C}_{\text{qq}}[1,2,1,5]=\frac{9034181}{5184000}-\frac{233 \zeta _2}{1440}\\
\mathcal{C}_{\text{qq}}[1,2,0,2]=\frac{31}{192}\\
\mathcal{C}_{\text{qq}}[1,2,0,3]=\frac{91}{192}\\
\mathcal{C}_{\text{qq}}[1,2,0,4]=\frac{2029}{3072}\\
\mathcal{C}_{\text{qq}}[1,2,0,5]=\frac{60109}{76800}\\
\mathcal{C}_{\text{qq}}[1,1,1,2]=-\frac{1}{48}\\
\mathcal{C}_{\text{qq}}[1,1,1,3]=-\frac{1}{16}\\
\mathcal{C}_{\text{qq}}[1,1,1,4]=-\frac{107}{1152}\\
\mathcal{C}_{\text{qq}}[1,1,1,5]=-\frac{671}{5760}\\
\mathcal{C}_{\text{qq}}[1,1,0,1]=-\frac{755 \zeta _2}{96}-\frac{125 \zeta _3}{16}+\frac{39757}{1728}\\
\mathcal{C}_{\text{qq}}[1,1,0,2]=-\frac{97 \zeta _2}{8}-\frac{375 \zeta _3}{32}+\frac{103001}{2304}\\
\mathcal{C}_{\text{qq}}[1,1,0,3]=-\frac{1027 \zeta _2}{48}-\frac{625 \zeta _3}{32}+\frac{188483}{2304}\\
\mathcal{C}_{\text{qq}}[1,1,0,4]=-\frac{19579 \zeta _2}{768}-\frac{1625 \zeta _3}{64}+\frac{5330431}{55296}\\
\mathcal{C}_{\text{qq}}[1,1,0,5]=-\frac{320311 \zeta _2}{11520}-\frac{5825 \zeta _3}{192}+\frac{1834676719}{17280000}\\
\mathcal{C}_{\text{qq}}[1,0,1,1]=\frac{\zeta _2}{12}-\frac{233}{216}\\
\mathcal{C}_{\text{qq}}[1,0,1,2]=\frac{\zeta _2}{8}-\frac{379}{288}\\
\mathcal{C}_{\text{qq}}[1,0,1,3]=\frac{5 \zeta _2}{24}-\frac{2359}{864}\\
\mathcal{C}_{\text{qq}}[1,0,1,4]=\frac{13 \zeta _2}{48}-\frac{2447}{768}\\
\mathcal{C}_{\text{qq}}[1,0,1,5]=\frac{233 \zeta _2}{720}-\frac{9034181}{2592000}\\
\mathcal{C}_{\text{qq}}[1,0,0,2]=-\frac{89}{192}\\
\mathcal{C}_{\text{qq}}[1,0,0,3]=-\frac{827}{576}\\
\mathcal{C}_{\text{qq}}[1,0,0,4]=-\frac{18019}{9216}\\
\mathcal{C}_{\text{qq}}[1,0,0,5]=-\frac{524983}{230400}\\
\mathcal{C}_{\text{qq}}[1,-1,1,2]=\frac{1}{24}\\
\mathcal{C}_{\text{qq}}[1,-1,1,3]=\frac{1}{8}\\
\mathcal{C}_{\text{qq}}[1,-1,1,4]=\frac{107}{576}\\
\mathcal{C}_{\text{qq}}[1,-1,1,5]=\frac{671}{2880}\\
\mathcal{C}_{\text{qq}}[1,-1,0,1]=\frac{679 \zeta _2}{96}+\frac{11 \zeta _3}{2}-\frac{27533}{1728}\\
\mathcal{C}_{\text{qq}}[1,-1,0,2]=\frac{329 \zeta _2}{32}+\frac{33 \zeta _3}{4}-\frac{72223}{2304}\\
\mathcal{C}_{\text{qq}}[1,-1,0,3]=\frac{1789 \zeta _2}{96}+\frac{55 \zeta _3}{4}-\frac{136697}{2304}\\
\mathcal{C}_{\text{qq}}[1,-1,0,4]=\frac{17165 \zeta _2}{768}+\frac{143 \zeta _3}{8}-\frac{3896183}{55296}\\
\mathcal{C}_{\text{qq}}[1,-1,0,5]=\frac{1416061 \zeta _2}{57600}+\frac{2563 \zeta _3}{120}-\frac{677578559}{8640000}\\
\mathcal{C}_{\text{qq}}[1,-2,1,1]=\frac{233}{432}-\frac{\zeta _2}{24}\\
\mathcal{C}_{\text{qq}}[1,-2,1,2]=\frac{379}{576}-\frac{\zeta _2}{16}\\
\mathcal{C}_{\text{qq}}[1,-2,1,3]=\frac{2359}{1728}-\frac{5 \zeta _2}{48}\\
\mathcal{C}_{\text{qq}}[1,-2,1,4]=\frac{2447}{1536}-\frac{13 \zeta _2}{96}\\
\mathcal{C}_{\text{qq}}[1,-2,1,5]=\frac{9034181}{5184000}-\frac{233 \zeta _2}{1440}\\
\mathcal{C}_{\text{qq}}[1,-2,0,2]=\frac{85}{192}\\
\mathcal{C}_{\text{qq}}[1,-2,0,3]=\frac{835}{576}\\
\mathcal{C}_{\text{qq}}[1,-2,0,4]=\frac{17777}{9216}\\
\mathcal{C}_{\text{qq}}[1,-2,0,5]=\frac{101797}{46080}\\
\mathcal{C}_{\text{qq}}[1,-3,1,2]=-\frac{1}{48}\\
\mathcal{C}_{\text{qq}}[1,-3,1,3]=-\frac{1}{16}\\
\mathcal{C}_{\text{qq}}[1,-3,1,4]=-\frac{107}{1152}\\
\mathcal{C}_{\text{qq}}[1,-3,1,5]=-\frac{671}{5760}\\
\mathcal{C}_{\text{qq}}[1,-3,0,1]=-\frac{67 \zeta _2}{32}-\frac{17 \zeta _3}{16}+\frac{189}{64}\\
\mathcal{C}_{\text{qq}}[1,-3,0,2]=-\frac{45 \zeta _2}{16}-\frac{51 \zeta _3}{32}+\frac{1535}{256}\\
\mathcal{C}_{\text{qq}}[1,-3,0,3]=-\frac{127 \zeta _2}{24}-\frac{85 \zeta _3}{32}+\frac{84911}{6912}\\
\mathcal{C}_{\text{qq}}[1,-3,0,4]=-\frac{1639 \zeta _2}{256}-\frac{221 \zeta _3}{64}+\frac{820645}{55296}\\
\mathcal{C}_{\text{qq}}[1,-3,0,5]=-\frac{410189 \zeta _2}{57600}-\frac{3961 \zeta _3}{960}+\frac{875637517}{51840000}\\
\mathcal{C}_{\text{qq}}[1,-4,0,2]=-\frac{9}{64}\\
\mathcal{C}_{\text{qq}}[1,-4,0,3]=-\frac{281}{576}\\
\mathcal{C}_{\text{qq}}[1,-4,0,4]=-\frac{5845}{9216}\\
\mathcal{C}_{\text{qq}}[1,-4,0,5]=-\frac{164329}{230400}\\
\mathcal{C}_{\text{qq}}[0,3,0,1]=-\frac{197 \zeta _2}{48}-\frac{131 \zeta _3}{48}-\frac{25 \zeta _4}{16}+\frac{33977}{3456}\\
\mathcal{C}_{\text{qq}}[0,3,0,2]=-\frac{453 \zeta _2}{64}-\frac{279 \zeta _3}{64}-\frac{75 \zeta _4}{32}+\frac{31855}{1536}\\
\mathcal{C}_{\text{qq}}[0,3,0,3]=-\frac{2365 \zeta _2}{192}-\frac{495 \zeta _3}{64}-\frac{125 \zeta _4}{32}+\frac{1560251}{41472}\\
\mathcal{C}_{\text{qq}}[0,3,0,4]=-\frac{134977 \zeta _2}{9216}-\frac{433 \zeta _3}{48}-\frac{325 \zeta _4}{64}+\frac{9788189}{221184}\\
\mathcal{C}_{\text{qq}}[0,3,0,5]=-\frac{18610607 \zeta _2}{1152000}-\frac{276937 \zeta _3}{28800}-\frac{1165 \zeta _4}{192}+\frac{3796028021}{77760000}\\
\mathcal{C}_{\text{qq}}[0,2,1,1]=\frac{7 \zeta _2}{24}+\frac{\zeta _3}{24}-\frac{55}{108}\\
\mathcal{C}_{\text{qq}}[0,2,1,2]=\frac{11 \zeta _2}{32}+\frac{\zeta _3}{16}-\frac{313}{384}\\
\mathcal{C}_{\text{qq}}[0,2,1,3]=\frac{173 \zeta _2}{288}+\frac{5 \zeta _3}{48}-\frac{5783}{3456}\\
\mathcal{C}_{\text{qq}}[0,2,1,4]=\frac{95 \zeta _2}{128}+\frac{13 \zeta _3}{96}-\frac{108079}{55296}\\
\mathcal{C}_{\text{qq}}[0,2,1,5]=\frac{73111 \zeta _2}{86400}+\frac{233 \zeta _3}{1440}-\frac{676881547}{311040000}\\
\mathcal{C}_{\text{qq}}[0,2,0,1]=-\frac{3 \zeta _3}{32}\\
\mathcal{C}_{\text{qq}}[0,2,0,2]=-\frac{25 \zeta _2}{128}-\frac{15 \zeta _3}{128}-\frac{493}{1152}\\
\mathcal{C}_{\text{qq}}[0,2,0,3]=-\frac{57 \zeta _2}{128}-\frac{49 \zeta _3}{384}-\frac{235}{192}\\
\mathcal{C}_{\text{qq}}[0,2,0,4]=-\frac{1931 \zeta _2}{3072}-\frac{205 \zeta _3}{1536}-\frac{56633}{36864}\\
\mathcal{C}_{\text{qq}}[0,2,0,5]=-\frac{3941 \zeta _2}{5120}-\frac{5309 \zeta _3}{38400}-\frac{22924469}{13824000}\\
\mathcal{C}_{\text{qq}}[0,1,1,2]=\frac{19}{288}\\
\mathcal{C}_{\text{qq}}[0,1,1,3]=\frac{53}{288}\\
\mathcal{C}_{\text{qq}}[0,1,1,4]=\frac{1049}{4608}\\
\mathcal{C}_{\text{qq}}[0,1,1,5]=\frac{3101}{12800}\\
\mathcal{C}_{\text{qq}}[0,1,0,1]=\frac{941 \zeta _2}{96}+\frac{647 \zeta _3}{96}+\frac{85 \zeta _4}{32}-\frac{78403}{3456}\\
\mathcal{C}_{\text{qq}}[0,1,0,2]=\frac{545 \zeta _2}{32}+\frac{167 \zeta _3}{16}+\frac{255 \zeta _4}{64}-\frac{72983}{1536}\\
\mathcal{C}_{\text{qq}}[0,1,0,3]=\frac{17729 \zeta _2}{576}+\frac{449 \zeta _3}{24}+\frac{425 \zeta _4}{64}-\frac{3636437}{41472}\\
\mathcal{C}_{\text{qq}}[0,1,0,4]=\frac{340037 \zeta _2}{9216}+\frac{5627 \zeta _3}{256}+\frac{1105 \zeta _4}{128}-\frac{68657525}{663552}\\
\mathcal{C}_{\text{qq}}[0,1,0,5]=\frac{28405097 \zeta _2}{691200}+\frac{18151 \zeta _3}{768}+\frac{3961 \zeta _4}{384}-\frac{29704217657}{259200000}\\
\mathcal{C}_{\text{qq}}[0,0,1,1]=-\frac{7 \zeta _2}{12}-\frac{\zeta _3}{12}+\frac{55}{54}\\
\mathcal{C}_{\text{qq}}[0,0,1,2]=-\frac{11 \zeta _2}{16}-\frac{\zeta _3}{8}+\frac{313}{192}\\
\mathcal{C}_{\text{qq}}[0,0,1,3]=-\frac{173 \zeta _2}{144}-\frac{5 \zeta _3}{24}+\frac{5783}{1728}\\
\mathcal{C}_{\text{qq}}[0,0,1,4]=-\frac{95 \zeta _2}{64}-\frac{13 \zeta _3}{48}+\frac{108079}{27648}\\
\mathcal{C}_{\text{qq}}[0,0,1,5]=-\frac{73111 \zeta _2}{43200}-\frac{233 \zeta _3}{720}+\frac{676881547}{155520000}\\
\mathcal{C}_{\text{qq}}[0,0,0,1]=\frac{3 \zeta _3}{32}\\
\mathcal{C}_{\text{qq}}[0,0,0,2]=\frac{37 \zeta _2}{128}+\frac{15 \zeta _3}{128}+\frac{1103}{1152}\\
\mathcal{C}_{\text{qq}}[0,0,0,3]=\frac{203 \zeta _2}{384}+\frac{49 \zeta _3}{384}+\frac{4985}{1728}\\
\mathcal{C}_{\text{qq}}[0,0,0,4]=\frac{6613 \zeta _2}{9216}+\frac{205 \zeta _3}{1536}+\frac{403535}{110592}\\
\mathcal{C}_{\text{qq}}[0,0,0,5]=\frac{13187 \zeta _2}{15360}+\frac{5309 \zeta _3}{38400}+\frac{55230241}{13824000}\\
\mathcal{C}_{\text{qq}}[0,-1,1,2]=-\frac{19}{144}\\
\mathcal{C}_{\text{qq}}[0,-1,1,3]=-\frac{53}{144}\\
\mathcal{C}_{\text{qq}}[0,-1,1,4]=-\frac{1049}{2304}\\
\mathcal{C}_{\text{qq}}[0,-1,1,5]=-\frac{3101}{6400}\\
\mathcal{C}_{\text{qq}}[0,-1,0,1]=-\frac{175 \zeta _2}{24}-\frac{127 \zeta _3}{24}-\frac{5 \zeta _4}{8}+\frac{54875}{3456}\\
\mathcal{C}_{\text{qq}}[0,-1,0,2]=-\frac{821 \zeta _2}{64}-\frac{499 \zeta _3}{64}-\frac{15 \zeta _4}{16}+\frac{50401}{1536}\\
\mathcal{C}_{\text{qq}}[0,-1,0,3]=-\frac{14173 \zeta _2}{576}-\frac{2729 \zeta _3}{192}-\frac{25 \zeta _4}{16}+\frac{2592121}{41472}\\
\mathcal{C}_{\text{qq}}[0,-1,0,4]=-\frac{275143 \zeta _2}{9216}-\frac{2163 \zeta _3}{128}-\frac{65 \zeta _4}{32}+\frac{49221349}{663552}\\
\mathcal{C}_{\text{qq}}[0,-1,0,5]=-\frac{116555507 \zeta _2}{3456000}-\frac{29473 \zeta _3}{1600}-\frac{233 \zeta _4}{96}+\frac{670254847}{8100000}\\
\mathcal{C}_{\text{qq}}[0,-2,1,1]=\frac{7 \zeta _2}{24}+\frac{\zeta _3}{24}-\frac{55}{108}\\
\mathcal{C}_{\text{qq}}[0,-2,1,2]=\frac{11 \zeta _2}{32}+\frac{\zeta _3}{16}-\frac{313}{384}\\
\mathcal{C}_{\text{qq}}[0,-2,1,3]=\frac{173 \zeta _2}{288}+\frac{5 \zeta _3}{48}-\frac{5783}{3456}\\
\mathcal{C}_{\text{qq}}[0,-2,1,4]=\frac{95 \zeta _2}{128}+\frac{13 \zeta _3}{96}-\frac{108079}{55296}\\
\mathcal{C}_{\text{qq}}[0,-2,1,5]=\frac{73111 \zeta _2}{86400}+\frac{233 \zeta _3}{1440}-\frac{676881547}{311040000}\\
\mathcal{C}_{\text{qq}}[0,-2,0,1]=\frac{3 \zeta _3}{32}\\
\mathcal{C}_{\text{qq}}[0,-2,0,2]=\frac{\zeta _2}{128}+\frac{15 \zeta _3}{128}-\frac{727}{1152}\\
\mathcal{C}_{\text{qq}}[0,-2,0,3]=\frac{107 \zeta _2}{384}+\frac{49 \zeta _3}{384}-\frac{3625}{1728}\\
\mathcal{C}_{\text{qq}}[0,-2,0,4]=\frac{4153 \zeta _2}{9216}+\frac{205 \zeta _3}{1536}-\frac{297373}{110592}\\
\mathcal{C}_{\text{qq}}[0,-2,0,5]=\frac{1819 \zeta _2}{3072}+\frac{5309 \zeta _3}{38400}-\frac{1667483}{552960}\\
\mathcal{C}_{\text{qq}}[0,-3,1,2]=\frac{19}{288}\\
\mathcal{C}_{\text{qq}}[0,-3,1,3]=\frac{53}{288}\\
\mathcal{C}_{\text{qq}}[0,-3,1,4]=\frac{1049}{4608}\\
\mathcal{C}_{\text{qq}}[0,-3,1,5]=\frac{3101}{12800}\\
\mathcal{C}_{\text{qq}}[0,-3,0,1]=\frac{51 \zeta _2}{32}+\frac{41 \zeta _3}{32}-\frac{15 \zeta _4}{32}-\frac{387}{128}\\
\mathcal{C}_{\text{qq}}[0,-3,0,2]=\frac{23 \zeta _2}{8}+\frac{55 \zeta _3}{32}-\frac{45 \zeta _4}{64}-\frac{3091}{512}\\
\mathcal{C}_{\text{qq}}[0,-3,0,3]=\frac{3539 \zeta _2}{576}+\frac{311 \zeta _3}{96}-\frac{75 \zeta _4}{64}-\frac{515935}{41472}\\
\mathcal{C}_{\text{qq}}[0,-3,0,4]=\frac{7787 \zeta _2}{1024}+\frac{3025 \zeta _3}{768}-\frac{195 \zeta _4}{128}-\frac{9928391}{663552}\\
\mathcal{C}_{\text{qq}}[0,-3,0,5]=\frac{30361843 \zeta _2}{3456000}+\frac{253577 \zeta _3}{57600}-\frac{233 \zeta _4}{128}-\frac{13192092551}{777600000}\\
\mathcal{C}_{\text{qq}}[0,-4,0,1]=-\frac{3 \zeta _3}{32}\\
\mathcal{C}_{\text{qq}}[0,-4,0,2]=-\frac{13 \zeta _2}{128}-\frac{15 \zeta _3}{128}+\frac{13}{128}\\
\mathcal{C}_{\text{qq}}[0,-4,0,3]=-\frac{139 \zeta _2}{384}-\frac{49 \zeta _3}{384}+\frac{755}{1728}\\
\mathcal{C}_{\text{qq}}[0,-4,0,4]=-\frac{4973 \zeta _2}{9216}-\frac{205 \zeta _3}{1536}+\frac{63737}{110592}\\
\mathcal{C}_{\text{qq}}[0,-4,0,5]=-\frac{10459 \zeta _2}{15360}-\frac{5309 \zeta _3}{38400}+\frac{1042367}{1536000}
\end{gather*} }
\endgroup

%%%%%%%%%%%%%%%%%%%%%%%%%%%%%%%%%%%%%%%%%%%%%
%%%%%%%%%%%%%%%%%%%%%%%%%%%%%%%%%%%%%%%%%%%%%

\subsection{The $qq^\prime$ channel}
\begingroup
\addtolength{\jot}{1em}
{\small \begin{gather*}
  \mathcal{C}_{qq'}[2,3,0,1]=\frac{1985}{384}-\frac{43 \zeta _2}{32}\\
  \mathcal{C}_{qq'}[2,3,0,2]=\frac{2403}{256}-\frac{129 \zeta _2}{64}\\
  \mathcal{C}_{qq'}[2,3,0,3]=\frac{37955}{2304}-\frac{215 \zeta _2}{64}\\
  \mathcal{C}_{qq'}[2,3,0,4]=\frac{182495}{9216}-\frac{559 \zeta _2}{128}\\
  \mathcal{C}_{qq'}[2,3,0,5]=\frac{14956}{675}-\frac{10019 \zeta _2}{1920}\\
  \mathcal{C}_{qq'}[2,2,1,1]=-\frac{29}{96}\\
  \mathcal{C}_{qq'}[2,2,1,2]=-\frac{23}{64}\\
  \mathcal{C}_{qq'}[2,2,1,3]=-\frac{121}{192}\\
  \mathcal{C}_{qq'}[2,2,1,4]=-\frac{595}{768}\\
  \mathcal{C}_{qq'}[2,2,1,5]=-\frac{50621}{57600}\\
  \mathcal{C}_{qq'}[2,1,0,1]=\frac{115 \zeta _2}{32}-\frac{4525}{384}\\
  \mathcal{C}_{qq'}[2,1,0,2]=\frac{345 \zeta _2}{64}-\frac{5463}{256}\\
  \mathcal{C}_{qq'}[2,1,0,3]=\frac{575 \zeta _2}{64}-\frac{88427}{2304}\\
  \mathcal{C}_{qq'}[2,1,0,4]=\frac{1495 \zeta _2}{128}-\frac{425861}{9216}\\
  \mathcal{C}_{qq'}[2,1,0,5]=\frac{5359 \zeta _2}{384}-\frac{89556961}{1728000}\\
  \mathcal{C}_{qq'}[2,0,1,1]=\frac{29}{48}\\
  \mathcal{C}_{qq'}[2,0,1,2]=\frac{23}{32}\\
  \mathcal{C}_{qq'}[2,0,1,3]=\frac{121}{96}\\
  \mathcal{C}_{qq'}[2,0,1,4]=\frac{595}{384}\\
  \mathcal{C}_{qq'}[2,0,1,5]=\frac{50621}{28800}\\
  \mathcal{C}_{qq'}[2,-1,0,1]=\frac{3095}{384}-\frac{101 \zeta _2}{32}\\
  \mathcal{C}_{qq'}[2,-1,0,2]=\frac{3717}{256}-\frac{303 \zeta _2}{64}\\
  \mathcal{C}_{qq'}[2,-1,0,3]=\frac{62989}{2304}-\frac{505 \zeta _2}{64}\\
  \mathcal{C}_{qq'}[2,-1,0,4]=\frac{304237}{9216}-\frac{1313 \zeta _2}{128}\\
  \mathcal{C}_{qq'}[2,-1,0,5]=\frac{32125921}{864000}-\frac{23533 \zeta _2}{1920}\\
  \mathcal{C}_{qq'}[2,-2,1,1]=-\frac{29}{96}\\
  \mathcal{C}_{qq'}[2,-2,1,2]=-\frac{23}{64}\\
  \mathcal{C}_{qq'}[2,-2,1,3]=-\frac{121}{192}\\
  \mathcal{C}_{qq'}[2,-2,1,4]=-\frac{595}{768}\\
  \mathcal{C}_{qq'}[2,-2,1,5]=-\frac{50621}{57600}\\
  \mathcal{C}_{qq'}[2,-3,0,1]=\frac{29 \zeta _2}{32}-\frac{185}{128}\\
  \mathcal{C}_{qq'}[2,-3,0,2]=\frac{87 \zeta _2}{64}-\frac{657}{256}\\
  \mathcal{C}_{qq'}[2,-3,0,3]=\frac{145 \zeta _2}{64}-\frac{12517}{2304}\\
  \mathcal{C}_{qq'}[2,-3,0,4]=\frac{377 \zeta _2}{128}-\frac{60871}{9216}\\
  \mathcal{C}_{qq'}[2,-3,0,5]=\frac{6757 \zeta _2}{1920}-\frac{12982241}{1728000}\\
  \mathcal{C}_{qq'}[1,3,0,1]=\frac{277 \zeta _2}{96}+\frac{27 \zeta _3}{8}-\frac{17327}{1728}\\
  \mathcal{C}_{qq'}[1,3,0,2]=\frac{149 \zeta _2}{32}+\frac{81 \zeta _3}{16}-\frac{44593}{2304}\\
  \mathcal{C}_{qq'}[1,3,0,3]=\frac{773 \zeta _2}{96}+\frac{135 \zeta _3}{16}-\frac{240269}{6912}\\
  \mathcal{C}_{qq'}[1,3,0,4]=\frac{7331 \zeta _2}{768}+\frac{351 \zeta _3}{32}-\frac{751631}{18432}\\
  \mathcal{C}_{qq'}[1,3,0,5]=\frac{66187 \zeta _2}{6400}+\frac{2097 \zeta _3}{160}-\frac{14463727}{324000}\\
  \mathcal{C}_{qq'}[1,2,1,1]=\frac{233}{432}-\frac{\zeta _2}{24}\\
  \mathcal{C}_{qq'}[1,2,1,2]=\frac{379}{576}-\frac{\zeta _2}{16}\\
  \mathcal{C}_{qq'}[1,2,1,3]=\frac{2359}{1728}-\frac{5 \zeta _2}{48}\\
  \mathcal{C}_{qq'}[1,2,1,4]=\frac{2447}{1536}-\frac{13 \zeta _2}{96}\\
  \mathcal{C}_{qq'}[1,2,1,5]=\frac{9034181}{5184000}-\frac{233 \zeta _2}{1440}\\
  \mathcal{C}_{qq'}[1,1,0,1]=-\frac{755 \zeta _2}{96}-\frac{125 \zeta _3}{16}+\frac{39757}{1728}\\
  \mathcal{C}_{qq'}[1,1,0,2]=-\frac{97 \zeta _2}{8}-\frac{375 \zeta _3}{32}+\frac{103001}{2304}\\
  \mathcal{C}_{qq'}[1,1,0,3]=-\frac{1027 \zeta _2}{48}-\frac{625 \zeta _3}{32}+\frac{188483}{2304}\\
  \mathcal{C}_{qq'}[1,1,0,4]=-\frac{19579 \zeta _2}{768}-\frac{1625 \zeta _3}{64}+\frac{5330431}{55296}\\
  \mathcal{C}_{qq'}[1,1,0,5]=-\frac{320311 \zeta _2}{11520}-\frac{5825 \zeta _3}{192}+\frac{1834676719}{17280000}\\
  \mathcal{C}_{qq'}[1,0,1,1]=\frac{\zeta _2}{12}-\frac{233}{216}\\
  \mathcal{C}_{qq'}[1,0,1,2]=\frac{\zeta _2}{8}-\frac{379}{288}\\
  \mathcal{C}_{qq'}[1,0,1,3]=\frac{5 \zeta _2}{24}-\frac{2359}{864}\\
  \mathcal{C}_{qq'}[1,0,1,4]=\frac{13 \zeta _2}{48}-\frac{2447}{768}\\
  \mathcal{C}_{qq'}[1,0,1,5]=\frac{233 \zeta _2}{720}-\frac{9034181}{2592000}\\
  \mathcal{C}_{qq'}[1,-1,0,1]=\frac{679 \zeta _2}{96}+\frac{11 \zeta _3}{2}-\frac{27533}{1728}\\
  \mathcal{C}_{qq'}[1,-1,0,2]=\frac{329 \zeta _2}{32}+\frac{33 \zeta _3}{4}-\frac{72223}{2304}\\
  \mathcal{C}_{qq'}[1,-1,0,3]=\frac{1789 \zeta _2}{96}+\frac{55 \zeta _3}{4}-\frac{136697}{2304}\\
  \mathcal{C}_{qq'}[1,-1,0,4]=\frac{17165 \zeta _2}{768}+\frac{143 \zeta _3}{8}-\frac{3896183}{55296}\\
  \mathcal{C}_{qq'}[1,-1,0,5]=\frac{1416061 \zeta _2}{57600}+\frac{2563 \zeta _3}{120}-\frac{677578559}{8640000}\\
  \mathcal{C}_{qq'}[1,-2,1,1]=\frac{233}{432}-\frac{\zeta _2}{24}\\
  \mathcal{C}_{qq'}[1,-2,1,2]=\frac{379}{576}-\frac{\zeta _2}{16}\\
  \mathcal{C}_{qq'}[1,-2,1,3]=\frac{2359}{1728}-\frac{5 \zeta _2}{48}\\
  \mathcal{C}_{qq'}[1,-2,1,4]=\frac{2447}{1536}-\frac{13 \zeta _2}{96}\\
  \mathcal{C}_{qq'}[1,-2,1,5]=\frac{9034181}{5184000}-\frac{233 \zeta _2}{1440}\\
  \mathcal{C}_{qq'}[1,-3,0,1]=-\frac{67 \zeta _2}{32}-\frac{17 \zeta _3}{16}+\frac{189}{64}\\
  \mathcal{C}_{qq'}[1,-3,0,2]=-\frac{45 \zeta _2}{16}-\frac{51 \zeta _3}{32}+\frac{1535}{256}\\
  \mathcal{C}_{qq'}[1,-3,0,3]=-\frac{127 \zeta _2}{24}-\frac{85 \zeta _3}{32}+\frac{84911}{6912}\\
  \mathcal{C}_{qq'}[1,-3,0,4]=-\frac{1639 \zeta _2}{256}-\frac{221 \zeta _3}{64}+\frac{820645}{55296}\\
  \mathcal{C}_{qq'}[1,-3,0,5]=-\frac{410189 \zeta _2}{57600}-\frac{3961 \zeta _3}{960}+\frac{875637517}{51840000}\\
  \mathcal{C}_{qq'}[0,3,0,1]=-\frac{197 \zeta _2}{48}-\frac{131 \zeta _3}{48}-\frac{25 \zeta _4}{16}+\frac{33977}{3456}\\
  \mathcal{C}_{qq'}[0,3,0,2]=-\frac{453 \zeta _2}{64}-\frac{279 \zeta _3}{64}-\frac{75 \zeta _4}{32}+\frac{31855}{1536}\\
  \mathcal{C}_{qq'}[0,3,0,3]=-\frac{2365 \zeta _2}{192}-\frac{495 \zeta _3}{64}-\frac{125 \zeta _4}{32}+\frac{1560251}{41472}\\
  \mathcal{C}_{qq'}[0,3,0,4]=-\frac{134977 \zeta _2}{9216}-\frac{433 \zeta _3}{48}-\frac{325 \zeta _4}{64}+\frac{9788189}{221184}\\
  \mathcal{C}_{qq'}[0,3,0,5]=-\frac{18610607 \zeta _2}{1152000}-\frac{276937 \zeta _3}{28800}-\frac{1165 \zeta _4}{192}+\frac{3796028021}{77760000}\\
  \mathcal{C}_{qq'}[0,2,1,1]=\frac{7 \zeta _2}{24}+\frac{\zeta _3}{24}-\frac{55}{108}\\
  \mathcal{C}_{qq'}[0,2,1,2]=\frac{11 \zeta _2}{32}+\frac{\zeta _3}{16}-\frac{313}{384}\\
  \mathcal{C}_{qq'}[0,2,1,3]=\frac{173 \zeta _2}{288}+\frac{5 \zeta _3}{48}-\frac{5783}{3456}\\
  \mathcal{C}_{qq'}[0,2,1,4]=\frac{95 \zeta _2}{128}+\frac{13 \zeta _3}{96}-\frac{108079}{55296}\\
  \mathcal{C}_{qq'}[0,2,1,5]=\frac{73111 \zeta _2}{86400}+\frac{233 \zeta _3}{1440}-\frac{676881547}{311040000}\\
  \mathcal{C}_{qq'}[0,1,0,1]=\frac{941 \zeta _2}{96}+\frac{647 \zeta _3}{96}+\frac{85 \zeta _4}{32}-\frac{78403}{3456}\\
  \mathcal{C}_{qq'}[0,1,0,2]=\frac{545 \zeta _2}{32}+\frac{167 \zeta _3}{16}+\frac{255 \zeta _4}{64}-\frac{72983}{1536}\\
  \mathcal{C}_{qq'}[0,1,0,3]=\frac{17729 \zeta _2}{576}+\frac{449 \zeta _3}{24}+\frac{425 \zeta _4}{64}-\frac{3636437}{41472}\\
  \mathcal{C}_{qq'}[0,1,0,4]=\frac{340037 \zeta _2}{9216}+\frac{5627 \zeta _3}{256}+\frac{1105 \zeta _4}{128}-\frac{68657525}{663552}\\
  \mathcal{C}_{qq'}[0,1,0,5]=\frac{28405097 \zeta _2}{691200}+\frac{18151 \zeta _3}{768}+\frac{3961 \zeta _4}{384}-\frac{29704217657}{259200000}\\
  \mathcal{C}_{qq'}[0,0,1,1]=-\frac{7 \zeta _2}{12}-\frac{\zeta _3}{12}+\frac{55}{54}\\
  \mathcal{C}_{qq'}[0,0,1,2]=-\frac{11 \zeta _2}{16}-\frac{\zeta _3}{8}+\frac{313}{192}\\
  \mathcal{C}_{qq'}[0,0,1,3]=-\frac{173 \zeta _2}{144}-\frac{5 \zeta _3}{24}+\frac{5783}{1728}\\
  \mathcal{C}_{qq'}[0,0,1,4]=-\frac{95 \zeta _2}{64}-\frac{13 \zeta _3}{48}+\frac{108079}{27648}\\
  \mathcal{C}_{qq'}[0,0,1,5]=-\frac{73111 \zeta _2}{43200}-\frac{233 \zeta _3}{720}+\frac{676881547}{155520000}\\
  \mathcal{C}_{qq'}[0,-1,0,1]=-\frac{175 \zeta _2}{24}-\frac{127 \zeta _3}{24}-\frac{5 \zeta _4}{8}+\frac{54875}{3456}\\
  \mathcal{C}_{qq'}[0,-1,0,2]=-\frac{821 \zeta _2}{64}-\frac{499 \zeta _3}{64}-\frac{15 \zeta _4}{16}+\frac{50401}{1536}\\
  \mathcal{C}_{qq'}[0,-1,0,3]=-\frac{14173 \zeta _2}{576}-\frac{2729 \zeta _3}{192}-\frac{25 \zeta _4}{16}+\frac{2592121}{41472}\\
  \mathcal{C}_{qq'}[0,-1,0,4]=-\frac{275143 \zeta _2}{9216}-\frac{2163 \zeta _3}{128}-\frac{65 \zeta _4}{32}+\frac{49221349}{663552}\\
  \mathcal{C}_{qq'}[0,-1,0,5]=-\frac{116555507 \zeta _2}{3456000}-\frac{29473 \zeta _3}{1600}-\frac{233 \zeta _4}{96}+\frac{670254847}{8100000}\\
  \mathcal{C}_{qq'}[0,-2,1,1]=\frac{7 \zeta _2}{24}+\frac{\zeta _3}{24}-\frac{55}{108}\\
  \mathcal{C}_{qq'}[0,-2,1,2]=\frac{11 \zeta _2}{32}+\frac{\zeta _3}{16}-\frac{313}{384}\\
  \mathcal{C}_{qq'}[0,-2,1,3]=\frac{173 \zeta _2}{288}+\frac{5 \zeta _3}{48}-\frac{5783}{3456}\\
  \mathcal{C}_{qq'}[0,-2,1,4]=\frac{95 \zeta _2}{128}+\frac{13 \zeta _3}{96}-\frac{108079}{55296}\\
  \mathcal{C}_{qq'}[0,-2,1,5]=\frac{73111 \zeta _2}{86400}+\frac{233 \zeta _3}{1440}-\frac{676881547}{311040000}\\
  \mathcal{C}_{qq'}[0,-3,0,1]=\frac{51 \zeta _2}{32}+\frac{41 \zeta _3}{32}-\frac{15 \zeta _4}{32}-\frac{387}{128}\\
  \mathcal{C}_{qq'}[0,-3,0,2]=\frac{23 \zeta _2}{8}+\frac{55 \zeta _3}{32}-\frac{45 \zeta _4}{64}-\frac{3091}{512}\\
  \mathcal{C}_{qq'}[0,-3,0,3]=\frac{3539 \zeta _2}{576}+\frac{311 \zeta _3}{96}-\frac{75 \zeta _4}{64}-\frac{515935}{41472}\\
  \mathcal{C}_{qq'}[0,-3,0,4]=\frac{7787 \zeta _2}{1024}+\frac{3025 \zeta _3}{768}-\frac{195 \zeta _4}{128}-\frac{9928391}{663552}\\
  \mathcal{C}_{qq'}[0,-3,0,5]=\frac{30361843 \zeta _2}{3456000}+\frac{253577 \zeta _3}{57600}-\frac{233 \zeta _4}{128}-\frac{13192092551}{777600000}
\end{gather*} }

\newpage
\section{Cross sections in the HXSWG recommended mass range}
\label{app:predictions}
We present here the gluon-fusion Higgs production cross-section at a proton-proton
collider for center-of-mass energies of 2, 7, 8, 13 and 14 TeV and for a Higgs boson of 
mass from 120 GeV to 130 GeV. The choice of these parameters follows 
the indications of the Higgs Cross Section Working Group~\cite{Denner:2047636}.

The components that enter (linearly) the theory uncertainty 
%, $\delta$(scale), $\delta$(trunc), $\delta$(PDF-TH), $\delta$(EW) and $\delta$($1/m_t$), 
have been discussed in the text.
To summarize the main points of that discussion, the scale variation uncertainty is assessed 
at each energy and Higgs mass through a scan over $\mu \in \left[{m_H}/{4},m_H\right]$. 
The uncertainties due to truncation, unknown N$^3$LO PDFs and unknown finite-mass effects 
are also evaluated every time, following the procedure described in the text. 
For the missing QCD-EW effects, we find that a reasonable 
estimate yields a 1\% uncertainty. At 2 TeV, however, we adopt \mbox{$\delta$(EW) = 0.8\%}, which 
is the most conservative estimate we obtain over the mass range analyzed.
Finally, we assign a 1\% uncertainty to missing finite-top mass effects at
NNLO~\cite{Pak:2009dg, Harlander:2009my}. 
We use the PDF set PDF4LHC15.

\begin{table}[!h]
\normalsize\setlength{\tabcolsep}{2pt}
\begin{center}
\begin{tabular}{rcrcrcrcr}
\multicolumn{9}{c}{$\sqrt{s}$ = 2 TeV}\\
\toprule
\multicolumn{1}{c}{$m_{H}$}& \phantom{a}&
\multicolumn{1}{c}{$\sigma$}& \phantom{a}&
\multicolumn{1}{c}{$\delta(\mbox{theory})$}& \phantom{a}&
\multicolumn{1}{c}{$\delta(\textrm{PDF})$}& \phantom{a}&
\multicolumn{1}{c}{$\delta(\alpha_s)$} 
\\ \midrule
120.0&&1.24 pb && ${}^{+0.05\textrm{pb}}_{-0.10\textrm{pb}}$(${}^{+4.18\%}_{-8.06\%}$)&& $\pm$ 0.04 pb ($\pm$ 3.06\%) && ${}^{+0.04\textrm{pb}}_{-0.05\textrm{pb}}$(${}^{+3.34\%}_{-3.65\%}$)  \\ \midrule 120.5&&1.23 pb && ${}^{+0.05\textrm{pb}}_{-0.10\textrm{pb}}$(${}^{+4.17\%}_{-8.04\%}$)&& $\pm$ 0.04 pb ($\pm$ 3.07\%) && ${}^{+0.04\textrm{pb}}_{-0.04\textrm{pb}}$(${}^{+3.34\%}_{-3.65\%}$)  \\ \midrule 121.0&&1.21 pb && ${}^{+0.05\textrm{pb}}_{-0.10\textrm{pb}}$(${}^{+4.16\%}_{-8.02\%}$)&& $\pm$ 0.04 pb ($\pm$ 3.08\%) && ${}^{+0.04\textrm{pb}}_{-0.04\textrm{pb}}$(${}^{+3.35\%}_{-3.66\%}$)  \\ \midrule 121.5&&1.20 pb && ${}^{+0.05\textrm{pb}}_{-0.10\textrm{pb}}$(${}^{+4.15\%}_{-8.\%}$)&& $\pm$ 0.04 pb ($\pm$ 3.09\%) && ${}^{+0.04\textrm{pb}}_{-0.04\textrm{pb}}$(${}^{+3.35\%}_{-3.66\%}$)  \\ \midrule 122.0&&1.18 pb && ${}^{+0.05\textrm{pb}}_{-0.09\textrm{pb}}$(${}^{+4.13\%}_{-7.99\%}$)&& $\pm$ 0.04 pb ($\pm$ 3.1\%) && ${}^{+0.04\textrm{pb}}_{-0.04\textrm{pb}}$(${}^{+3.35\%}_{-3.66\%}$)  \\ \midrule 122.5&&1.17 pb && ${}^{+0.05\textrm{pb}}_{-0.09\textrm{pb}}$(${}^{+4.12\%}_{-7.97\%}$)&& $\pm$ 0.04 pb ($\pm$ 3.12\%) && ${}^{+0.04\textrm{pb}}_{-0.04\textrm{pb}}$(${}^{+3.35\%}_{-3.67\%}$)  \\
\end{tabular}
\end{center}
\end{table} 
\vspace*{4cm}
\begin{table}
\begin{center} 
\begin{tabular}{rcrcrcrcr}
\toprule
\multicolumn{1}{c}{$m_{H}$}& \phantom{a}&
\multicolumn{1}{c}{$\sigma$}& \phantom{a}&
\multicolumn{1}{c}{$\delta(\mbox{theory})$}& \phantom{a}&
\multicolumn{1}{c}{$\delta(\textrm{PDF})$}& \phantom{a}&
\multicolumn{1}{c}{$\delta(\alpha_s)$} 
\\ \midrule125.1&&1.10 pb && ${}^{+0.04\textrm{pb}}_{-0.09\textrm{pb}}$(${}^{+4.05\%}_{-7.88\%}$)&& $\pm$ 0.03 pb ($\pm$ 3.18\%) && ${}^{+0.04\textrm{pb}}_{-0.04\textrm{pb}}$(${}^{+3.36\%}_{-3.69\%}$)  \\ \midrule 125.2&&1.09 pb && ${}^{+0.04\textrm{pb}}_{-0.09\textrm{pb}}$(${}^{+4.05\%}_{-7.88\%}$)&& $\pm$ 0.03 pb ($\pm$ 3.18\%) && ${}^{+0.04\textrm{pb}}_{-0.04\textrm{pb}}$(${}^{+3.36\%}_{-3.69\%}$)  \\ \midrule 125.3&&1.09 pb && ${}^{+0.04\textrm{pb}}_{-0.09\textrm{pb}}$(${}^{+4.05\%}_{-7.87\%}$)&& $\pm$ 0.03 pb ($\pm$ 3.18\%) && ${}^{+0.04\textrm{pb}}_{-0.04\textrm{pb}}$(${}^{+3.36\%}_{-3.69\%}$)  \\ \midrule 125.4&&1.09 pb && ${}^{+0.04\textrm{pb}}_{-0.09\textrm{pb}}$(${}^{+4.05\%}_{-7.87\%}$)&& $\pm$ 0.03 pb ($\pm$ 3.18\%) && ${}^{+0.04\textrm{pb}}_{-0.04\textrm{pb}}$(${}^{+3.36\%}_{-3.69\%}$)  \\ \midrule 125.5&&1.09 pb && ${}^{+0.04\textrm{pb}}_{-0.09\textrm{pb}}$(${}^{+4.04\%}_{-7.87\%}$)&& $\pm$ 0.03 pb ($\pm$ 3.18\%) && ${}^{+0.04\textrm{pb}}_{-0.04\textrm{pb}}$(${}^{+3.36\%}_{-3.69\%}$)  \\ \midrule 125.6&&1.08 pb && ${}^{+0.04\textrm{pb}}_{-0.09\textrm{pb}}$(${}^{+4.04\%}_{-7.86\%}$)&& $\pm$ 0.03 pb ($\pm$ 3.19\%) && ${}^{+0.04\textrm{pb}}_{-0.04\textrm{pb}}$(${}^{+3.36\%}_{-3.69\%}$)  \\ \midrule 125.7&&1.08 pb && ${}^{+0.04\textrm{pb}}_{-0.08\textrm{pb}}$(${}^{+4.04\%}_{-7.86\%}$)&& $\pm$ 0.03 pb ($\pm$ 3.19\%) && ${}^{+0.04\textrm{pb}}_{-0.04\textrm{pb}}$(${}^{+3.36\%}_{-3.69\%}$)  \\
\end{tabular}
\end{center}
\caption{Gluon-fusion Higgs production cross-section at a proton-proton collider for $\sqrt{s}$ = 2 TeV. 
%The theory error includes scale variation, truncation, missing-N$^3$LO PDFs, electroweak, missing light-quark effects and missing NNLO finite-top mass uncertainties. We use the PDF4LHC set. 
Details on the calculation of the theory error are given at the beginning of this Appendix and in the main text.
}
\label{tab:HXSWG2TeV}
\end{table}
\vspace*{4cm}\newpage

\begin{table}[!h]
\normalsize\setlength{\tabcolsep}{2pt}
\begin{center}
\begin{tabular}{rcrcrcrcr}
\multicolumn{9}{c}{$\sqrt{s}$ = 7 TeV}\\
\toprule
\multicolumn{1}{c}{$m_{H}$}& \phantom{a}&
\multicolumn{1}{c}{$\sigma$}& \phantom{a}&
\multicolumn{1}{c}{$\delta(\mbox{theory})$}& \phantom{a}&
\multicolumn{1}{c}{$\delta(\textrm{PDF})$}& \phantom{a}&
\multicolumn{1}{c}{$\delta(\alpha_s)$} 
\\ \midrule
120.0&&18.31 pb && ${}^{+0.83\textrm{pb}}_{-1.30\textrm{pb}}$(${}^{+4.51\%}_{-7.12\%}$)&& $\pm$ 0.35 pb ($\pm$ 1.89\%) && ${}^{+0.49\textrm{pb}}_{-0.49\textrm{pb}}$(${}^{+2.69\%}_{-2.67\%}$)  \\ \midrule 120.5&&18.16 pb && ${}^{+0.82\textrm{pb}}_{-1.29\textrm{pb}}$(${}^{+4.50\%}_{-7.11\%}$)&& $\pm$ 0.34 pb ($\pm$ 1.89\%) && ${}^{+0.49\textrm{pb}}_{-0.48\textrm{pb}}$(${}^{+2.69\%}_{-2.67\%}$)  \\ \midrule 121.0&&18.00 pb && ${}^{+0.81\textrm{pb}}_{-1.28\textrm{pb}}$(${}^{+4.49\%}_{-7.09\%}$)&& $\pm$ 0.34 pb ($\pm$ 1.89\%) && ${}^{+0.48\textrm{pb}}_{-0.48\textrm{pb}}$(${}^{+2.68\%}_{-2.66\%}$)  \\ \midrule 121.5&&17.85 pb && ${}^{+0.8\textrm{pb}}_{-1.26\textrm{pb}}$(${}^{+4.48\%}_{-7.07\%}$)&& $\pm$ 0.34 pb ($\pm$ 1.89\%) && ${}^{+0.48\textrm{pb}}_{-0.48\textrm{pb}}$(${}^{+2.68\%}_{-2.66\%}$)  \\ \midrule 122.0&&17.71 pb && ${}^{+0.79\textrm{pb}}_{-1.25\textrm{pb}}$(${}^{+4.47\%}_{-7.06\%}$)&& $\pm$ 0.33 pb ($\pm$ 1.89\%) && ${}^{+0.47\textrm{pb}}_{-0.47\textrm{pb}}$(${}^{+2.68\%}_{-2.66\%}$)  \\ \midrule 122.5&&17.56 pb && ${}^{+0.78\textrm{pb}}_{-1.24\textrm{pb}}$(${}^{+4.46\%}_{-7.04\%}$)&& $\pm$ 0.33 pb ($\pm$ 1.89\%) && ${}^{+0.47\textrm{pb}}_{-0.47\textrm{pb}}$(${}^{+2.68\%}_{-2.66\%}$)  \\
\end{tabular}
\end{center}
\end{table} 
\vspace*{4cm}
\begin{table}
\begin{center} 
\begin{tabular}{rcrcrcrcr}
\toprule
\multicolumn{1}{c}{$m_{H}$}& \phantom{a}&
\multicolumn{1}{c}{$\sigma$}& \phantom{a}&
\multicolumn{1}{c}{$\delta(\mbox{theory})$}& \phantom{a}&
\multicolumn{1}{c}{$\delta(\textrm{PDF})$}& \phantom{a}&
\multicolumn{1}{c}{$\delta(\alpha_s)$} 
\\ \midrule125.1&&16.82 pb && ${}^{+0.74\textrm{pb}}_{-1.17\textrm{pb}}$(${}^{+4.41\%}_{-6.96\%}$)&& $\pm$ 0.32 pb ($\pm$ 1.89\%) && ${}^{+0.45\textrm{pb}}_{-0.45\textrm{pb}}$(${}^{+2.67\%}_{-2.66\%}$)  \\ \midrule 125.2&&16.79 pb && ${}^{+0.74\textrm{pb}}_{-1.17\textrm{pb}}$(${}^{+4.40\%}_{-6.96\%}$)&& $\pm$ 0.32 pb ($\pm$ 1.89\%) && ${}^{+0.45\textrm{pb}}_{-0.45\textrm{pb}}$(${}^{+2.67\%}_{-2.66\%}$)  \\ \midrule 125.3&&16.77 pb && ${}^{+0.74\textrm{pb}}_{-1.17\textrm{pb}}$(${}^{+4.40\%}_{-6.96\%}$)&& $\pm$ 0.32 pb ($\pm$ 1.89\%) && ${}^{+0.45\textrm{pb}}_{-0.45\textrm{pb}}$(${}^{+2.67\%}_{-2.66\%}$)  \\ \midrule 125.4&&16.74 pb && ${}^{+0.74\textrm{pb}}_{-1.16\textrm{pb}}$(${}^{+4.40\%}_{-6.95\%}$)&& $\pm$ 0.32 pb ($\pm$ 1.89\%) && ${}^{+0.45\textrm{pb}}_{-0.44\textrm{pb}}$(${}^{+2.67\%}_{-2.66\%}$)  \\ \midrule 125.5&&16.71 pb && ${}^{+0.74\textrm{pb}}_{-1.16\textrm{pb}}$(${}^{+4.40\%}_{-6.95\%}$)&& $\pm$ 0.32 pb ($\pm$ 1.89\%) && ${}^{+0.45\textrm{pb}}_{-0.44\textrm{pb}}$(${}^{+2.67\%}_{-2.66\%}$)  \\ \midrule 125.6&&16.68 pb && ${}^{+0.73\textrm{pb}}_{-1.16\textrm{pb}}$(${}^{+4.40\%}_{-6.95\%}$)&& $\pm$ 0.32 pb ($\pm$ 1.89\%) && ${}^{+0.45\textrm{pb}}_{-0.44\textrm{pb}}$(${}^{+2.67\%}_{-2.66\%}$)  \\ \midrule 125.7&&16.66 pb && ${}^{+0.73\textrm{pb}}_{-1.16\textrm{pb}}$(${}^{+4.40\%}_{-6.94\%}$)&& $\pm$ 0.31 pb ($\pm$ 1.89\%) && ${}^{+0.44\textrm{pb}}_{-0.44\textrm{pb}}$(${}^{+2.67\%}_{-2.66\%}$)  \\
\end{tabular}
\end{center}
\caption{Gluon-fusion Higgs production cross-section at a proton-proton collider for $\sqrt{s}$ = 7 TeV. %The theory error includes scale variation, truncation, missing-N$^3$LO PDFs, electroweak, missing light-quark effects and missing NNLO finite-top mass uncertainties. We use the PDF4LHC set.
Details on the calculation of the theory error are given at the beginning of this Appendix and in the main text.
}
\label{tab:HXSWG7TeV}
\end{table}
\vspace*{4cm}\newpage

\begin{table}[!h]
\normalsize\setlength{\tabcolsep}{2pt}
\begin{center}
\begin{tabular}{rcrcrcrcr}
\multicolumn{9}{c}{$\sqrt{s}$ = 8 TeV}\\
\toprule
\multicolumn{1}{c}{$m_{H}$}& \phantom{a}&
\multicolumn{1}{c}{$\sigma$}& \phantom{a}&
\multicolumn{1}{c}{$\delta(\mbox{theory})$}& \phantom{a}&
\multicolumn{1}{c}{$\delta(\textrm{PDF})$}& \phantom{a}&
\multicolumn{1}{c}{$\delta(\alpha_s)$} 
\\ \midrule
120.0&&23.22 pb && ${}^{+1.05\textrm{pb}}_{-1.64\textrm{pb}}$(${}^{+4.54\%}_{-7.05\%}$)&& $\pm$ 0.44 pb ($\pm$ 1.88\%) && ${}^{+0.62\textrm{pb}}_{-0.61\textrm{pb}}$(${}^{+2.67\%}_{-2.64\%}$)  \\ \midrule 120.5&&23.03 pb && ${}^{+1.04\textrm{pb}}_{-1.62\textrm{pb}}$(${}^{+4.53\%}_{-7.04\%}$)&& $\pm$ 0.43 pb ($\pm$ 1.88\%) && ${}^{+0.61\textrm{pb}}_{-0.61\textrm{pb}}$(${}^{+2.66\%}_{-2.64\%}$)  \\ \midrule 121.0&&22.85 pb && ${}^{+1.03\textrm{pb}}_{-1.60\textrm{pb}}$(${}^{+4.52\%}_{-7.02\%}$)&& $\pm$ 0.43 pb ($\pm$ 1.87\%) && ${}^{+0.61\textrm{pb}}_{-0.60\textrm{pb}}$(${}^{+2.66\%}_{-2.64\%}$)  \\ \midrule 121.5&&22.66 pb && ${}^{+1.02\textrm{pb}}_{-1.59\textrm{pb}}$(${}^{+4.51\%}_{-7.01\%}$)&& $\pm$ 0.42 pb ($\pm$ 1.87\%) && ${}^{+0.60\textrm{pb}}_{-0.60\textrm{pb}}$(${}^{+2.66\%}_{-2.63\%}$)  \\ \midrule 122.0&&22.48 pb && ${}^{+1.01\textrm{pb}}_{-1.57\textrm{pb}}$(${}^{+4.50\%}_{-6.99\%}$)&& $\pm$ 0.42 pb ($\pm$ 1.87\%) && ${}^{+0.60\textrm{pb}}_{-0.59\textrm{pb}}$(${}^{+2.66\%}_{-2.63\%}$)  \\ \midrule 122.5&&22.30 pb && ${}^{+1.\textrm{pb}}_{-1.55\textrm{pb}}$(${}^{+4.49\%}_{-6.97\%}$)&& $\pm$ 0.42 pb ($\pm$ 1.87\%) && ${}^{+0.59\textrm{pb}}_{-0.59\textrm{pb}}$(${}^{+2.66\%}_{-2.63\%}$)  \\
\end{tabular}
\end{center}
\end{table} 
\vspace*{4cm}
\begin{table}
\begin{center} 
\begin{tabular}{rcrcrcrcr}
\toprule
\multicolumn{1}{c}{$m_{H}$}& \phantom{a}&
\multicolumn{1}{c}{$\sigma$}& \phantom{a}&
\multicolumn{1}{c}{$\delta(\mbox{theory})$}& \phantom{a}&
\multicolumn{1}{c}{$\delta(\textrm{PDF})$}& \phantom{a}&
\multicolumn{1}{c}{$\delta(\alpha_s)$} 
\\ \midrule125.1&&21.39 pb && ${}^{+0.95\textrm{pb}}_{-1.47\textrm{pb}}$(${}^{+4.43\%}_{-6.89\%}$)&& $\pm$ 0.40 pb ($\pm$ 1.87\%) && ${}^{+0.57\textrm{pb}}_{-0.56\textrm{pb}}$(${}^{+2.65\%}_{-2.63\%}$)  \\ \midrule 125.2&&21.36 pb && ${}^{+0.95\textrm{pb}}_{-1.47\textrm{pb}}$(${}^{+4.43\%}_{-6.89\%}$)&& $\pm$ 0.40 pb ($\pm$ 1.87\%) && ${}^{+0.57\textrm{pb}}_{-0.56\textrm{pb}}$(${}^{+2.65\%}_{-2.62\%}$)  \\ \midrule 125.3&&21.32 pb && ${}^{+0.94\textrm{pb}}_{-1.47\textrm{pb}}$(${}^{+4.43\%}_{-6.89\%}$)&& $\pm$ 0.40 pb ($\pm$ 1.87\%) && ${}^{+0.57\textrm{pb}}_{-0.56\textrm{pb}}$(${}^{+2.65\%}_{-2.62\%}$)  \\ \midrule 125.4&&21.29 pb && ${}^{+0.94\textrm{pb}}_{-1.47\textrm{pb}}$(${}^{+4.43\%}_{-6.88\%}$)&& $\pm$ 0.40 pb ($\pm$ 1.87\%) && ${}^{+0.56\textrm{pb}}_{-0.56\textrm{pb}}$(${}^{+2.65\%}_{-2.62\%}$)  \\ \midrule 125.5&&21.26 pb && ${}^{+0.94\textrm{pb}}_{-1.46\textrm{pb}}$(${}^{+4.42\%}_{-6.88\%}$)&& $\pm$ 0.40 pb ($\pm$ 1.87\%) && ${}^{+0.56\textrm{pb}}_{-0.56\textrm{pb}}$(${}^{+2.65\%}_{-2.62\%}$)  \\ \midrule 125.6&&21.22 pb && ${}^{+0.94\textrm{pb}}_{-1.46\textrm{pb}}$(${}^{+4.42\%}_{-6.88\%}$)&& $\pm$ 0.40 pb ($\pm$ 1.87\%) && ${}^{+0.56\textrm{pb}}_{-0.56\textrm{pb}}$(${}^{+2.65\%}_{-2.62\%}$)  \\ \midrule 125.7&&21.19 pb && ${}^{+0.94\textrm{pb}}_{-1.46\textrm{pb}}$(${}^{+4.42\%}_{-6.87\%}$)&& $\pm$ 0.40 pb ($\pm$ 1.87\%) && ${}^{+0.56\textrm{pb}}_{-0.56\textrm{pb}}$(${}^{+2.65\%}_{-2.62\%}$)  \\
\end{tabular}
\end{center}
\caption{Gluon-fusion Higgs production cross-section at a proton-proton collider for $\sqrt{s}$ = 8 TeV. %The theory error includes scale variation, truncation, missing-N$^3$LO PDFs, electroweak, missing light-quark effects and missing NNLO finite-top mass uncertainties. We use the PDF4LHC set.
Details on the calculation of the theory error are given at the beginning of this Appendix and in the main text.
}
\label{tab:HXSWG8TeV}
\end{table}
\vspace*{5cm}

\begin{table}[!h]
\normalsize\setlength{\tabcolsep}{2pt}
\begin{center}
\begin{tabular}{rcrcrcrcr}
\multicolumn{9}{c}{$\sqrt{s}$ = 13 TeV}\\
\toprule
\multicolumn{1}{c}{$m_{H}$}& \phantom{a}&
\multicolumn{1}{c}{$\sigma$}& \phantom{a}&
\multicolumn{1}{c}{$\delta(\mbox{theory})$}& \phantom{a}&
\multicolumn{1}{c}{$\delta(\textrm{PDF})$}& \phantom{a}&
\multicolumn{1}{c}{$\delta(\alpha_s)$} 
\\ \midrule
120.0&&52.22 pb && ${}^{+2.44\textrm{pb}}_{-3.59\textrm{pb}}$(${}^{+4.67\%}_{-6.88\%}$)&& $\pm$ 0.98 pb ($\pm$ 1.87\%) && ${}^{+1.37\textrm{pb}}_{-1.36\textrm{pb}}$(${}^{+2.63\%}_{-2.61\%}$)  \\ \midrule 120.5&&51.84 pb && ${}^{+2.41\textrm{pb}}_{-3.56\textrm{pb}}$(${}^{+4.65\%}_{-6.86\%}$)&& $\pm$ 0.97 pb ($\pm$ 1.87\%) && ${}^{+1.36\textrm{pb}}_{-1.35\textrm{pb}}$(${}^{+2.63\%}_{-2.60\%}$)  \\ \midrule 121.0&&51.46 pb && ${}^{+2.39\textrm{pb}}_{-3.52\textrm{pb}}$(${}^{+4.64\%}_{-6.85\%}$)&& $\pm$ 0.96 pb ($\pm$ 1.87\%) && ${}^{+1.35\textrm{pb}}_{-1.34\textrm{pb}}$(${}^{+2.63\%}_{-2.60\%}$)  \\ \midrule 121.5&&51.08 pb && ${}^{+2.37\textrm{pb}}_{-3.49\textrm{pb}}$(${}^{+4.64\%}_{-6.83\%}$)&& $\pm$ 0.95 pb ($\pm$ 1.86\%) && ${}^{+1.34\textrm{pb}}_{-1.33\textrm{pb}}$(${}^{+2.62\%}_{-2.60\%}$)  \\ \midrule 122.0&&50.71 pb && ${}^{+2.34\textrm{pb}}_{-3.46\textrm{pb}}$(${}^{+4.62\%}_{-6.82\%}$)&& $\pm$ 0.94 pb ($\pm$ 1.86\%) && ${}^{+1.33\textrm{pb}}_{-1.32\textrm{pb}}$(${}^{+2.62\%}_{-2.60\%}$)  \\ \midrule 122.5&&50.35 pb && ${}^{+2.32\textrm{pb}}_{-3.42\textrm{pb}}$(${}^{+4.61\%}_{-6.80\%}$)&& $\pm$ 0.94 pb ($\pm$ 1.86\%) && ${}^{+1.32\textrm{pb}}_{-1.31\textrm{pb}}$(${}^{+2.62\%}_{-2.59\%}$)  \\
\end{tabular}
\end{center}
\end{table} 
\vspace*{4cm}
\begin{table}
\begin{center} 
\begin{tabular}{rcrcrcrcr}
\toprule
\multicolumn{1}{c}{$m_{H}$}& \phantom{a}&
\multicolumn{1}{c}{$\sigma$}& \phantom{a}&
\multicolumn{1}{c}{$\delta(\mbox{theory})$}& \phantom{a}&
\multicolumn{1}{c}{$\delta(\textrm{PDF})$}& \phantom{a}&
\multicolumn{1}{c}{$\delta(\alpha_s)$} 
\\ \midrule125.1&&48.51 pb && ${}^{+2.21\textrm{pb}}_{-3.26\textrm{pb}}$(${}^{+4.56\%}_{-6.72\%}$)&& $\pm$ 0.90 pb ($\pm$ 1.86\%) && ${}^{+1.27\textrm{pb}}_{-1.25\textrm{pb}}$(${}^{+2.61\%}_{-2.58\%}$)  \\ \midrule 125.2&&48.44 pb && ${}^{+2.21\textrm{pb}}_{-3.25\textrm{pb}}$(${}^{+4.56\%}_{-6.72\%}$)&& $\pm$ 0.90 pb ($\pm$ 1.86\%) && ${}^{+1.26\textrm{pb}}_{-1.25\textrm{pb}}$(${}^{+2.61\%}_{-2.58\%}$)  \\ \midrule 125.3&&48.37 pb && ${}^{+2.20\textrm{pb}}_{-3.25\textrm{pb}}$(${}^{+4.55\%}_{-6.71\%}$)&& $\pm$ 0.90 pb ($\pm$ 1.86\%) && ${}^{+1.26\textrm{pb}}_{-1.25\textrm{pb}}$(${}^{+2.61\%}_{-2.58\%}$)  \\ \midrule 125.4&&48.30 pb && ${}^{+2.20\textrm{pb}}_{-3.24\textrm{pb}}$(${}^{+4.55\%}_{-6.71\%}$)&& $\pm$ 0.90 pb ($\pm$ 1.86\%) && ${}^{+1.26\textrm{pb}}_{-1.25\textrm{pb}}$(${}^{+2.61\%}_{-2.58\%}$)  \\ \midrule 125.5&&48.23 pb && ${}^{+2.20\textrm{pb}}_{-3.23\textrm{pb}}$(${}^{+4.55\%}_{-6.71\%}$)&& $\pm$ 0.89 pb ($\pm$ 1.85\%) && ${}^{+1.26\textrm{pb}}_{-1.24\textrm{pb}}$(${}^{+2.61\%}_{-2.58\%}$)  \\ \midrule 125.6&&48.16 pb && ${}^{+2.19\textrm{pb}}_{-3.23\textrm{pb}}$(${}^{+4.55\%}_{-6.70\%}$)&& $\pm$ 0.89 pb ($\pm$ 1.85\%) && ${}^{+1.26\textrm{pb}}_{-1.24\textrm{pb}}$(${}^{+2.61\%}_{-2.58\%}$)  \\ \midrule 125.7&&48.10 pb && ${}^{+2.19\textrm{pb}}_{-3.22\textrm{pb}}$(${}^{+4.55\%}_{-6.70\%}$)&& $\pm$ 0.89 pb ($\pm$ 1.85\%) && ${}^{+1.25\textrm{pb}}_{-1.24\textrm{pb}}$(${}^{+2.61\%}_{-2.58\%}$)  \\
\end{tabular}
\end{center}
\caption{Gluon-fusion Higgs production cross-section at a proton-proton collider for $\sqrt{s}$ = 13 TeV. %The theory error includes scale variation, truncation, missing-N$^3$LO PDFs, electroweak, missing light-quark effects and missing NNLO finite-top mass uncertainties. We use the PDF4LHC set.
Details on the calculation of the theory error are given at the beginning of this Appendix and in the main text.
}
\end{table}
\vspace*{5cm}

\begin{table}[!h]
\normalsize\setlength{\tabcolsep}{2pt}
\begin{center}
\begin{tabular}{rcrcrcrcr}
\multicolumn{9}{c}{$\sqrt{s}$ = 14 TeV}\\
\toprule
\multicolumn{1}{c}{$m_{H}$}& \phantom{a}&
\multicolumn{1}{c}{$\sigma$}& \phantom{a}&
\multicolumn{1}{c}{$\delta(\mbox{theory})$}& \phantom{a}&
\multicolumn{1}{c}{$\delta(\textrm{PDF})$}& \phantom{a}&
\multicolumn{1}{c}{$\delta(\alpha_s)$} 
\\ \midrule
120.0&&58.71 pb && ${}^{+2.75\textrm{pb}}_{-4.03\textrm{pb}}$(${}^{+4.69\%}_{-6.86\%}$)&& $\pm$ 1.10 pb ($\pm$ 1.87\%) && ${}^{+1.54\textrm{pb}}_{-1.53\textrm{pb}}$(${}^{+2.63\%}_{-2.61\%}$)  \\ \midrule 120.5&&58.29 pb && ${}^{+2.73\textrm{pb}}_{-3.99\textrm{pb}}$(${}^{+4.68\%}_{-6.85\%}$)&& $\pm$ 1.09 pb ($\pm$ 1.87\%) && ${}^{+1.53\textrm{pb}}_{-1.52\textrm{pb}}$(${}^{+2.63\%}_{-2.61\%}$)  \\ \midrule 121.0&&57.87 pb && ${}^{+2.70\textrm{pb}}_{-3.95\textrm{pb}}$(${}^{+4.67\%}_{-6.83\%}$)&& $\pm$ 1.08 pb ($\pm$ 1.87\%) && ${}^{+1.52\textrm{pb}}_{-1.51\textrm{pb}}$(${}^{+2.62\%}_{-2.60\%}$)  \\ \midrule 121.5&&57.45 pb && ${}^{+2.67\textrm{pb}}_{-3.92\textrm{pb}}$(${}^{+4.66\%}_{-6.81\%}$)&& $\pm$ 1.07 pb ($\pm$ 1.87\%) && ${}^{+1.51\textrm{pb}}_{-1.49\textrm{pb}}$(${}^{+2.62\%}_{-2.60\%}$)  \\ \midrule 122.0&&57.04 pb && ${}^{+2.65\textrm{pb}}_{-3.88\textrm{pb}}$(${}^{+4.65\%}_{-6.80\%}$)&& $\pm$ 1.07 pb ($\pm$ 1.87\%) && ${}^{+1.49\textrm{pb}}_{-1.48\textrm{pb}}$(${}^{+2.62\%}_{-2.60\%}$)  \\ \midrule 122.5&&56.64 pb && ${}^{+2.63\textrm{pb}}_{-3.84\textrm{pb}}$(${}^{+4.64\%}_{-6.78\%}$)&& $\pm$ 1.06 pb ($\pm$ 1.87\%) && ${}^{+1.48\textrm{pb}}_{-1.47\textrm{pb}}$(${}^{+2.62\%}_{-2.60\%}$)  \\
\end{tabular}
\end{center}
\end{table} 
\vspace*{4cm}
\begin{table}
\begin{center} 
\begin{tabular}{rcrcrcrcr}
\toprule
\multicolumn{1}{c}{$m_{H}$}& \phantom{a}&
\multicolumn{1}{c}{$\sigma$}& \phantom{a}&
\multicolumn{1}{c}{$\delta(\mbox{theory})$}& \phantom{a}&
\multicolumn{1}{c}{$\delta(\textrm{PDF})$}& \phantom{a}&
\multicolumn{1}{c}{$\delta(\alpha_s)$} 
\\ \midrule125.1&&54.60 pb && ${}^{+2.50\textrm{pb}}_{-3.66\textrm{pb}}$(${}^{+4.58\%}_{-6.70\%}$)&& $\pm$ 1.02 pb ($\pm$ 1.86\%) && ${}^{+1.42\textrm{pb}}_{-1.41\textrm{pb}}$(${}^{+2.61\%}_{-2.58\%}$)  \\ \midrule 125.2&&54.52 pb && ${}^{+2.50\textrm{pb}}_{-3.65\textrm{pb}}$(${}^{+4.58\%}_{-6.70\%}$)&& $\pm$ 1.01 pb ($\pm$ 1.86\%) && ${}^{+1.42\textrm{pb}}_{-1.41\textrm{pb}}$(${}^{+2.61\%}_{-2.58\%}$)  \\ \midrule 125.3&&54.45 pb && ${}^{+2.49\textrm{pb}}_{-3.65\textrm{pb}}$(${}^{+4.58\%}_{-6.70\%}$)&& $\pm$ 1.01 pb ($\pm$ 1.86\%) && ${}^{+1.42\textrm{pb}}_{-1.41\textrm{pb}}$(${}^{+2.61\%}_{-2.58\%}$)  \\ \midrule 125.4&&54.37 pb && ${}^{+2.49\textrm{pb}}_{-3.64\textrm{pb}}$(${}^{+4.58\%}_{-6.69\%}$)&& $\pm$ 1.01 pb ($\pm$ 1.86\%) && ${}^{+1.42\textrm{pb}}_{-1.4\textrm{pb}}$(${}^{+2.61\%}_{-2.58\%}$)  \\ \midrule 125.5&&54.29 pb && ${}^{+2.48\textrm{pb}}_{-3.63\textrm{pb}}$(${}^{+4.58\%}_{-6.69\%}$)&& $\pm$ 1.01 pb ($\pm$ 1.86\%) && ${}^{+1.42\textrm{pb}}_{-1.4\textrm{pb}}$(${}^{+2.61\%}_{-2.58\%}$)  \\ \midrule 125.6&&54.22 pb && ${}^{+2.48\textrm{pb}}_{-3.63\textrm{pb}}$(${}^{+4.57\%}_{-6.69\%}$)&& $\pm$ 1.01 pb ($\pm$ 1.86\%) && ${}^{+1.41\textrm{pb}}_{-1.4\textrm{pb}}$(${}^{+2.61\%}_{-2.58\%}$)  \\ \midrule 125.7&&54.14 pb && ${}^{+2.48\textrm{pb}}_{-3.62\textrm{pb}}$(${}^{+4.57\%}_{-6.68\%}$)&& $\pm$ 1.01 pb ($\pm$ 1.86\%) && ${}^{+1.41\textrm{pb}}_{-1.4\textrm{pb}}$(${}^{+2.61\%}_{-2.58\%}$)  \\
\end{tabular}
\end{center}
\caption{Gluon-fusion Higgs production cross-section at a proton-proton collider for $\sqrt{s}$ = 14 TeV. 
%The theory error includes scale variation, truncation, missing-N$^3$LO PDFs, electroweak, missing light-quark effects and missing NNLO finite-top mass uncertainties. We use the PDF4LHC set.
Details on the calculation of the theory error are given at the beginning of this Appendix and in the main text.}
\label{tab:HXSWG14TeV}
\end{table}
\vspace*{5cm}
\end{document}